\documentclass[12pt]{article}
\pdfoutput=1
\usepackage[a4paper]{geometry}

\usepackage{jheppub, amsmath,amssymb,amsfonts,amsxtra, mathrsfs, makeidx,graphics,graphicx,amsthm,epsfig,bm,longtable,float, color,tikz,mathtools,xfrac,footnote,rotating, lscape, makecell, environ,mathtools, empheq}

\usetikzlibrary{positioning}
\usetikzlibrary{chains}
\usetikzlibrary{arrows,fit,decorations.pathreplacing}
\tikzstyle{every picture}+=[remember picture]
\tikzstyle{na} = [baseline]

\usetikzlibrary{arrows, decorations.markings, calc, fadings, decorations.pathreplacing, patterns, decorations.pathmorphing, positioning}

\tikzstyle{every picture}+=[remember picture]
\tikzstyle{na} = [baseline=-.5ex]

\usepackage{bbm}
\usepackage{subfigure}

\newcommand{\Vol}{\mathrm{Vol}}

\newcommand{\Arg}{\mathrm{Arg}}

\newcommand{\rr}{\mathbb{R}}
\newcommand{\cc}{\mathbb{C}}

\newcommand{\None}{{\mathcal{N}=1}}
\newcommand{\Ntwo}{{\mathcal{N}=2}}
\newcommand{\Nfour}{{\mathcal{N}=4}}

\renewcommand{\Re}{\mathrm{Re}}

\DeclareMathOperator{\tr}{tr}
\DeclareMathOperator{\Tr}{Tr}
\DeclareMathOperator{\SU}{SU}
\DeclareMathOperator{\U}{U}

\numberwithin{equation}{section}

\newcommand{\be}{\begin{equation}}
\newcommand{\ee}{\end{equation}}
\newcommand{\bea}{\begin{equation} \begin{aligned}}
\newcommand{\eea}{\end{aligned} \end{equation}}

\newcommand\ddfrac[2]{\frac{\displaystyle #1}{\displaystyle #2}}

\usepackage[en-US,showdow,showzone]{datetime2}
\DTMlangsetup[en-US]{abbr=true,showyear=false,zone=atlantic}

\newcommand{\todo}[1]{}
\renewcommand{\todo}[1]{{\color{red} TODO: {#1}}}
\newcommand{\redtd}[1]{}
\renewcommand{\redtd}[1]{{\color{red} {#1}}}
\newcommand{\bluetd}[1]{}
\renewcommand{\bluetd}[1]{{\color{blue} {#1}}}

\def\tilde{\widetilde}

\def\hat{\widehat}

\def\rt2{\sqrt{2}}

\def\Re{\mathop{\rm Re}}

\def\Tr{\mathop{\rm Tr}}
\def\tr{\mathop{\rm tr}}

\def\CF{{\cal F}}

\def\1{{\ds 1}}

\def\SU{\mathrm{SU}}

\def\repa{\raise4pt\hbox{$\square$}\mkern-14mu\raise-4pt\hbox{$\square$}}
\def\repab{\overline{\raise4pt\hbox{$\square$}\mkern-14mu\raise-4pt\hbox{$\square$}\mkern-1mu}}

\def\smileface{\ensuremath{\hbox{\large$\bigcirc$}\mkern-15mu\raise-1pt\hbox{\scriptsize$\smallsmile$}%
\mkern-10mu\raise4pt\hbox{..}\mkern4mu}}
\def\frownface{\ensuremath{\hbox{\large$\bigcirc$}\mkern-15mu\raise-1pt\hbox{\scriptsize$\smallfrown$}%
\mkern-10mu\raise4pt\hbox{..}\mkern4mu}}

\DeclareMathOperator{\sign}{sign}

\newcommand{\ba}{\begin{array}}
\newcommand{\ea}{\end{array}}
\newcommand{\bi}{\begin{itemize}}
\newcommand{\ei}{\end{itemize}}
\def\vec#1{\bm{#1}}
\def\bea#1\eea{\allowdisplaybreaks \begin{align}#1\end{align}}
 \newcommand{\ben}{\begin{enumerate}}
\newcommand{\een}{\end{enumerate}}
\newcommand{\bean}{\begin{eqnarray*}}
\newcommand{\eean}{\end{eqnarray*}}
\newcommand{\eref}[1]{(\ref{#1})}

\newcommand{\PE}{\mathop{\rm PE}}

\newcommand{\BC}{\mathbb{C}}

\newcommand{\BZ}{\mathbb{Z}}

\newcommand{\comment}[1]{}

\newcommand{\fflat}{\mathcal{F}^\flat}

\definecolor{light-gray}{gray}{0.7}

\def\aup#1 {\overset{#1}{\uparrow} \, \overset{\tilde{#1}}{\downarrow}}

\tikzset{snake it/.style={decorate, decoration={snake, amplitude=.4mm, segment length=2mm,
                       post length=0mm,pre length=0mm}}}

\begin{document}


\begin{titlepage}

\begin{center}

\vskip .3in \noindent

{\Large \bf{New 3d $\mathcal{N}=2$ SCFT's with $N^{3/2}$ scaling}}

\bigskip

Antonio Amariti$^a$, Marco Fazzi$^b$, Noppadol Mekareeya$^{c,d}$, and Anton Nedelin$^b$

\bigskip

\bigskip
{\small 

$^a$ INFN, sezione di Milano, Via Celoria 16, I-20133 Milan, Italy\\
\vspace{.25cm}
$^b$ Department of Physics, Technion, 32000 Haifa, Israel\\
\vspace{.25cm}
$^c$ INFN, sezione di Milano-Bicocca, Piazza della Scienza 3, I-20126 Milan, Italy \\
\vspace{.25cm}
$^d$ Department of Physics, Faculty of Science, Chulalongkorn University, \\ Phayathai Road, Pathumwan, Bangkok 10330, Thailand	
}

\vskip .3cm
{\small \tt \href{mailto: antonio.amariti@mi.infn.it}{ antonio.amariti@mi.infn.it} \hspace{.5cm} \href{mailto:mfazzi@physics.technion.ac.il}{mfazzi@physics.technion.ac.il} \hspace{.5cm} \href{mailto:n.mekareeya@gmail.com}{n.mekareeya@gmail.com} \hspace{.5cm} \href{mailto:anton.nedelin@physics.uu.se}{anton.nedelin@physics.uu.se}
}

\vskip .6cm
     	{\bf Abstract }
\vskip .1in
\end{center}
We construct several novel examples of 3d $\mathcal{N}=2$ models whose free energy scales as $N^{3/2}$ at large $N$. This is the first step towards the identification of field theories with an M-theory dual. Furthermore, we match the volumes extracted from the free energy with the ones computed from the Hilbert series. We perform a similar analysis for the 4d $\None$ parents of the 3d models, matching the volume extracted from the $a$ central charge to that obtained from the Hilbert series. For some of the 4d models, we show the existence of a Sasaki--Einstein metric on the internal space of the candidate type IIB gravity dual.
\noindent
\vfill
\eject
\end{titlepage}
\tableofcontents

\section{Introduction}
\label{sec:intro}

The number of degrees of freedom of a stack of 
$N$ coincident M2-branes is expected to scale as $N^{3/2}$ at large $N$
\cite{Klebanov:1996un}. 
Finding conformal field theories with such a scaling is
a smoking gun in the search of a gravitational dual description
as predicted by the AdS$_4$/CFT$_3$ correspondence.
On the field theory side, the large-$N$ scaling is extracted from 
the free energy on a three-sphere, $F_{S^3}$.
Many examples of models with the $N^{3/2}$  scaling have by now been found in the literature (see e.g. 
\cite{Drukker:2010nc,Drukker:2011zy,Hanany:2008fj, Davey:2009sr, Davey:2009qx, Davey:2009et, Herzog:2010hf,Martelli:2011qj,Cheon:2011vi,Jafferis:2011zi,Amariti:2011uw,Marino:2011eh,Gulotta:2011aa,Gulotta:2011vp, Davey:2011mz, Gulotta:2012yd,Crichigno:2012sk,Amariti:2012tj}).
They correspond to supersymmetric quiver gauge theories with Chern--Simons (CS) interactions.\footnote{There exist other realizations of the AdS$_4$/CFT$_3$
correspondence in terms of CS quivers with a $N^{5/3}$ scaling \cite{Aharony:2010af} that we will not consider here. Neither will we comment on the  $N^2 \log N$ scaling of \cite{Assel:2012cp}.}
The models obtained so far can be organized in three classes.
\begin{itemize}
\item The first class corresponds to quiver gauge theories with a \emph{vector-like} field content. 
This means that each pair of gauge nodes is connected by a bifundamental and an 
anti-bifundamental matter field.
These quivers  have the same structure of the ones engineering the $L^{aba}$ singularities in four dimensions \cite{Benvenuti:2005ja,Butti:2005sw,Franco:2005sm}.
The difference is that in the 3d action there are CS terms.
The gauge group is $\prod_{a=1}^r \U(N_a)_{k_a}$, with $N_a =N$ and $\sum_{a=1}^{r} k_a = 0$. 

\item The second class of models corresponds to flavored vector-like quivers. The flavors
are in the fundamental and in the anti-fundamental representation of the gauge group and
 the vector-like structure is not imposed on them (but rather on the gauge bifundamentals). 
The product gauge group is still $\prod_{a=1}^r \U(N_a)_{k_a}$ with $N_a =N$. In this case the sum of the \emph{bare} levels $\sum_{a=1}^r k_a \neq 0$, whereas that of the effective levels vanishes \cite{benini-closset-cremonesi1}.
\item 
A third class of models corresponds to quivers with an ADE structure, studied in  \cite{Gulotta:2011vp,Crichigno:2012sk,Jain:2019lqb}.
In this case the links in the (affine) Dynkin diagram correspond to pairs of 
bifundamental and anti-bifundamental fields 
and the gauge group is $ \prod_{a=1}^r \U(N \theta_a)_{k_a}$,
where $\theta_a$ corresponds to the 
Coxeter label for each node of the Dynkin diagram, and 
$\sum_{a=1}^r \theta_a k_a = 0$. Observe that in this case there are no
adjoint matter fields at any gauge node.
\end{itemize}
As discussed above, in order to expand the AdS$_4$/CFT$_3$ landscape,
it is desirable to find further classes of models with $N^{3/2}$
scaling.
Indeed this is a necessary step to extend the AdS$_4$/CFT$_3$ 
correspondence to gravity solutions that do not have 
yet a field theoretic counterpart. Moreover, it can guide the
search for new solutions on the supergravity side.

Motivated by such a necessity, in this paper we identify a new class of 
models exhibiting the sought-after $N^{3/2}$ scaling.
It is comprised of CS-matter quiver gauge theories with a vector-like field content,
gauge group $\prod_{a=1}^r \U(N_a)_{k_a} $, and
\begin{equation}\label{eq:ranklevel}
\sum_{a=1}^r {N_a k_a} = 0\ , \quad N_a = n_a N\ .
\end{equation}
This is similar to the ADE cases, with the important difference that here we will also have adjoint matter fields.
We show that the free energy of such models scales as $N^{3/2}$ and we provide an algorithm to compute it in the presence of extra adjoint fields.
This algorithm generalizes  the one  of  \cite{Gulotta:2011vp,Crichigno:2012sk},
where it was first noticed that 
a split in the eigenvalues of the matrix model
is induced by the varying ranks of the gauge groups.
 We will show that in presence of charged adjoint matter 
the prescription of  \cite{Gulotta:2011vp,Crichigno:2012sk} requires some 
modifications.
We  apply our new algorithm to a series of models of increasing complexity,
showing the $N^{3/2}$ scaling for each of them.

In many of the models under investigation  the R-symmetry is not fixed to its superconformal value,
and the free energy is a function of the R-charges of the various fields,
satisfying only the superpotential constraints.
The exact R-symmetry is obtained by maximizing the free energy in terms of 
these R-charges \cite{Jafferis:2010un}. 
Under the holographic correspondence this maximization translates into 
the minimization of the volume of the dual internal manifold \cite{martelli-sparks-yau-volmin}.
The volume can be computed from the field theory data 
by calculating the Hilbert series \cite{Cremonesi:2016nbo} and by extracting 
its leading order behavior w.r.t. the 
fugacity of the R-symmetry \cite{martelli-sparks-yau-volmin} (see also \cite{Nekrasov:2003rj}).
This calculation, when applied to models with an M-theory  holographic dual,
gives the volume of the fourfold probed by the M2-branes 
\cite{Hanany:2008fj}.
In the examples that we study here we observe that the volume extracted from the
Hilbert series coincides with the one obtained from the free energy, 
suggesting a holographic interpretation for all the quivers under investigation.

This paper is organized as follows.
In Section \ref{sec:genproblem} we discuss the general aspects of the calculation of the
three-sphere free energy at large $N$, summarizing our algorithm.
In Section \ref{sec:laufer} we apply the algorithm to a quiver (known as Laufer's theory) with two gauge groups and adjoint matter,
and show the matching between the volume extracted from this calculation 
and the one obtained from the Hilbert series.
In Section \ref{sec:CS} we repeat the analysis for many other models, with 
varying ranks and charged adjoint matter.
In each case we find agreement between the field theory calculation of the free energy and the volume obtained from the Hilbert series.
In Section \ref{4dholo} we discuss a possible 4d holographic interpretation
of the models discussed in Sections \ref{sec:laufer} and \ref{sec:CS}. We show the agreement between the volumes extracted from the Hilbert series and from the $a$ conformal anomaly, the absence of leading $N^2$ contributions to the gravitational anomaly $c-a$, and prove, when possible, the existence of a Sasaki--Einstein (SE) metric on the internal space $B_5$ of the putative AdS$_5 \times B_5$ gravity dual. We present further directions of investigation and some speculations in Section \ref{sec:further}.

In Appendix \ref{families} we discuss possible generalizations of our construction
to infinite families of quivers with varying gauge group ranks and adjoint matter. In Appendix \ref{AppAnton} we provide further details on the matrix models discussed
in the main body of the paper. 
In Appendix \ref{apptor} we compute the volumes of a 3d model discussed 
in Section \ref{sec:CS}, whose 4d parent is obtained via a Seiberg duality transformation.
After applying the duality, the model can in fact be studied with the usual geometric techniques of toric geometry, and we can check our results against the ones obtained with this well-known approach. 
In Appendix \ref{appNoppie} we discuss various aspects and details of the calculations of the Hilbert series appearing in the paper.

\section{The matrix model}
\label{sec:genproblem}

The counting of the scaling properties of
the degrees of freedom of QFT's in even spacetime dimensions
is associated with the calculation of the conformal anomalies, as
the Virasoro central charge in 2d or the coefficient of the Euler
density in 4d known as $a$. In odd dimensions a more sophisticated quantity
is needed, because of the absence of conformal anomalies.
In 3d a good candidate is provided by the free energy computed on
the three sphere. However this is a rather complex quantity
and its calculation is very nontrivial for interacting QFT's.
A breakthrough, in the supersymmetric case, has been made
by applying exact mathematical techniques, commonly referred to as
supersymmetric localization \cite{Pestun:2007rz}.

It was first shown in \cite{Kapustin:2009kz} that the $S^3$ partition function
can be reduced to a matrix integral in the case of $\mathcal{N}\geq 3$
supersymmetry.
The $\mathcal{N}=2$ case was then tackled in \cite{Jafferis:2010un,Hama:2010av}.
At large $N$ these matrix integrals further simplify for
some specific classes of models, corresponding to
vector-like quivers with adjoint matter fields
and the extra condition $\sum_a k_a =0$.
For these classes of models the free energy scales as $N^{3/2}$
at large $N$, and they have been shown in \cite{Herzog:2010hf,Jafferis:2011zi} to satisfy the relation
\begin{equation}\label{volF}
\Vol (\text{SE}_7) = \frac{2 \pi^6}{27} \frac{1}{F_{S^3}^{2}}\ ,
\end{equation}
where Vol(SE$_7$) is the volume of the seven-dimensional manifold probed by $N$ coincident M2-branes at a singularity,
in the AdS/CFT correspondence.\footnote{More precisely, the Calabi--Yau fourfold singularity admits a conical metric,
hence its seven-dimensional base $B_7$ admits a Sasaki--Einstein metric, whose associated volume is computed by the field theory free energy under the AdS/CFT correspondence.}

The analysis has been performed also for  the ADE models, where the gauge ranks are not all coincident.
In this case, as shown in \cite{Crichigno:2012sk,Gulotta:2011vp,Jain:2015kzj},  the matrix model techniques of \cite{Herzog:2010hf,Jafferis:2011zi} have to be modified,
taking into account the so-called \emph{bifurcation of the eigenvalues}.
In this Section we will reconsider the approach of \cite{Crichigno:2012sk,Gulotta:2011vp,Jain:2015kzj}, extending the analysis
from the ADE class  to other classes of models,
with different gauge ranks and crucially with adjoint matter fields of arbitrary R-charge.

We will see that the presence of adjoint matter fields requires a modification of the algorithm of \cite{Crichigno:2012sk,Gulotta:2011vp},
which will play an important role in our analysis.\footnote{Another modification, that can be easily generalized to ours, was introduced very recently in \cite{Jain:2019lqb} in order to consider
${\cal N}=2$ $D$-type quivers.}

\subsection{The setup}
\label{subsec:setting}

We consider solutions to the matrix models of 3d ${\cal N}=2$ quiver gauge theories with varying ranks of the gauge nodes.
In particular we consider theories with gauge group $\prod_a \U\left( n_a N \right)$ (i.e. the ranks of the gauge groups are characterized by having $\gcd=N$). The technique we will present here is applicable at large $N$.
We consider bifundamental matter fields connecting two nodes, denoted $a$ and $b$, with corresponding R-charge $\Delta_{ab}$, and adjoint matter fields for some of the gauge nodes with R-charge $\Delta$.


The $S^3$ partition function of a general ${\cal N}=2$ theory can be written, using methods of supersymmetric localization
\cite{Pestun:2007rz,Pestun:2016zxk,Willett:2016adv,Kapustin:2009kz}, in the form of the following matrix model
\begin{equation}
Z=\int\prod\limits_{a=1}^r\prod\limits_{i=1}^{n_a N}d\lambda_{a,i}\exp\left( -F\left[ \{\lambda\} \right]\right)\equiv e^{-F}\,,
\label{partition:function:gen}
\end{equation}
where $r$ is total number of nodes in the quiver, and in the last equality we define the free energy $F$ of theory.
The form of the free energy functional $F\left[ \{\lambda\} \right]$ depends on the
theory under investigation. The various contributions to the free energy, the numerical
analysis, the so-called long-range force cancellation and the large $N$ approximation have been studied in \cite{Jafferis:2011zi}. We review the main properties of their analysis in Appendix \ref{AppAnton} for the convenience of the reader.

In the following we will show the general algorithm for solving the matrix model.
In the case with equal ranks it reduces to the one of \cite{Jafferis:2011zi}; if the nodes have different ranks but no adjoint matter fields are present, it can be shown to coincide with the one of \cite{Gulotta:2011vp,Crichigno:2012sk,Jain:2019lqb}.

\subsection{Solution to the matrix model}
\label{sub:gen}

In this Section we will briefly explain the general algorithm for solving the matrix model, as reviewed in Appendix \ref{AppAnton}.

In the large-$N$ approximation, we evaluate the matrix integral (\ref{partition:function:gen}) by solving the saddle-point
   equation
   \be
     \frac{d F\left[ \{\lambda\} \right]}{d\lambda_{a,i}}=0\ .
     \label{saddle:point:body}
   \ee
   Solving this equation can be thought of as finding the equilibrium positions of $\sum_a N_a$ particles, with the eigenvalues $\lambda_i^{(a)}$ representing the positions. In this language, the derivatives of the free energy functional
        in the saddle-point equation (\ref{saddle:point:body}) translate into forces acting on the eigenvalues. There can be external (or central potential)
        forces coming from the CS term, as well as interaction forces (see \eqref{eigenvalue:forces}). We are interested in theories
          with $N^{3/2}$ behavior of the free energy as discussed above. It is known \cite{Jafferis:2011zi} that such a scaling takes place when the \emph{long-range force cancellation} is achieved in the matrix model. More concretely, in the limit of large separations of eigenvalues $|\lambda_i^{(a)}-\lambda_j^{(b)}|\gg1$, the leading order of the forces appearing in the saddle-point equation cancels, and we should focus on the subleading terms. As we show in Appendix \ref{AppAnton}, for the class of theories we consider
          in this paper the long-range force cancellation condition leads to the following Ansatz for the eigenvalues $\lambda_i^{(a)}$ of the matrix model:
         \be
         \lambda_i^{(a,I)}=N^\alpha\, x_i+iy_i^{(a,I)}\,,\quad a=1,\dots,r\,,~ ~I=1,\dots,n_a\,,~ ~i=1,\dots,N\ .
         \label{eigenvalue:ansatz:body}
         \ee
        Here we split the eigenvalues of each node $a$ into $n_a$ groups of eigenvalues of size $N$ (labeled by the index $I$). The above Ansatz then tells us that the eigenvalues of each group have the same real part, which scales with $N$, but different imaginary parts $y_i$'s which do not scale with $N$. Then the long-range force cancellation becomes possible and it translates into a simple algebraic equation for the parameters of the quiver:
      \begin{equation}
      \sum\limits_{b\in (a,b)}n_b\left( \Delta_{ab}+\Delta_{ba} \right)+2n_a\sum\limits_{i\in \mathrm{adj}~a}\Delta_i^{(a)}=2\left[ n_a\,\left(n^{(a)}_{\mathrm{adj}}-1\right) +\sum\limits_{b\in (a,b)}n_b\right]\ .
      \label{long:range:force:condition:body}
      \end{equation}
Moreover, we will be interested in solving the saddle-point equation \eqref{saddle:point:body} in the \emph{M-theory limit}, whereby one takes $N$ to be large while keeping the CS levels $k_a$ fixed. In this limit we can introduce the continuous variable $s=i/N$ and eigenvalue distributions $x_i=x\left( i/N \right),\, y^{(a,I)}_i=y\left( i/N \right)$.
      Then the normalized eigenvalue density is given by
      \begin{equation}
      \rho(x)=\frac{d s}{d x}\,,\quad  \int dx \rho(x)=1\,.
      \label{eigenvalue:density:bulk}
      \end{equation}
Once the long-range force cancellation in the continuous limit is achieved, we are left with expressions \eqref{free:energy:CS:largeN}-\eqref{free:energy:vec:largeN}, giving the contributions of different terms to the free energy.
   At the saddle point, the central forces coming from the CS terms is balanced by the interaction forces coming from the various multiplets. We can see from those expressions that this balance is possible only when $\alpha=1/2$, implying that the contributions coming from CS terms and multiplets are of the same order $N^{3/2}$. This is the common scaling for all theories that achieve long-range force cancellation.

\begin{figure}
      \includegraphics[width=0.5\linewidth]{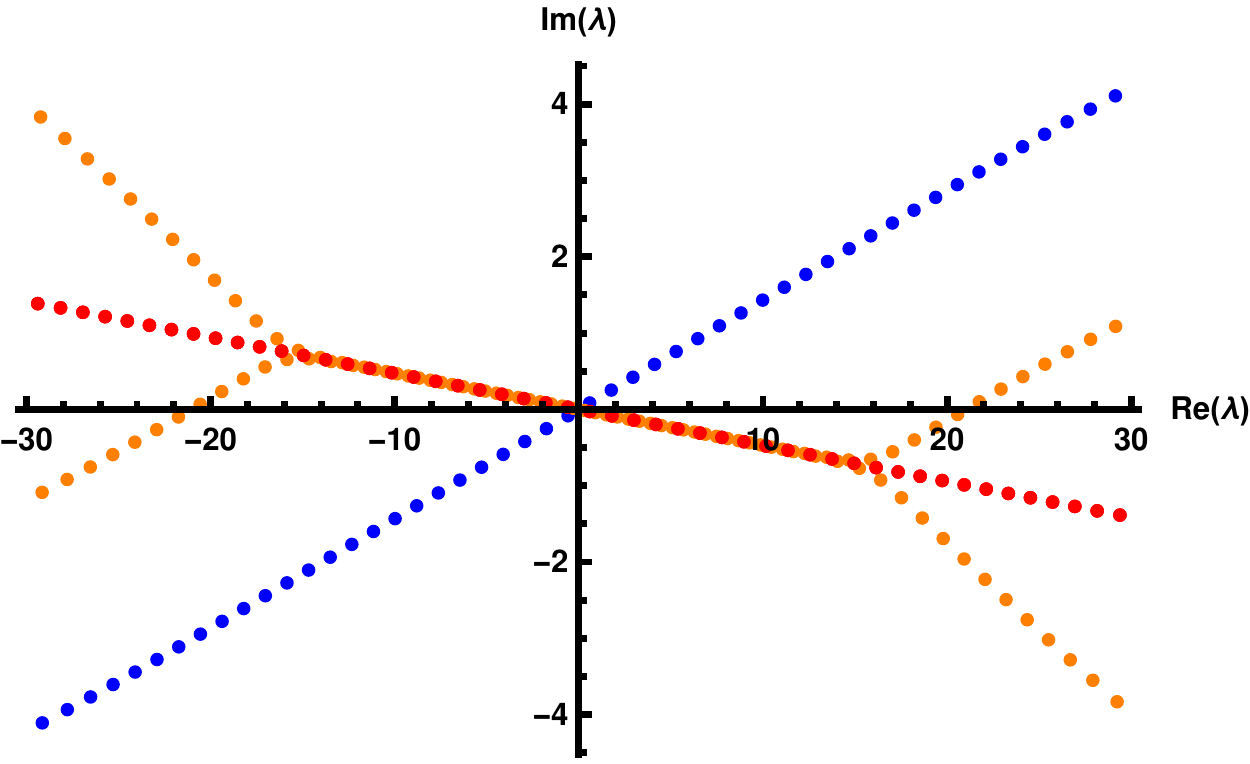}
      \includegraphics[width=0.5\linewidth]{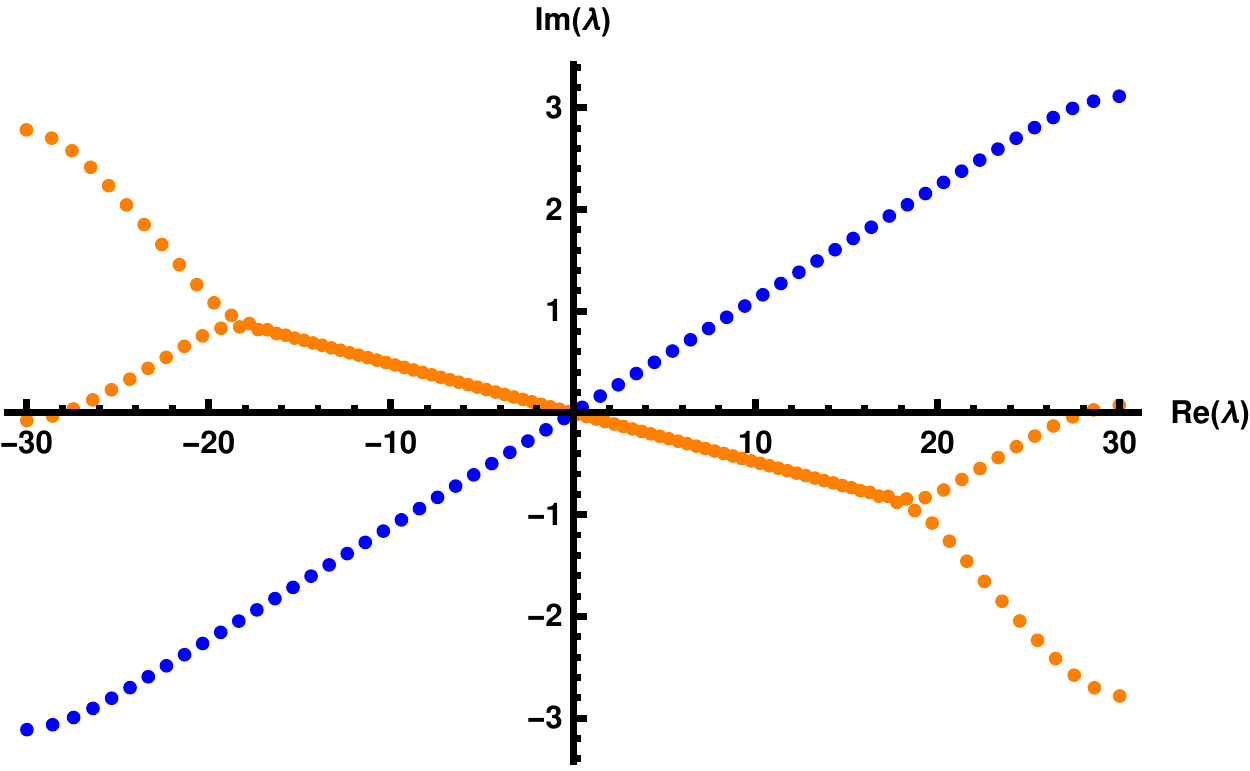}
      \caption{Examples of the eigenvalue distributions for $D_4$ (left) and Laufer's (right) theory (that we will study in Section \ref{sec:laufer}). In both plots the bifurcation of the orange cut (corresponding to the eigenvalues of one of the nodes) takes place.}
      \label{bifurcation-example}
    \end{figure}

Another useful observation was made in \cite{Gulotta:2011vp,Crichigno:2012sk}. The authors noticed that the matrix
   models coming from quiver theories with varying ranks of the gauge groups develop a \emph{bifurcation of the cuts}. This means that the eigenvalues corresponding to the same node $a$ organize themselves along curves that bifurcate at certain points. Some examples of numerical solutions
   showing this behavior are shown in Figure \ref{bifurcation-example}. In particular we show the example of the $D_4$ theory considered in \cite{Gulotta:2011vp} and of Laufer's theory, to which we will devote Section \ref{sec:laufer}.
    In both cases one of the nodes (whose eigenvalues are shown in orange in the plots) has rank twice bigger than the other nodes. As can be seen from the plots, at a certain point the cut corresponding to the eigenvalues of this node splits into two. This happens when, for some pair of eigenvalue groups, $\delta y_{(aI,bJ)}\equiv |y^{(a,I)}-y^{(b,J)}|$ goes outside the principal value of $\Arg$ functions contributing to the free energy (see Appendix \ref{AppAnton} for more details). The particular boundary value of this
   difference depends on a particular pair of fields, as well as their R-charges. In what follows, we will specify these boundary values of (the imaginary part of the) differences for each theory we consider.

We are now ready to formulate the general algorithm necessary to solve the matrix models.
   \begin{enumerate}
     \item First, motivated by the picture obtained for the numerical solution, we write down the free energy functional using expressions \eqref{free:energy:CS:largeN}-\eqref{free:energy:vec:largeN} assuming that all differences $\delta y_{(aI,bJ)}$ are inside the principal value region, and also that $\delta y_{(aI,aJ)}=0$ for all nodes $a$. We also add a chemical potential term, so as to get
        \begin{equation}
        F_\mu\left[\rho,\{y\},\{\Delta\},\mu \right]\equiv F\left[ \rho,\{y\},\{\Delta\} \right]-\frac{\mu}{2\pi}\int dx\,\rho(x)\,.
        \label{free:energy:chem}
        \end{equation}
The free energy functional above depends on the eigenvalue density $\rho(x)$, and the \emph{differences} of the eigenvalue imaginary parts $\delta y_{(aI,bJ)}(x)$.

\item Then we solve the saddle-point equations by varying the free energy functional \eqref{free:energy:chem} w.r.t. the eigenvalue density $\rho(x)$ and functions $y_{(a,I)}(x)$:
        \begin{equation}
          \frac{\delta F_\mu}{\delta \rho}=0\,,\quad \frac{\delta F_\mu}{\delta y_{(aI,bJ)}}=0\,,\quad I,J=1,\dots,n_a,\,~ ~ a,b=1,\dots,r\ .
        \end{equation}
Moreover we impose the constraint \eqref{eig:dens:norm}. This allows us to determine one of the endpoints $x^*_1$ or $x^*_2$ of the distribution support in terms of the chemical potential $\mu$.

    \item Next we should substitute the obtained solution back into the free energy functional $F\left[ \rho,\{y\},\{\Delta\} \right]$ to determine its on-shell value as a function of $\mu$ and the undetermined endpoint of the distribution support.\footnote{Often, forces acting on the eigenvalues are symmetric w.r.t. the $x \to -x$ reflection, which results in a symmetric distribution of the eigenvalues. In this case the normalization condition (\ref{eig:dens:norm}) fixes the position of the endpoints as a function of $\mu$, and no other parameters are left.} We extremize the on-shell free energy $F_\text{on-shell}$
        w.r.t. the chemical potential $\mu$ and the position of the undetermined endpoint, say $x_1^*$:
        \begin{equation}
          \frac{dF_{\text{on-shell}}}{d\mu}=0\,,\quad \frac{dF_\text{on-shell}}{d x^*_1}=0\,.
        \end{equation}

        Notice that the opposite can equivalently be done. We can express $\mu$ in terms of $x^*_{1,2}$. Then the free energy is just a function of the endpoint positions.
      \item Using the extremal value of $\mu$ and $x^*_1$ we can now find the position of the second endpoint $x^*_2$. Then we check if at any of the endpoints one (or more) among the $\delta y_{(aI,bJ)}$ goes outside the principal value of the $\Arg$ function. In case it does not, this is the final solution.
      \item In case one or more of the $\delta y_{(aI,bJ)}$ is outside the principal value of the $\Arg$ functions, we find the point, call it $x_1$, where it saturates the principal value. E.g. let us say the free energy integrand contains the function $\Arg\left( e^{i\delta y_{(aI,bJ)}} \right)$. (This is the simplest possible case). Then the point where the condition $|\delta y_{(aI,bJ)}(x_1)|=\pi$ is attained saturates the principal value. We assume that for $x<x_1$   the solutions found previously are correct, and for $x>x_1$ we assume that $\delta y_{(aI,bJ)}$ saturates the
        principal value of $\Arg$, i.e. $\delta y_{(aI,bJ)}=\pi$. Also inspired by the numerics, we conjecture that for $x>x_1$ the bifurcation of the cuts corresponding to the nodes $a$ and $b$ (or at least for one of them) happens and $\delta y_{(aI,aJ)}\neq 0$, $\delta y_{(bI,bJ)}\neq 0$, i.e. the bifurcation of the cuts takes place at these points.

        Using these assumptions we write down the free energy functional once again, and subsequently go through steps 1-4.

      \item The algorithm should be applied up to the point where extremization will result in the endpoint position where none of $\delta y_{(aI,bJ)}$ go outside the principal value of the $\Arg$ functions contributing to the free energy. This can also happen when all of $\delta y_{(aI,bJ)}$ saturate these values, and hence will be just constant functions in the last iteration of the computation.
   \end{enumerate}

   It is worth noticing that our algorithm is more complicated than the one used for $D$-type quivers in \cite{Gulotta:2011vp,Crichigno:2012sk,Jain:2015kzj,Jain:2019lqb}. In the latter case the authors do not go through the step-wise extremization procedure, and just terminate the eigenvalue density at the point where all of the $\delta y_{(aI,bJ)}$ saturate the boundaries of the principal value regions.
   However we have observed that, if the theory includes adjoint multiplets, the eigenvalue distribution can go beyond the point where all of the $\delta y_{(aI,bJ)}$ are saturated, and terminates only when the eigenvalue density becomes zero. On the other hand the algorithm we presented above is a direct generalization of the one used in 
 \cite{Jafferis:2011zi} and is universal, i.e. it does not depend on the matter and gauge content of the quiver. We should also mention that the extremization procedure can be technically difficult, and
 for complicated quivers the simplified algorithm similar to the one used in \cite{Gulotta:2011vp,Crichigno:2012sk,Jain:2015kzj,Jain:2019lqb} can be applied. However for all of the theories studied in this paper we have been successful in exploiting the prescription summarized above.


\section{A detailed example: Laufer's theory} 
\label{sec:laufer}

\begin{figure}[ht!]
       \begin{center}
       \includegraphics[width=0.5\linewidth]{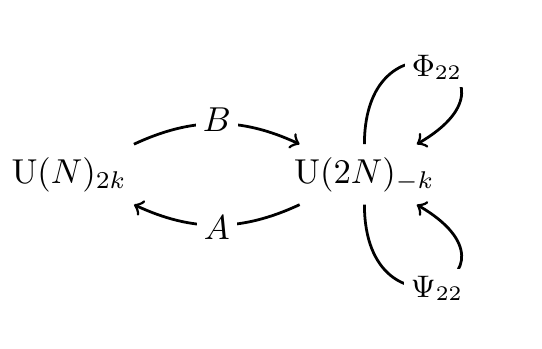}
      \caption{Quiver for Laufer's 3d theory.}
      \end{center}
      \label{fig:laufer3d}
    \end{figure}

In this Section we study the free energy at large $N$ of the CS-matter quiver depicted in Figure \ref{fig:laufer3d}, with $\U(N)_{2k}\times \U(2N)_{-k}$  gauge group.  The superpotential can be written schematically as follows:\footnote{In Appendix \ref{LauferApp} we will provide a more detailed
structure of this superpotential discussing the presence of
multi-trace terms.}
\begin{equation}
W=BA\Phi_{22}^2-\frac{1}{4}\Phi_{22}^4+\Phi_{22}\Psi_{22}^2-\frac{1}{2}\left( BA \right)^2\ .
      \label{superpotential:laufer}
\end{equation}
$A$ and $B$ are two bifundamental fields of R-charges $\Delta_A$ and $\Delta_B$ connecting the nodes of the quiver, and $\Phi_{22},\Psi_{22}$ are two adjoints of R-charges $\Delta_{1,2}$ respectively, based at the second node of the quiver.
We refer to this model as \emph{Laufer's theory}. Observe that the polynomial in the adjoint fields appearing in this superpotential is of the same form
as the one discussed in \cite{Brodie:1996vx} for the generalization of 4d Seiberg duality to the case with two adjoints. We note that it would be interesting to study the three-dimensional version of such a duality applied to Laufer's theory.

This quiver was obtained in \cite{aspinwall-morrison} following a procedure that associates a quiver with relations (the F-terms of the superpotential \eqref{superpotential:laufer}) to a CY$_3$ singularity. In Laufer's case, the singularity is the hypersurface \cite{laufer}
\begin{equation}\label{eq:laufer}
x^2+y^3+w z^2+w^3 y =0\ \subset\ \cc^4\ ,
\end{equation}
which is singular at the origin. As customary, the singularity can be probed by $N$ D3-branes, and the 4d field theory is the one living on the worldvolume of the stack. Aspects of this 4d $\None$ quiver gauge theory (i.e. without CS interactions), such as the classical moduli space and how the geometry is encoded in the quiver, were thoroughly analyzed in \cite{collinucci-fazzi-valandro}.\footnote{The same theory can also be obtained by having D5-branes wrap a certain vanishing cycle of the local threefold geometry $\cc^2/D_4 \hookrightarrow  \text{CY}_3 \to \cc$, where the local K3 singularity is nontrivially fibered over the complex line \cite{Cachazo:2001gh}.}

Here we will study the 3d $\Ntwo$ version of the theory, obtained by replacing the standard kinetic terms for the vector multiplets with CS interactions, such that condition \eqref{eq:ranklevel} holds. The 3d theory contains a CY$_4$ among the various branches of its moduli space, which we study in Appendix \ref{appNoppie}. It is not a complete intersection, and is generated by a quantum relation satisfied by the monopole operators of GNO charge $\pm 1$. (A generalization of this theory with three adjoint fields will be studied in Section \ref{sub:laufer-n}.)

\subsection{The free energy}

Let us apply  the algorithm presented in Section \ref{sub:gen} to Laufer's theory, and extract the large $N$ behavior of the free energy, which will exhibit the $N^{3/2}$ scaling.
       Notice that the form of the superpotential written above forces the R-charges to satisfy
      the following conditions:
      \begin{equation}
      \Delta_A=\frac{1}{2}-b\,,\quad \Delta_B=\frac{1}{2}+b\,,\quad \Delta_1\equiv \Delta_{\Phi_{22}}=\frac{1}{2}\,,\quad \Delta_2\equiv \Delta_{\Psi_{22}}=\frac{3}{4}\,,
      \label{r:charges:lauf:cond}
      \end{equation}
It is important to emphasise that $b$ turns out to be zero, as we show explicitly in Appendix \ref{appNoppie} around \eref{volmin}.  If the gauge group were taken to be $\SU(N) \times \SU(2N)$, there would be a baryonic symmetry whose charge is denoted by $b$.  However, when the gauge group is taken to be $\U(N) \times \U(2N)$, as in our present discussion, such a symmetry can be reabsorbed into the topological symmetry in the same way as discussed in \cite{Aharony:2008ug}; hence there is no baryonic symmetry.  The R-charges of $A$ and $B$ therefore take their canonical values, i.e. $1/2$.

\subsubsection*{Long-range force cancellation}

      First of all let us check that the long-range forces cancel and hence the free energy of the theory exhibits the expected $N^{3/2}$ scaling.
The conditions (\ref{long:range:force:condition:body}) translate into:
\begin{equation}
\begin{split}
& \text{node 1:} \quad \Delta_A+\Delta_B=1\ , \\
& \text{node 2:} \quad \Delta_A+\Delta_B+4\left( \Delta_1+\Delta_2 \right)=6\ .
\end{split}
\end{equation}
We can easily see that the conditions written above are consistent with the constraints (\ref{r:charges:lauf:cond}) on the R-charges coming
      from the superpotential (\ref{superpotential:laufer}). Hence the long-range force cancellation is achieved.

\subsubsection*{Free energy functional}

      Since the long-range force cancellation takes place we can list all terms of order $N^{3/2}$ contributing to the free energy:
\begin{align}      \label{free:energy:lauf:gen}
\frac{F_{S^3}}{N^{3/2}} =&\!\! \int\!\! dx \! \left[\frac{kx}{2\pi}\rho(x)\left( \delta y_{1,2}+\delta y_{1,3} \right)\! + \frac{1}{2}\rho(x)^2\! \left( \frac{29}{8}\pi^2- \Arg\left(e^{-i \delta y_{1,2}}\right) - \Arg\left(e^{-i \delta y_{1,3}}\right)\!+ \right. \right. \nonumber  \\
&+ 2\Arg\left(e^{i\pi+i \delta y_{3,2}}\right)+\Arg\left(e^{i \delta y_{3,2}}\right)+\frac{1}{6\pi}\left(2\Arg\left(e^{\frac{i\pi}{2}-i\delta y_{3,2} }\right)
           +\frac{3\pi}{2}+3\delta y_{3,2}\right)\cdot \nonumber \\
&\cdot \left(\pi^2-\Arg\left(e^{\frac{i\pi}{2}-i\delta y_{3,2} }\right)^2 \right) +\frac{1}{6\pi}\left(2\Arg\left(e^{\frac{i\pi}{2}+i\delta y_{3,2} }\right)
           +\frac{3\pi}{2}-3\delta y_{3,2}\right) \cdot \nonumber \\
&\cdot \left.\left. \left(\pi^2-\Arg\left(e^{\frac{i\pi}{2}+i\delta y_{3,2} }\right)^2 \right)\right) \right]\ .
\end{align}      
We have introduced the notation
      \begin{equation}
 \delta y_{1,2}\equiv \delta y_{(11,21)}(x)\,,~ \delta y_{1,3}\equiv y_{(11,22)}(x)\,,~ \delta y_{3,2}\equiv y_{(22,21)}(x)\,.
      \end{equation}
       In order for the arguments of the $\Arg$ functions to stay in the principal value region, the following relations for the $\delta y$'s should be satisfied:
       \begin{equation}
       |\delta y_{1,2}|\leq \pi\,,\quad |\delta y_{1,3}|\leq \pi\,,\quad |\delta y_{3,2}|\leq \pi\ .
       \label{lauf:1/2:dy:rel}
       \end{equation}
In fact this can be directly checked in the expressions contributing to the free energy functional (\ref{free:energy:lauf:gen}).

\subsubsection*{Solution}

We can now find an extremal value of the free energy functional (\ref{free:energy:lauf:gen}) using the algorithm presented in Section \ref{sub:gen}.
The free energy functional (\ref{free:energy:lauf:gen}) at $\Delta=1/2$ has the symmetry $\rho(-x)\to \rho(x)\,,\quad \delta y(-x)\to -\delta y(x)$. Therefore this should also be a symmetry of the solution for this $\Delta$, as can be seen from the numerics in Figure \ref{bifurcation-example}. Below we consider only the
      solution for positive $x$, since the solution for negative $x$ can be reconstructed using this symmetry.

\begin{enumerate}
\item As explained in Section \ref{sub:gen}, we start by assuming $\delta y_{3,2}=0$ so that $\delta y_{1,3}=\delta y_{1,2}$. In this case the free energy functional
       depends only on $\rho(x)$ and $\delta y_{1,2}(x)$, and equals
       \begin{equation}
       \frac{F_1}{N^{3/2}}=\int dx\left[\frac{kx}{\pi}\rho(x) \delta y_{1,2}+\rho(x)^2\left(\frac{21 \pi^2}{8}-\delta y_{1,2}  \right) \right]\,.
       \label{free:en:lauf:12:1}
       \end{equation}
       The solution to the extremization problem for the functional $F_\mu=F_1-\frac{\mu}{2\pi}\int dx\rho(x)$ gives
       \begin{equation}
       \rho_1(x)=\frac{2\mu}{21 \pi^3}\,,\quad \delta y_{1,2}(x)=\frac{21 k \pi^2 x}{4\mu}\,.
       \label{rho:lauf:12:1}
       \end{equation}
       The normalization condition (\ref{eig:dens:norm}) leads to the following endpoints:
       \begin{equation}
       x_2^*=-x_1^*=\frac{21 \pi^3}{4\mu}\, .
        \label{endp:lauf:12:1}
       \end{equation}
       Substituting the solution (\ref{rho:lauf:12:1}) into the free energy functional (\ref{free:en:lauf:12:1}) and integrating between the points
       (\ref{endp:lauf:12:1}) we get the free energy as a function of the chemical potential $\mu$:
       \begin{equation}
       \frac{F_1}{N^{3/2}}=\frac{3087 k^2 \pi^7}{128 \mu^3}+\frac{\mu}{4\pi}\,.
       \label{free:en:os:lauf:12:1}
       \end{equation}
       Finally, minimizing w.r.t. $\mu$ we get:
       \begin{equation}
       \mu=\frac{21^{3/4} \sqrt{k}\pi^2}{2^{5/4}}\,.
       \label{lauf:12:extremmu:1}
       \end{equation}
       With this $\mu$ we obtain $\delta y_{1,2}(x_2^*)=\sqrt{\frac{21}{8}}\pi$ which is larger than the boundary value $\delta y_{1,2}=\pi$. Hence we need
       to confine our solution inside
       \begin{equation}
       |x|<\frac{4\mu}{21 \pi k}\,,
        \label{lauf:1/2:region1}
       \end{equation}
       and find a new solution for $x$ outside this region.

\item
       We focus on the region $|x|>\frac{4\mu}{21\pi k}$. We assume that at the boundary point the bifurcation takes place as can be motivated from the results of the numerics shown in Figure \ref{bifurcation-example}. Hence in this region we assume that $\delta y_{3,2}\neq 0$ and $\delta y_{1,2}=\pi$. With these assumptions
       the free energy functional (\ref{free:energy:lauf:gen}) becomes:
       \begin{equation}
       \frac{F_2}{N^{3/2}}=\int dx\left[\frac{4 k x}{\pi}\rho(x)\left( 2\pi-\delta y_{3,2}\right) +\rho(x)^2\left(\frac{13 \pi^2}{8}+\pi \delta y_{3,2}-
       \frac{1}{4}\delta y_{3,2}^2  \right) \right]\,,
       \label{free:en:lauf:12:2}
       \end{equation}
       where we have also assumed that $\delta y_{3,2}<0$.\footnote{It can be checked that with the opposite assumption the solution is not continuous at $x=\frac{4\mu}{21 \pi k}$.} The solution to the extremization problem for the functional $F_\mu=F_2-\frac{\mu}{2\pi}\int dx\rho(x)$ then gives
       \begin{equation}
       \rho_2(x)=\frac{2\mu}{21 \pi^3}\,,\quad \delta y_{3,2}(x)=2\pi-\frac{21 k \pi^2 x}{2\mu}\,.
       \label{rho:lauf:12:2}
       \end{equation}
A good check of the solution's validity is continuity condition with solution (\ref{rho:lauf:12:1}), which is valid in the previous region. It can be checked
       that indeed at the border of the two regions (i.e. at $x=\frac{4\mu}{21 \pi k}$) all the solutions are continuous. The normalization condition (\ref{eig:dens:norm}) leads to
       \begin{equation}
       x_2^*=-x_1^*=\frac{21 \pi^3}{4\mu}\, .
        \label{endp:lauf:12:2}
       \end{equation}
       Substituting the solutions \eqref{rho:lauf:12:2}, \eqref{rho:lauf:12:1} into the free energy functionals \eqref{free:en:lauf:12:2}, \eqref{free:en:lauf:12:1}, 
       and integrating between the points (\ref{endp:lauf:12:2}) we get the free energy as a function of the chemical potential $\mu$:
       \begin{equation}
       \frac{F_2}{N^{3/2}}=\frac{3087 k^2 \pi^7}{128 \mu^3}+\frac{\mu}{4\pi}\,,
       \label{free:en:os:lauf:12:2}
       \end{equation}
       which appears to be exactly the same as the on-shell free energy in the previous step. Therefore it will lead
       to the same extremum value of $\mu$, which was given in (\ref{lauf:12:extremmu:1}).
       With this $\mu$ we get $\delta y_{3,2}(x_2^*)=(4-\sqrt{42}\pi/2)$ which is smaller then the boundary value $\delta y_{3,2}=-\pi$. Hence we need
       to confine our solution inside
       \begin{equation}
       \frac{4\mu}{21 \pi k}<x<\frac{2\mu}{7\pi k}\,,
        \label{lauf:1/2:region2}
       \end{equation}
       and find a new solution for $x$ outside this region once again.

\item
       Now we go to the region $x> \frac{2\mu}{7\pi k}$ where we assume that all $\delta y$'s are fixed. In particular $\delta y_{1,2}=\pi$, $\delta y_{2,3}=-\pi$.
       Then the free energy functional (\ref{free:energy:lauf:gen}) depends only on the eigenvalue density $\rho(x)$ and is given by
        \begin{equation}
        \frac{F_3}{N^{3/2}}=\int dx\left[\frac{3 k x}{2}\rho(x)+\frac{3\pi^2 \rho(x)^2}{8}\right]\, .
       \label{free:en:lauf:12:3}
       \end{equation}
The solution to the extremization problem for the functional $F_\mu=F_3-\frac{\mu}{2\pi}\int dx\rho(x)$ then gives
       \begin{equation}
       \rho_3(x)=\frac{2\mu-6\pi k x}{3 \pi^3}\,.
       \label{rho:lauf:12:3}
       \end{equation}

       Once again we check the continuity conditions with solution (\ref{rho:lauf:12:2}) in the previous region. It can be checked
       that indeed at the border of the two regions $x=\frac{2\mu}{7 \pi k}$ the solution is continuous. The normalization condition (\ref{eig:dens:norm}) leads to
       \begin{equation}
       x_2^*=-x_1^*=\frac{14\mu-\sqrt{52\mu^2-882 k\pi^4 }}{42 \pi k}\, .
        \label{endp:lauf:12:3}
       \end{equation}
       Substituting the solutions (\ref{rho:lauf:12:3}), (\ref{rho:lauf:12:2}) and (\ref{rho:lauf:12:1}) into the  free energy functionals
       (\ref{free:en:lauf:12:3}), (\ref{free:en:lauf:12:2}), (\ref{free:en:lauf:12:1}) respectively,
       and integrating between the points (\ref{endp:lauf:12:3}) we get the free energy as a function of the chemical potential $\mu$:
       \begin{equation}
       \frac{F}{N^{3/2}}=\frac{13 \mu ^2 \left(\sqrt{52 \mu ^2-882 \pi ^4 k}-14 \mu \right)}{18522 \pi ^5 k}-\frac{\sqrt{52 \mu ^2-882 \pi ^4 k}-42 \mu }{84 \pi }\,.
       \label{free:en:os:lauf:12:3}
       \end{equation}
        Minimizing this expression w.r.t. $\mu$ we obtain:
        \begin{equation}
        \mu=21\pi^2\sqrt{\frac{k}{26}}\,.
        \label{muextrem:lauf:12:3}
        \end{equation}
        One needs to check if with this choice of $\mu$ the eigenvalue density at the endpoints is real and non-negative. Indeed it appears that 
        \begin{equation}
        \rho(x_2^*)=0 \quad \text{at}\quad x_2^*=-x_1^*=\frac{7\pi}{\sqrt{26 k}}\ .
        \label{rho:lauf:12:endp}
        \end{equation}
        So the endpoints of the solution appear to be exactly the zeros of the eigenvalue density. We remark that a detailed analysis of this and many other cases considered in Section \ref{sec:CS} shows that this happens every time we have an adjoint matter field.\footnote{Therefore in more complicated cases, when the prescription presented above cannot be implemented, one should be able to  avoid the  step-wise extremization procedure. Instead it is reasonable to assume that the saturation always happens, and the solution continues from one region to another up to the point where the eigenvalue density becomes zero.}
\end{enumerate}

\subsubsection*{Summary}
\begin{figure}[ht!]
    \centering
      \includegraphics[width=0.45\linewidth]{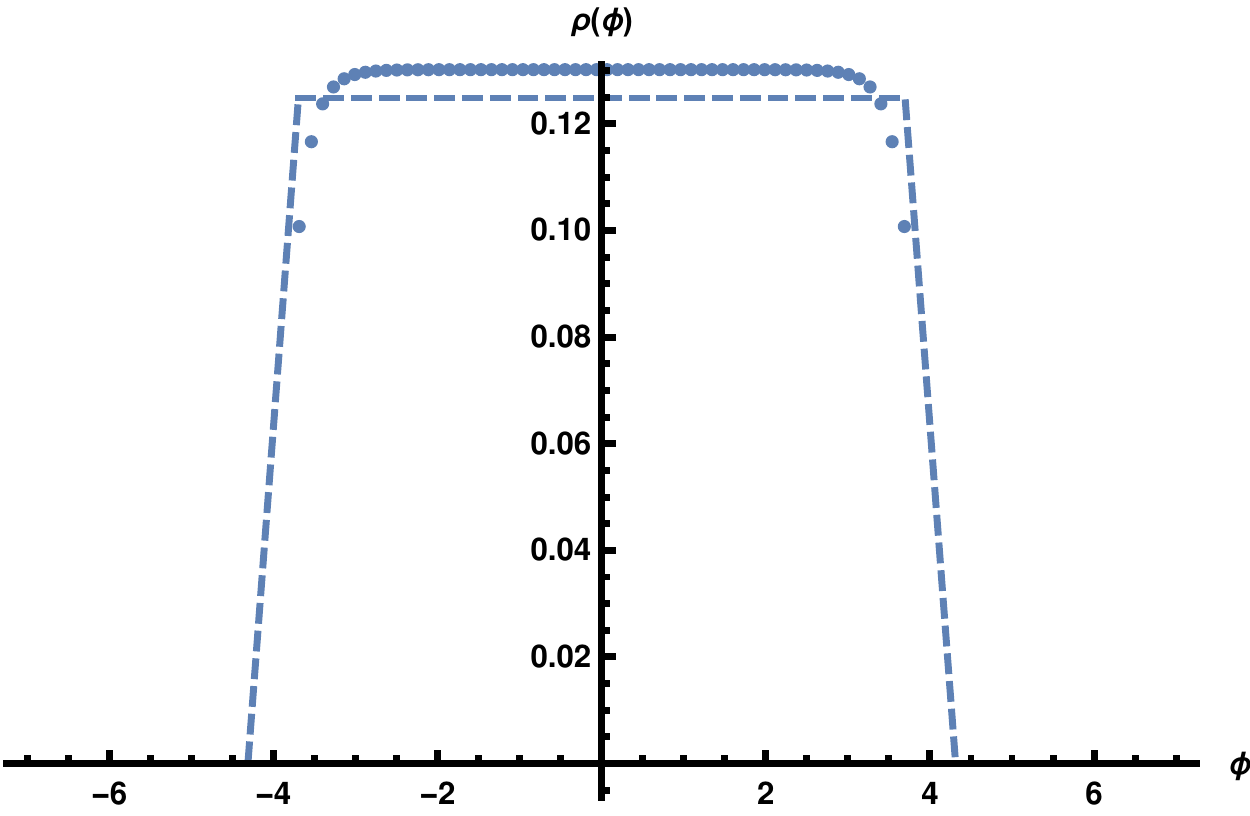}
      \includegraphics[width=0.45\linewidth]{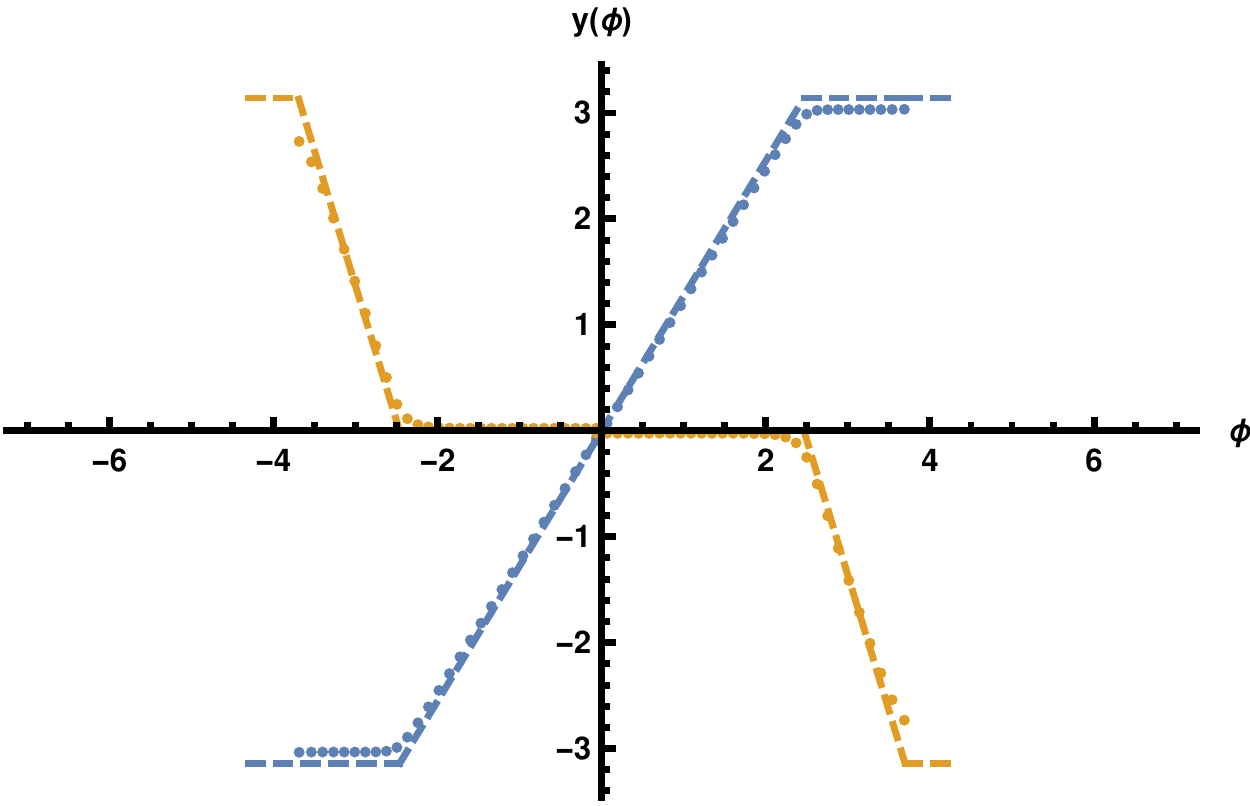}
      \caption{Numerical (dots) and analytical (dashed lines) solutions for Laufer's theory for $\Delta=\tfrac{1}{2}$, $k=1$, and $N=60$. On the left we show the eigenvalue density $\rho(x)$, while on the right we show $\delta y_{1,2}$ (in blue) and $\delta y_{3,2}$ (in orange).}
      \label{fig:lauf:12}
    \end{figure}

        Using the extremum value (\ref{muextrem:lauf:12:3}) of the chemical potential $\mu$ we can write the final expression combining solutions
        (\ref{rho:lauf:12:1}), (\ref{rho:lauf:12:2}) and (\ref{rho:lauf:12:3}) in different regions:

  \begin{equation}
  \begin{array}{c|c|c|c}
  \text{region}
  &\rho(x) & \delta y_{12} & \delta y_{32}\\
  \hline
 -\frac{7\pi}{\sqrt{26 k}}<x<  -\frac{3 \pi \sqrt 2}{\sqrt{13 k}}
 &
\frac{7}{\pi}\sqrt{\frac{2k}{13}}+\frac{2kx}{\pi^2},
  &
-\pi
  &
\pi
  \\
  -\frac{3 \pi \sqrt 2}{\sqrt{13 k}}<x<  -\frac{2 \pi \sqrt 2}{\sqrt{13 k}}
&
\frac{1}{\pi}\sqrt{\frac{2k}{13}}
&
-\pi
&
-2\pi-x\sqrt{\frac{13k}{2}}
\\
 -\frac{2 \pi \sqrt 2}{\sqrt{13 k}}<x< \phantom{-} \frac{2 \pi \sqrt 2}{\sqrt{13 k}}
 &
 \frac{1}{\pi}\sqrt{\frac{2k}{13}}
 &\frac{x}{2}\sqrt{\frac{13k}{2}}
 &0\\
\phantom{-}
\frac{2 \pi \sqrt 2}{\sqrt{13 k}}<x<
\phantom{-}
\frac{3 \pi \sqrt 2}{\sqrt{13 k}}
&
\frac{1}{\pi}\sqrt{\frac{2k}{13}}
&
\pi
&-2\pi+x\sqrt{\frac{13k}{2}}
\\
\phantom{-}\frac{3 \pi \sqrt 2}{\sqrt{13 k}}<x<
\phantom{-}
\frac{7\pi}{\sqrt{26 k}}
&
\frac{7}{\pi}\sqrt{\frac{2k}{13}}-\frac{2kx}{\pi^2}
&\pi
&-\pi
\end{array}
\end{equation}
We can also check this solution against the results of the numerical calculation. The results of this match are shown in Figure \ref{fig:lauf:12}. Due to the relatively small value of $N$ (we took $N=60$), the regions close to the endpoints of the distribution are not so well pronounced (since very few eigenvalues can be found there).\footnote{This effect is especially evident in the plot of $\delta y_{3,2}$, which only starts curving towards the endpoints of the distribution.} However up to this subtlety at the endpoints of the distribution the numerical result correctly reproduces the analytical solution.

We can finally write down the free energy of Laufer's theory corresponding to the extremized value (\ref{free:en:os:lauf:12:3}):
    \begin{equation}
    F_{S^3}=N^{3/2}\,7\pi\sqrt{\frac{k}{26}}\ .
    \label{free:en:lauf:12:fin}
    \end{equation}
    From this value we can extract an expression for the volume of the SE$_7$ base of the CY$_4$ at large $N$:
 \begin{equation}
 \label{volLAUF}
\Vol(\text{SE}_7) =\frac{\pi^4}{k} \, \frac{2^2 \, 13}{3^3 \, 7^2} ~.
    \end{equation}
We have also computed the Hilbert series of the CY$_4$ in Appendix \ref{LauferApp}. For $k=1$ at the fixed point we obtain:
    \begin{equation}
H(t) = \frac{1-t^2-t^3+t^4+t^5+t^6-t^8+t^9-t^{11}-t^{12}-t^{13}+t^{14}+t^{15}-t^{17}}
{(1-t^2) (1-t^3)(1-t^6)(1-t^7)^2}\ .
    \end{equation}
(We emphasize that in this expression $t$ is the fugacity that counts the R-charge in the unit of $1/4$.)
 By considering the limit $\frac{\pi^4}{48} \lim_{s \rightarrow 0} s^4\,H(e^{-\frac{1}{4}s})$,
 we can reproduce  (\ref{volLAUF}), corroborating the holographic interpretation.

\section{Case studies}
\label{sec:CS}

In this Section we extend the analysis of the free energy at large $N$ for 
other  quiver gauge theories with bifundamental and 
adjoint matter fields, and with product gauge group
$\prod_{a=1}^{r} \U(N_a)_{k_a}$ satisfying
\begin{equation}
\sum_{a=1}^{r} N_a k_a=0\ , \quad N_a = n_a N\ .
\end{equation}
We show that the free energy of these models scales as $N^{3/2}$.
Furthermore, in order to test the holographic interpretation 
we extract the volume by using formula (\ref{volF}) and 
test it against the volume obtained from the Hilbert series.
We refer the reader to Appendix \ref{appNoppie} for the details on the 
Hilbert series and its relation with the volumes.
In order to simplify the reading we collect our results in Table \ref{tab:3dmodels}, where we show the quiver, the ranks and the levels, and the off-shell behavior of the free energy for the various models.
\begin{table}[ht!]
\centering
\begin{tabular}{c|ccc}
Section & \begin{tabular}{c} Quiver \& \\ ranks $\{N_a\}_{a=1}^r$ \end{tabular} & CS levels & Free energy $ F_{S^3}/N^{3/2}$ \\[15pt]
\hline
\rule{0pt}{30pt} \ref{sec:laufer} &
\begin{tabular}{c}
\includegraphics[scale=.5]{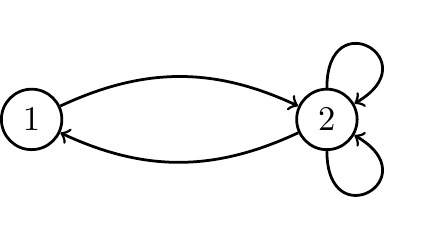} \\ $(N,2N)$
\end{tabular}
&
$(2k,-k)$
&     
$ 7\pi \sqrt{\frac{k}{26}}$
\\[15pt]
\hline
\rule{0pt}{30pt} \ref{C3Z2Z2}
 &
\begin{tabular}{c}
\!\!\!\!\!\!\!
\includegraphics[scale=.5]{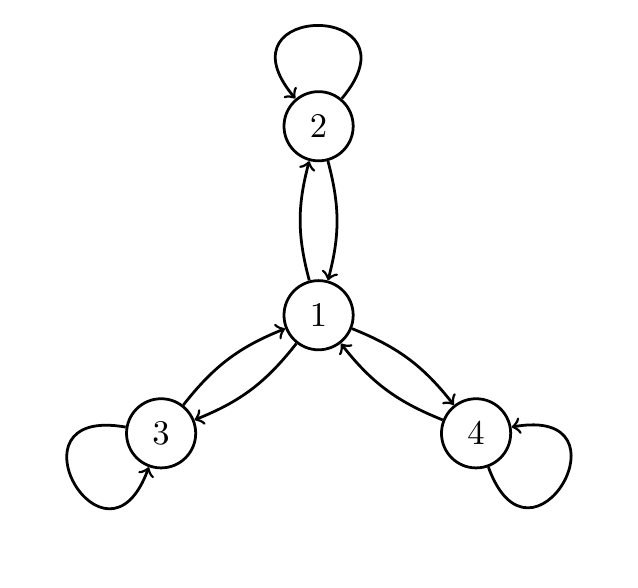}
\!\!\!\!\!\!\!\!\!\!
\end{tabular}
&   
$(k,-2k,0,0)$
& 
$
 \frac{2\pi  \Delta _3 \Delta _4}{3}  \sqrt{\frac{2 \left(4-\Delta _3-\Delta _4\right) k}{\Delta _3+\Delta _4}}
$    
\\ 
& $(2N,N,N,N)$ 
&  
$(k,0,-k,-k)$
&
$ \frac{\pi  (4-\Delta_3-\Delta_4) \sqrt{k \Delta_3\Delta_4  (\Delta_3+\Delta_4)} }{3}  
$
\\[5pt]
\hline
\rule{0pt}{30pt} \ref{genlaufer}
 &
\begin{tabular}{c}
\includegraphics[scale=.5]{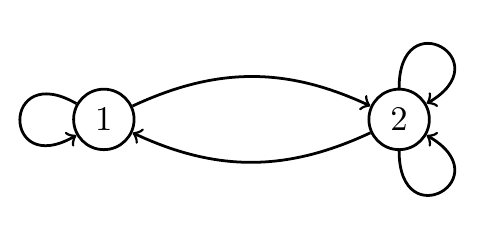} \\ $(N,2N)$
\end{tabular}
&
$(2k,-k)$
&     
$ \frac{2  \pi  (6-5 \Delta ) (2-\Delta ) }{3} \sqrt{\frac{2 \Delta  k}{10-7 \Delta }}$
\\[15pt]
 \hline
\rule{0pt}{30pt} \ref{case1}
 &
\begin{tabular}{c}
\includegraphics[scale=.5]{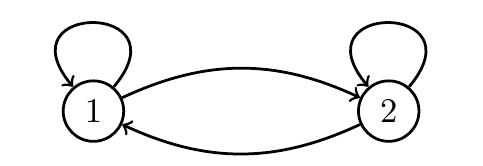} \\ $(N,2N)$
\end{tabular}
& 
$(2k,-k)$
 &    
$ \frac{2  \pi  (2-\Delta ) (4-3 \Delta )  }{3}\sqrt{\frac{2 \Delta  k}{8-5 \Delta }}$                 
\\[15pt]
\hline   
\rule{0pt}{20pt} &
 &  
$(2k,-k,0)$
 &     
$ \frac{16  \pi  (1-\Delta ) (2-\Delta ) }{3}  \sqrt{\frac{\Delta  k}{6-5 \Delta }}$
\\
\rule{0pt}{30pt} \ref{case2}
&
\begin{tabular}{c}
\includegraphics[scale=.45]{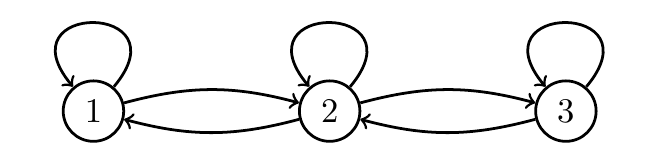} \\ $(N,2N,N)$
\end{tabular}  
& 
$(k,-k,k)$
&     
$\frac{8  \pi  (1-\Delta ) (2-\Delta ) }{3}\sqrt{\frac{2 \Delta  k}{4-3 \Delta }}$
\\ 
 &
&           
$(k,0,-k)$
&     
$ \frac{8  \pi  (1-\Delta ) (2-\Delta ) }{3}  \sqrt{\frac{2 \Delta  k}{4-3 \Delta }}$
\\[5pt]
\hline
\rule{0pt}{30pt} \ref{sub:laufer-n}
 &
\begin{tabular}{c}
\includegraphics[scale=.5]{CASEIV.pdf} \\ $(N,2N)$
\end{tabular}
&
$(2k,-k)$
&     
$\frac{2\pi(2n+1) (6n+1)}{3(1+n)^2}  \sqrt{\frac{2k}{(3+10 n)}}$
\end{tabular}
\caption{Summary of results for the new 3d $\Ntwo$ models introduced in the paper. We show the quiver, the different ranks $N_a$ (the progressive labeling of the nodes in the quiver is given by $a$), the various CS assignments $k_a$ we considered such that $\sum_a N_a k_a=0$, and finally the three-sphere free energy exhibiting the $N^{3/2}$ scaling. The latter was obtained analytically by solving the matrix model according to the rules explained in Section \ref{sec:genproblem}.}
\label{tab:3dmodels}
\end{table}

\subsection{A model with 4d toric parent}
\label{C3Z2Z2}

The first model that we study corresponds to a non-toric dual phase of a 4d toric quiver gauge theory.
The 4d parent (obtained by forgetting the CS interactions) corresponds to a $\mathbb{Z}_2 \times \mathbb{Z}_2$ orbifold of $\mathcal{N}=4$ SYM.
The toric gauge theory corresponds to a product of four $\SU(N)$
gauge groups connected by bifundamental matter fields
$Q_{ij}$ with $i,j=1,4$ and $i \neq j$. (Henceforth we will refer to a bifundamental connecting the $i$-th and
$j$-th node as $Q_{ij}$, and to an adjoint based at the $i$-th node as $\Phi_{ii}$.) The superpotential reads:
\begin{align}
W &= Q_{12} Q_{23} Q_{31}
-Q_{31} Q_{14} Q_{43} 
+Q_{14} Q_{42} Q_{21}
 -Q_{21} Q_{13} Q_{32} \ +
 \nonumber \\
  &\quad +
 Q_{13} Q_{34} Q_{41}
 -Q_{41}  Q_{12} Q_{24}
-Q_{23} Q_{34} Q_{42}
+Q_{32} Q_{24} Q_{43}\ .
 \end{align}
 Here we are interested in a 3d version of this model, where we
 consider $\U(N)$ instead of $\SU(N)$ gauge factors. 
\begin{figure}[hb!]
\centering
\includegraphics[scale=.8]{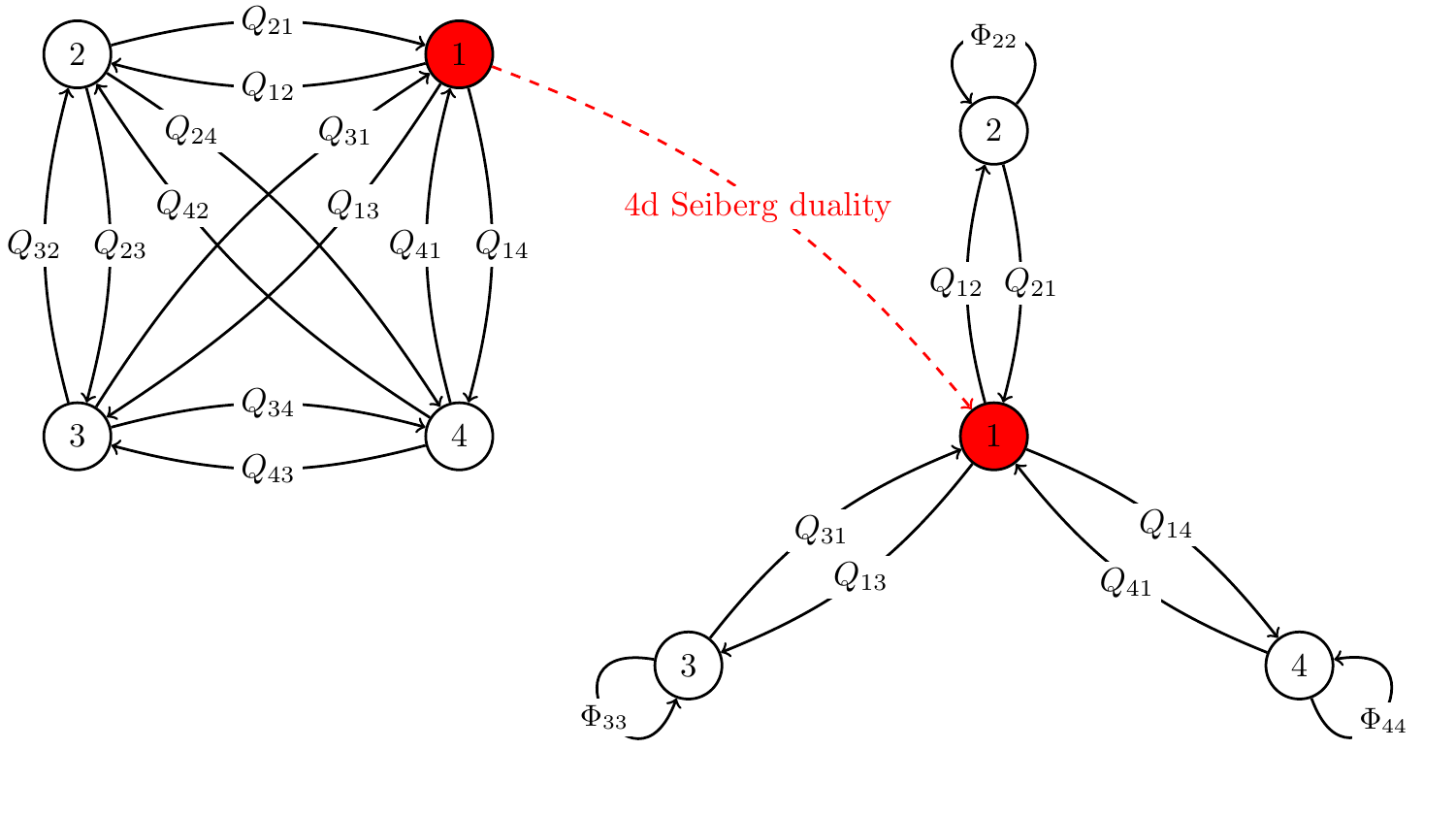}
\caption{Seiberg duality for the toric $\mathbb{Z}_2 \times \mathbb{Z}_2$ orbifold 
of 4d $\mathcal{N}=4$ SYM. The dual phase is represented by the non-toric quiver with three adjoints.}
\label{fig:SD}
\end{figure}%
The dual phase we are interested in can be obtained by applying the rules of Seiberg duality (performed e.g. on node 1 in Figure \ref{fig:SD}). The dual model is a $\U(2N) \times \U(N) \times \U(N) \times \U(N)$ quiver gauge theory with superpotential
\begin{equation}
W  = Q_{12} \Phi_{22} Q_{21} 
+ Q_{13} \Phi_{33} Q_{31}
+ Q_{14} \Phi_{44} Q_{41}
+
 Q_{12}Q_{21} [Q_{13} Q_{31}, Q_{14} Q_{41}]\ .
\end{equation}%
In 3d we add CS interactions to the gauge nodes enforcing, in the phase with equal ranks,  the constraint $k_1+k_2+k_3+k_4=0$.
By applying the rules of three-dimensional Seiberg duality \cite{Giveon:2008zn,Amariti:2009rb}, in the dual phase the gauge group becomes $\U(2N)_{-k_1} \times \U(N)_{k_1+k_2} \times \U(N)_{k_1+k_3}  \times \U(N)_{k_1+k_4}$. This 3d model is then of the type discussed in Section \ref{sec:genproblem}: indeed it has different ranks, adjoint matter fields, and the condition on the levels is that $\sum_{a=1}^4 N_a k_a = 0$.

In this case the $N^{3/2}$ scaling is expected because of Seiberg duality and 
we can use this example as a test of the rules discussed in Appendix \ref{AppAnton}.
In the following we will study two explicit cases (obtained via two different CS level assignments), and compare the results against the ones obtained in Appendix \ref{apptor} in the toric phase.

\subsubsection*{First CS assignment}
We parameterize the R-charges of the adjoints as 
\begin{equation} \label{defDeltai}
\Delta_i \equiv \Delta_{\Phi_{ii}} \quad
\text{with}
\quad
\Delta_2 +\Delta_3 + \Delta_4 = 4\ ,
\end{equation}
where the constraint is imposed by the superpotential.
We choose the levels as $k_1=k_3=-k,\,k_2=k_4=k$ in the toric phase. This boils down to 
the choice $k_1 = k$, $k_2 = -2k$ and $k_3=k_4=0$ after the duality.
The free energy in this cases is 
\begin{equation}
F_{S^3} =N^{3/2}\, \frac{2\pi}{3}  \Delta _3 \Delta _4 \sqrt{\frac{2k \left(4-\Delta _3-\Delta _4\right)}{\Delta _3+\Delta _4}}\ ,
\label{FS3:nontoric:2}
\end{equation}
where we have used the same notation as in \eref{defDeltai}.
We checked this result against the geometric computations in Appendix \ref{apptor}
finding a perfect agreement.
By maximizing the free energy we obtain
\be
\Delta_3=\Delta_4 =\frac{3}{2}~.
\ee
It follows that there are no singlets hitting the unitarity bound, consistently with the claim
that the model is superconformal and that it describes the moduli space probed by a stack of M2-branes. 

Again we computed the Hilbert series as explained in Appendix \ref{appC2}.
In the case with $\Delta_3=\Delta_4=2-\Delta/2$, we found 
\begin{equation}
H(t;\Delta) = \frac{1+ 3 t^{2-{\Delta }/{2}}+3 t^2+t^{4-{\Delta }/{2}}}{\left(1-t^{2-{\Delta }/{2}}\right)^3 \left(1-t^{\Delta }\right)}\ .
\end{equation}
The volume extracted from this formula is 
\begin{equation}
\Vol (\text{SE}_7)= \frac{\pi^4}{48}  \lim_{s \rightarrow 0} s^4 H(t \mapsto e^{-s};\Delta) 
=
\frac{4 \pi ^4}{3 (4-\Delta )^3 \Delta }\ ,
\end{equation}
which coincides with the one computed from the free energy using formula 
(\ref{volF}). {This attains its minimum $\frac{4 \pi ^4}{81}$ at $\Delta=1$.}

\subsubsection*{Second CS assignment}
In this case we choose the levels as  $k_1=-k,\,k_2=k,\,k_3=k_4=0$ for the toric phase. This boils down to  the choice $k_1=k$, $k_2=0$ and $k_3=k_4=-k$ after the duality.
%
The free energy in this case is given by
\begin{equation}
F_{S^3} = N^{3/2}\frac{\pi}{3}  (4-\Delta_3-\Delta_4)
\sqrt{k \Delta_3\Delta_4  (\Delta_3+\Delta_4)}\ .
\label{FS3:nontoric:1}
\end{equation}
We checked this result against the geometric computations presented in Appendix \ref{apptor}, 
finding a perfect agreement.
By maximizing the free energy we have 
\be
\Delta_3=\Delta_4 = \frac{6}{5}~.
\ee
It follows that 
there are no singlets hitting the bound of unitarity, consistently with the claim
that the model is superconformal and that it describes the moduli space probed by a stack of M2-branes. 

Moreover we computed the Hilbert series along the lines of discussion in  Appendix
\ref{appC2}.
In the case with $\Delta_3 = \Delta_4 = 2(1-\Delta)$ we found 
\begin{equation}
H(t;\Delta) = \frac{(1-t^2) (1+t^{2-2\Delta})}{(1-t^{2 \Delta })^2 (1-t^{2-2 \Delta })^3}\ .
\end{equation}
The volume extracted from this formula is 
\begin{equation} \label{toricvolfirstassignment}
\Vol (\text{SE}_7)= \frac{\pi^4}{48}  \lim_{s \rightarrow 0} s^4 H(t \mapsto e^{-s};\Delta) 
=
\frac{\pi ^4}{384 (1-\Delta)^3 \Delta ^2}\ ,
\end{equation}
which coincides with the one computed from the free energy using formula 
(\ref{volF}). {This attains its minimum $\tfrac{3125}{41472} \pi^4$ at $\Delta=\frac{2}{5}$.}

\subsection{A UV completion of Laufer's theory}
\label{genlaufer}

In this Section we discuss a model closely related to Laufer's theory (studied in Section \ref{sec:laufer}). It is a $\U(N)_{2k} \times \U(2N)_{-k}$ theory with superpotential
\begin{equation}
W = 
Q_{21} \Phi_{11} Q_{12}
+
Q_{12} {\Phi_{22}}^2 Q_{21} 
+
\Phi_{22}{\Psi_{22}}^2\ .
\end{equation}
By deforming this model with a holomorphic mass term for the adjoint $\Phi_{11}$
and with a quartic deformation for the adjoint $\Phi_{22}$ we can show that it 
flows to Laufer's theory.

The interesting aspect of the model studied here is the presence of a non-baryonic
global symmetry that can mix with the R-symmetry. 
This becomes explicit once we parameterize the R-charges as
\begin{equation}
R_{Q_{21}} = R_{Q_{12}} = 1-\Delta\ ,
\quad
R_{\Phi_{11}} = 2 \Delta\ ,
\quad
R_{\Phi_{22}} = \Delta\ ,
\quad
R_{\Psi_{22}} = 1- \frac{\Delta}{2}\ ,
\end{equation} 
where $\Delta$ is undetermined so far. 

In order to determine the mixing we need to compute the 
free energy at large $N$ for this model and maximize it w.r.t. $\Delta$. The free energy in this case is 
\begin{equation}
\label{eq:LaufGenOffShell}
F_{S^3} = N^{3/2}\, \frac{2  \pi }{3}  (6-5 \Delta ) (2-\Delta ) \sqrt{\frac{2 \Delta  k}{10-7 \Delta }}\ ,
\end{equation}
which attains its maximum for
\begin{equation}
\Delta = 
\frac{79}{70}-\frac{1}{70} \sqrt[3]{23561-140 \sqrt{14885}}-\frac{641}{70 \sqrt[3]{23561-140 \sqrt{14885}}}
\approx .37\ .
\end{equation}
The free energy at the fixed point becomes 
\begin{equation}
F_{S^3}^* \approx 4.478\, \sqrt k N^{3/2}\ ,
\end{equation}
which is larger than $F_{S^3}^{\text{Laufer}}$ provided in  \eqref{free:en:lauf:12:fin}.
It follows that the claim of an RG flow between this model and Laufer's theory is  consistent with the $F$-theorem.

We observe that $\Tr \Phi_{22}$ is a singlet in the chiral ring with R-charge 
$\Delta < \frac{1}{2}$, i.e. below the unitarity bound.
This implies the existence of an accidental symmetry that can mix with the 
R-symmetry at the fixed point.
This mixing can be taken into account by adding the superpotential deformation
$S\Tr \Phi_{22}$. At the level of the partition function one adds the contribution 
of the extra singlet $S$ by hand and extremizes again.
Observe that the singlet does not affect the partition function at large $N$ though, and this procedure should not modify the extremization. 
However it is needed in order to set the operator $\Tr \Phi_{22}$ to zero in the chiral ring. 

Selecting $k=1$, the Hilbert series (see Appendix \ref{appC3} for details) reads
\begin{equation}
\label{HLaufgen}
H(t;\Delta) 
=
\frac{1-t^{1-\frac{\Delta }{2}}
\!+\! 
t^{3-\frac{3 \Delta }{2}}
\!+\!
t^{3-\frac{5 \Delta }{2}}
\!-\!
2 t^{5-\frac{7 \Delta }{2}}
\!-\!
t^{7-\frac{9 \Delta }{2}}
\!+\!
2 t^{2-\Delta }
\!-\!
t^{4-2 \Delta }
\!-\!
t^{4-3 \Delta }
\!+\!
t^{6-4 \Delta }}{\left(1-t^{3-{5 \Delta }/{2}}\right)^2 \left(1-t^{2-\Delta }\right) \left(1-t^{1-{\Delta }/{2}}\right) \left(1-t^{2 \Delta }\right)}\ .
\end{equation}
It is tempting to extract the most divergent contribution of this expression for 
$t \rightarrow 1$, in order to compare with the free energy. We obtain 
\begin{equation}
\label{eqthreearmsVL}
\frac{\pi^4}{48}  \lim_{s \rightarrow 0} s^4 H(t \mapsto e^{-s};\Delta) 
=
\frac{\pi ^4 (10-7 \Delta )}{12 (2-\Delta )^2 \Delta  (6-5 \Delta )^2}\ ,
\end{equation}
which matches with the expression (\ref{volF}) upon plugging in (\ref{eq:LaufGenOffShell}).

\subsection{A model with two gauge groups}
\label{case1}

The model corresponds to a $\U(N)_{2k} \times \U(2N)_{-k}$ 
quiver gauge theory with superpotential
\begin{equation}\label{eq:Wtwonode}
W = Q_{21} \Phi_{11} Q_{12} + Q_{12} \Phi_{22}^4 Q_{21}\ .
\end{equation}
We parameterize the R-charges as
\begin{equation}\label{eq:Rtwonode}
R_{Q_{21}} = R_{Q_{12}} = 1-\Delta\ ,
\quad
R_{\Phi_{11}} = 2 \Delta\ ,
\quad
R_{\Phi_{22}} = \frac{\Delta}{2}\ ,
\end{equation} 
and the free energy at large $N$ for this model is
\begin{equation} \label{FS3case1}
F_{S^3} = N^{3/2}\frac{2  \pi  }{3} (2-\Delta ) (4-3 \Delta ) \sqrt{\frac{2 \Delta  k}{8-5 \Delta }}\ ,
\end{equation}
which attains its maximum for
\begin{equation}
\Delta = \frac{1}{9} \left(11-13 \left(\frac{5}{311-18 \sqrt{129}}\right)^\frac{1}{3}-\left(\frac{311-18 \sqrt{129}}{5}\right)^\frac{1}{3}\right)
\approx .39\ .
\end{equation}
Again there are accidental symmetries, because
$\Tr \Phi_{22}$ and $\Tr \Phi_{22}^2$ hit the unitarity bound. We can cure this by adding the terms $S_1\Tr \Phi_{22}$ and  $S_2 \Tr \Phi_{22}^2$
to the superpotential, where $S_1$ and $S_2$ are singlets. 
Unfortunately, in this case, we are not able to provide the Hilbert series in order to compare with the above result.  The reason is given in Appendix \ref{appC4}.

\subsection{A model with three gauge groups}
\label{case2}

The model corresponds to a $\U(N)_{k_1} \times \U(2N)_{k_2} \times \U(N)_{k_3}$ quiver gauge theory, with $k_1+2k_2+k_3=0$ and  superpotential
\begin{equation}
W = Q_{21} \Phi_{11} Q_{12} + Q_{12} \Phi_{22}^2 Q_{21}
+ Q_{32} \Phi_{22}^2 Q_{23}
+Q_{23} \Phi_{33} Q_{32} \ .
\end{equation}
We parameterize the R-charges as
\begin{equation}
R_{Q_{21}} = R_{Q_{12}} 
=
R_{Q_{32}} = R_{Q_{23}} 
= 
1-\Delta\ ,
\quad
R_{\Phi_{11}} = R_{\Phi_{33}}  = 2 \Delta
\quad
R_{\Phi_{22}} = \Delta\ ,
\end{equation} 
and distinguish two cases.
\begin{itemize}
\item In the first case we choose CS levels
$k_1=-2k_2=2k$, while $k_3=0$.
This choice is compatible with the constraint $\sum_a k_a N_a = 0$, 
and we can compute the free energy at large $N$
as above.
Before extremization the free energy at large $N$ reads
\begin{equation}
\label{eqthreearmsF1}
F_{S^3} = N^{3/2} \frac{16   \pi}{3}   (1-\Delta ) (2-\Delta ) \sqrt{\frac{\Delta  k}{6-5 \Delta }}\ .
\end{equation}
In this case $F$-maximization gives 
\begin{equation}
\Delta = 
1-\frac{\left(5-\sqrt{15}\right)^\frac{1}{3}}{10^{2/3}}-\frac{1}{\left(10 \left(5-\sqrt{15}\right)\right)^\frac{1}{3}} \approx .33\ .
\end{equation}
The operator $\Tr \Phi_{22}$ hits the bound of unitarity and we
have to add a superpotential term proportional to 
$S \Tr \Phi_{22}$ to cure the presence of accidental symmetries.  

We compute the Hilbert series
as explained in the Appendix \ref{appC5}, arriving at the following expression:
\begin{equation}
\label{eqthreearmsH1}
H(t;\Delta) =
\frac{1+t^{2-\Delta }+3 t^{4-3 \Delta }+t^{4-4 \Delta }-t^{6-4 \Delta }-3 t^{6-5 \Delta }-t^{8-7 \Delta }-t^{10-8 \Delta }}{\left(1-t^{4-4\Delta}\right)^2 \left(1-t^{2-\Delta }\right)^2 \left(1-t^{2 \Delta }\right)}\ .
\end{equation}
Extracting the most divergent contribution of this expression for 
$t \rightarrow 1$, we obtain 
\begin{equation}
\label{eqthreearmsV1}
\frac{\pi^4}{48}  \lim_{s \rightarrow 0} s^4 H(t \mapsto e^{-s};\Delta) 
=
\frac{\pi ^4 (6-5 \Delta )}{384 (\Delta -2)^2 (\Delta -1)^2 \Delta }\ ,
\end{equation}
which matches with the expression (\ref{volF}) if we plug in (\ref{eqthreearmsF1}).

\item 
In the second case we choose CS levels
$k_1=-k_2=k_3=k$.
This choice is compatible with the constraint $\sum_a k_a N_a = 0$
and we can compute the free energy at large $N$ as a function of the R-charge
$\Delta$. We obtain
\begin{equation}
\label{eqthreearmsF2}
F_{S^3} = N^{3/2}\,\frac{8  \pi }{3}   (1-\Delta ) (2-\Delta ) \sqrt{\frac{2 \Delta  k}{4-3 \Delta }}\ .
\end{equation}
In this case $F$-maximization gives 
\begin{equation}
\Delta = 
\frac{1}{18}\left( 19-(431-18 \sqrt{417})^\frac{1}{3}
-\frac{37}{\left(431-18 \sqrt{417}\right)^\frac{1}{3}}\right) \approx .32\ .
\end{equation}
The operator $\Tr \Phi_{22}$ is in the chiral ring and it hits the bound of unitarity.
We cure the accidental symmetry by adding a superpotential interaction  
proportional to $S\Tr \Phi_{22}$.

The Hilbert series is again discussed in Appendix \ref{appC5}, and is given by
\begin{equation}
\label{eqthreearmsH2}
H(t;\Delta) = \frac{\left(1+t^{2-\Delta }\right) \left(1-t^{4-3 \Delta }\right)}{\left(1-t^{2-\Delta }\right)^2 \left(1-t^{2-2\Delta}\right)^2 \left(1-t^{2 \Delta }\right)  }\ .
\end{equation}
Extracting the most divergent contribution for $t \rightarrow 1$ we arrive at the expression  
\begin{equation}
\label{eqthreearmsV2}
\frac{\pi^4}{48}  \lim_{s \rightarrow 0} s^4 H(t \mapsto e^{-s};\Delta) 
=
\frac{\pi ^4 (4-3 \Delta )}{192 (\Delta -2)^2 (\Delta -1)^2 \Delta }\ ,
\end{equation}
which matches with the expression (\ref{volF}) if we plug in (\ref{eqthreearmsF2}).

We also considered a third assignment of CS levels, corresponding to 
 $k_1=-k_3=k$ and $k_2=0$.
We checked that in this case the large-$N$ free energy and the Hilbert series for $N=1$ match with the ones obtained for the CS assignment $k_1=-k_2=k_3=k$.
We have written down the generators of the moduli space in each case when $k=1$ (see \eqref{eq:genCS1} and \eqref{eq:genCS2} respectively), and the two relations they satisfy (which are equal in both cases).
It is then tempting to speculate about a finite-$N$ duality for such models.
Indeed, if we focus solely on the CS levels, this duality can be obtained by 
applying the rules discussed in 
\cite{aharony-bergman-jafferis,Giveon:2008zn,Amariti:2009rb}.
Nevertheless the rank of the third node is unchanged here, while,
by applying the rules of \cite{aharony-bergman-jafferis,Giveon:2008zn,Amariti:2009rb}, it should be shifted by $|k_3|$.
Furthermore it is unclear how to dualize the adjoint matter field
involved in the  $\mathcal{N}=4$ superpotential interaction. 
We have also tried to study the relation between
the superconformal indices of $\U(N)_k \times \U(2N)_{-k} \times \U(N)_k$ and $\U(N)_k \times \U(2N)_{0} \times \U(N)_{k}$ for $N=1$ and $k=1$, and we have observed that they disagree.
Denoting by $\mathcal{I}_{(k_1,k_2,k_3)}$ the superconformal index for CS assignment $\vec k = (k_1,k_2,k_3)$, for $\Delta=\tfrac{1}{3}$ we obtained the expansions:
\begin{align}
\mathcal{I}_{(1,-1,1)} &= 1 + x^{1/3} + 2 x^{2/3} + \tfrac{u + w + 2 u w \omega + u^2 w \omega^2 + u w^2 \omega^2}{u w \omega} x + \ldots \ , \\
\mathcal{I}_{(1,0,-1)} &= 1 + \tfrac{1+\omega + \omega^2}{\omega} x^{1/3} + \tfrac{1+2\omega+3 \omega ^2 +2 \omega ^3+\omega ^4}{\omega^2} x^{2/3} + \tfrac{2(1+2\omega+2 \omega ^2 +2 \omega ^3+\omega ^4)}{\omega^2} x + \ldots\ . \nonumber
\end{align}
Here $x$ is an R-symmetry fugacity, and $u,w,\omega$ are fugacities for the three topological symmetries.
\end{itemize}

\subsection[Higher-$n$ Laufer quivers: three adjoints]{\boldmath Higher-$n$ Laufer quivers: three adjoints}
\label{sub:laufer-n}

\begin{figure}[ht!]
\centering
\includegraphics[scale=1]{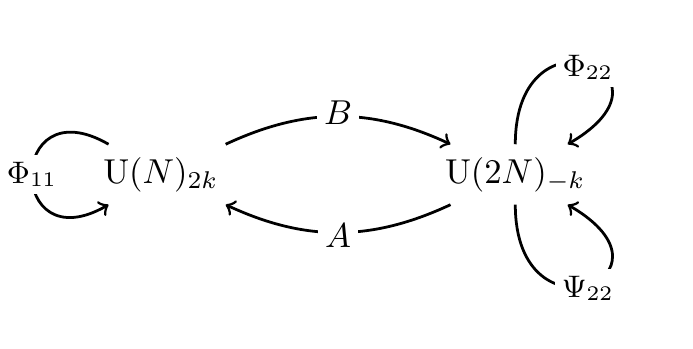}
\caption{The 3d model given by the higher-$n$ Laufer quiver.}
\label{fig:laufer-n}
\end{figure}

Consider the theory with quiver as in Figure \ref{fig:laufer-n} and superpotential below \cite{aspinwall-morrison}:
\begin{align}\label{eq:lauf-nsuperpot}
W &= B \Phi_{11} A + BA \Phi_{22}^2 + \frac{1}{n+1} \Phi_{11}^{n+1} + 
 \frac{(-1)^n}{2n+2} \Phi_{22}^{2n+2}+\Phi_{22} \Psi_{22}^2  \nonumber \\
&= \Tr \left[A \Phi_{11} B\right] +  \Tr\left[ A  \Phi_{22}  B\right] \Tr \Phi_{22} + \frac{1}{n+1} \left(\Tr \Phi_{11}\right)^{n+1} +  \frac{(-1)^n}{2n+2}\left( \Tr \Phi_{22}\right)^{2n+2} + \nonumber \\ &\quad\, +  \Tr[\Phi_{22} \Psi_{22} ]\Tr\Phi_{22}\ ,
\end{align}
where $\Tr$ denotes the trace over the $\U(2N)$ gauge group generators. Notice that for $n=1$, the third summand is just a mass term, hence the adjoint field $\Phi_{11}$ can be integrated out, leaving us with the model in Figure \ref{fig:laufer3d} (i.e. Laufer's theory).\footnote{The 4d parent theory (obtained by forgetting the CS interactions and replacing the $\U$ gauge groups with $\SU$) is the worldvolume theory of $N$ D3-branes probing the singularity 
\begin{equation}\label{eq:laufer-n}
x^2 + y^3 +wz^2 +w^{2n+1}y =0 \ \subset\ \cc^4\ , \nonumber
\end{equation}
which for $n=1$ is nothing but Laufer's threefold \eqref{eq:laufer}. We will rederive this result in Appendix \ref{appc5}.} 
The fields are charged under the global $\U(1)_R$ and the local $\U(N)_{2k} \times \U(2N)_{-k}$ symmetries according to the following table:
\begin{equation}\label{tab:lauf-ncharges}
\begin{array}{c|ccc}
&\U(N)_{2k} & \U(2N)_{-k} & \U(1)_R\\[2.5pt]
\hline
\rule{0pt}{15pt} B           &    \overline{\mathbf{N}}     &  \mathbf{2N}  &  {n}/{(n+1)} \\
\rule{0pt}{15pt}A           &    \mathbf{N}     &  \overline{\mathbf{2N}}  &  {n}/{(n+1)} \\
\rule{0pt}{15pt} \Phi_{11}   &    \mathbf{Adj}     &  \mathbf{1}  & {2}/{(n+1)} \\
\rule{0pt}{15pt} \Phi_{22}   &    \mathbf{1}     &  \mathbf{Adj}  &  {1}/{(n+1)} \\
\rule{0pt}{15pt} \Psi_{22}   &    \mathbf{1}     &  \mathbf{Adj}  & {(2n+1)}/{(2n+2)}
\end{array}
\end{equation}
The three-sphere free energy at the fixed point is found to be
\begin{equation}
\label{F3laufgen}
F_{S^3} =N^{3/2}\, \frac{2\pi}{3}\frac{(2n+1) (6n+1)}{(1+n)^2}  \sqrt{\frac{2k}{(3+10 n)}}\ .
\end{equation}
The Hilbert series in this case is discussed in Appendix \ref{appc5} and
is given by the expression 
\begin{equation}
\label{H2genlauf}
H(t) = \frac{1+t^{\frac{2 n+1}{n+1}}-t^{\frac{6 n+2}{n+1}}-t^{\frac{8 n+3}{n+1}}+t^{\frac{6 n+1}{2 n+2}}+3 t^{\frac{6 n+3}{2 n+2}}-3 t^{\frac{10 n+3}{2 n+2}}-t^{\frac{10 n+5}{2 n+2}}}{\left(1-t^{\frac{2}{n+1}}\right) \left(1-t^{\frac{2 n+1}{n+1}}\right)^2 \left(1-t^{\frac{6 n+1}{2 n+2}}\right)^2}\ .
\end{equation}
By extracting the most divergent contribution for 
$t \rightarrow 1$ we arrive at the expression  
\begin{equation}\label{V2Lauf}
\frac{\pi^4}{48}  \lim_{s \rightarrow 0} s^4 H(t \mapsto e^{-s})  = \frac{(1+n)^4 (3+10n)\pi^4}{12 (2n+1)^2(6n+1)^2}\ ,
\end{equation}
which matches with (\ref{volF}) upon plugging in (\ref{F3laufgen}).

\section{Four-dimensional models and AdS/CFT}
\label{4dholo}

In this Section we will consider the 4d $\None$ models one obtains from the 3d quivers studied in the previous sections upon replacing the CS interactions with standard kinetic terms for the vector multiplets and $\U$ gauge groups with $\SU$ ones. The superpotential is not modified.

We will then compute the Hilbert series, and the $a$ and $c$ central charges of the 4d models  in field theory. The central charges agree at large $N$, signaling the absence of a gravitational anomaly at leading order. This suggests that a holographic interpretation  is possible: the 4d gauge theory is the worldvolume theory of $N$ D3-branes probing a CY$_3$ singularity; the near-horizon geometry should be a type IIB AdS$_5 \times B_5$ vacuum, which is the gravity dual of the quiver gauge theory. (That the latter flows to a CFT in the IR can be argued for by showing that all NSVZ gauge coupling beta functions vanish, again implying $a=c$ at large $N$ \cite{benvenuti-hanany}). As usual in AdS/CFT, the CY metric on the threefold should be  \emph{conical}, $ds^2_\text{CY$_3$} = dr^2 + r^2 ds^2_{B_5}$, for some appropriate coordinate $r$. (The vector field $r\partial_r$ is identified with radial rescalings of AdS in the near-horizon limit.) This condition is equivalent to the existence of a so-called Sasaki--Einstein (SE) metric on the 5d internal space $B_5$ of the type IIB vacuum (called \emph{base} of the CY$_3$) \cite{sparks-rev}. 

Under the assumption of existence of an AdS/CFT pair, the volume of the SE internal space can be computed in two independent ways: from the $a$ central charge, and from the large-$N$ Hilbert series of the gauge theory \cite{Benvenuti:2006qr,ginzburg,Eager:2010yu}. We will show that in all 4d models obtained from the 3d ones the two formulae for the volume match precisely, strongly corroborating the holographic interpretation. 

Actually, we can push the argument further. The CY$_3$ moduli space obtained from the $N=1$ Hilbert series 
of a 4d gauge theory can readily be used to prove the existence of a SE metric on its base.
The proof simply requires computing the \emph{Futaki invariant} of the singularity \cite{donaldson}, and showing it is strictly positive. This condition is indeed equivalent to \emph{K-stability} \cite{collins-szekelyhidi-Ksemi} of the threefold singularity, which implies the existence of a Ricci-flat 
 K\"ahler cone metric on the latter \cite[Thm. 1.1]{collins-szekelyhidi}, i.e.~a SE metric on its base.\footnote{We refer the reader interested in these topics to \cite{collins-phd} for a complete and rigorous overview of the subject. Here we will only introduce the few notions we need.} (The role of K-stability in the holographic context has recently been stressed in \cite{Collins:2016icw,Xie:2019qmw,Fazzi:2019gvt}, and in the field theory context in \cite{Benvenuti:2017lle,Aghaei:2017xqe}.)

From the Futaki invariant one can also extract the volume of the base, showing equivalence with the one computed in field theory. Although proving the existence of a SE$_5$ metric is not on the same level as exhibiting the full supergravity background, it is the first important step in establishing the validity of the AdS$_5$/CFT$_4$ proposal from the gravity side. We will explicitly carry out this procedure in the case of Laufer's theory, its generalization with three adjoints, and finally for the 4d parent of the model discussed in Section \ref{case2}. Indeed in these three cases the CY$_3$ is a complete intersection defined by a single equation, which simplifies the calculations.\footnote{\label{foot:ind}In the case of multiple equations, we simply need to apply \cite[Prop. 4.3.10]{collins-phd} to compute the index character $I(s;\xi,\epsilon)$ introduced below, and thus extract the Futaki invariant. Notice that all CY$_3$'s associated to our 4d models are complete intersections, so our arguments always apply.}

We have collected in Table \ref{tab:4dmodels} the large-$N$ Hilbert series, the volume of the candidate gravity dual, the $N^2$ term of the $a$ central charge, and the gravitational anomaly (i.e. $c-a=-\tfrac{1}{16}\Tr R$, which is always of order $N^0$) for the 4d parents of the models discussed in Section \ref{sec:CS}.
In this case the gauge groups are all special unitary because the 
$\U(1)$ centers of $\U(N_a)$ are IR-free.
In all cases the volume extracted from the Hilbert series coincides with the one
obtained from the $a$ central charge.
\begin{table}[ht!]
\centering
\begin{tabular}{l|cccc}
\!\!\!\!\!\!
\begin{tabular}{c}
\!\!\!3d \!\!\! \\
\!\!\!theory \!\!\!
\end{tabular}
\!\!\! \!\!\!  \!\!\! 
& $H(t;\Delta)$ &
$\Vol(\text{SE}_5)$&
$a(N)/N^2$ & $c-a $
\\[5pt]
\hline
\rule{0pt}{20pt}
 \ref{sec:laufer}
&
$\ddfrac{1 -t^{3/4}+ t^{3/2}}{\left(1-t^{3/4}\right) (1-t) \left(1-t^{7/4}\right)}$
&
$\frac{2^7}{3^4\,7} \pi^3$
&
$\frac{567}{512}$
&
$\frac{5}{64}$
\\
\rule{0pt}{20pt}
\ref{genlaufer}
&
$
\ddfrac{1+t^{3-3 \Delta/2}}
{(1-t^{3-{5\Delta}/{2}}  ) (1-t^{2-\Delta }) (1-t^{2 \Delta })}
$
&
$
\frac{16 \pi ^3}{27 \Delta (\Delta -2)  (5 \Delta -6)}
$
&
$
\frac{27 \Delta  (\Delta -2)(5 \Delta -6) }{64}  
$
&
$
\frac{5 \Delta}{32} 
$
\\[5pt]
\rule{0pt}{20pt}
\ref{case1}
&
$
\ddfrac{1+t^{2-\Delta/2}}{(1\!-\!t^{2 \Delta }) (1\!-\!t^{2-{3 \Delta }/{2}}) 
(1\!-\!t^{2-\Delta })}
$
&
$
\frac{16 \pi ^3}{27\Delta (\Delta -2 ) (3 \Delta -4)}
$
&
$
\frac{27   \Delta (\Delta -2)  (3 \Delta -4) }{64}
$
&
$\frac{5 \Delta }{32}$
\\[5pt]
\rule{0pt}{20pt}
\ref{case2}
&
$
\ddfrac{1+t^{4\!-\!3 \Delta }}{(1\!-\!t^{2 \Delta }) (1\!-\!t^{4-4\Delta}) (1\!-\!t^{2-\Delta })}
$
&
$
\frac{2 \pi ^3}{27 \Delta(\Delta -1) (\Delta -2)  }
$
&
$
\frac{27  \Delta(\Delta -1) (\Delta -2) }{8} 
$
&
$
\frac{5 \Delta }{16}
$
\\[5pt]
\rule{0pt}{20pt}
\ref{sub:laufer-n} &
$
\ddfrac{1+t^{\frac{3 (2 n+1)}{2 (n+1)}}}{\big(1\!-\!t^{\frac{2}{n+1}}\big) \big(1\!-\!t^{\frac{2 n+1}{n+1}}\!\big) \big(1\!-\!t^{\frac{6 n+1}{2 (n+1)}}\!\big)}
$
&
$
\frac{16 \pi ^3 (n+1)^3}{27 (2 n+1) (6 n+1)}
$
&
$
\frac{27  (2 n+1)(6n+1)}{64 (n+1)^3}
$
&
$
\frac{5}{32 (n+1)}
$
\end{tabular}
\caption{We show the large-$N$ Hilbert series, the volume of the candidate gravity dual, the
leading $N^2$ contribution to the central charge $a$, and the gravitational anomaly $c-a$ for the 4d parents of the 3d models discussed in Section \ref{sec:CS}, and summarized in Table \ref{tab:3dmodels}. (Each 3d model is discussed separately in a subsection, which is here indicated in the first column.)
We omit in this analysis the 4d parent of the 3d model studied in Section \ref{C3Z2Z2}, since for that case the holographic correspondence has already been established in the literature due to its toric nature (see e.g. \cite{feng-hanany-he}).  Here $t$ is the fugacity for the R-symmetry of gauge-invariant combinations in the theory.}
\label{tab:4dmodels}
\end{table}

\subsection{4d Laufer's theory}

\begin{figure}[ht!]
\centering
       \includegraphics[width=0.5\linewidth]{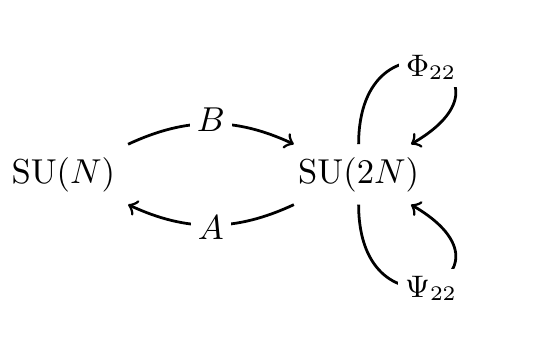}
      \caption{Quiver for 4d Laufer's theory.}
      \label{fig:4dlaufer}
    \end{figure}

Consider the 4d $\None$ model in Figure \ref{fig:4dlaufer}.
The superpotential is given in \eqref{superpotential:laufer} and the R-charges are given in \eref{r:charges:lauf:cond}. We have:
\begin{equation}\label{eq:aclaufer}
a_\text{Laufer} = \left( \frac{567}{512} - \frac{27}{16} b^2 \right)N^2-\frac{831}{2048}\ , \quad c_\text{Laufer} = \left( \frac{567}{512} - \frac{27}{16} b^2 \right)N^2-\frac{671}{2048}\ .
\end{equation}
The central charges are equal at order $N^2$, as it should be for a holographic theory (i.e. one for which $c-a$ vanishes at large $N$, signaling the absence of gravitational anomalies).   Observe that $b$ only appears at order $N^2$; thus upon taking a derivative of $a$ w.r.t $b$ we obtain an expression that is linear in $b$.  This conforms to the expectation that the baryonic symmetry can be absorbed into other symmetries by an appropriate reparametrization \cite{Butti:2005vn, Bertolini:2004xf}.  Indeed, upon using $a$-maximization on \eref{eq:aclaufer}, we find that $b=0$.
We can now extract the volume of the SE$_5$ internal space according to the standard formula \cite{gubser-vol,henningson-skenderis}
\begin{equation}\label{eq:laufvol4d}
\Vol(B_5) = \Vol(S^5)\,\frac{a_{\Nfour\,\text{SYM}}}{a} =\frac{\pi^3}{4} \frac{N^2}{a(N)} =\frac{2^7}{3^4\,7} \pi^3\ .
\end{equation}
In the above, only the leading $N^2$ term of the $a$ central charge should be kept (since we are performing a holographic check at large $N$), and $a_{\Nfour\,\text{SYM}}=\frac{N^2}{4}$ for $\SU(N)$ gauge group.

On the other hand we can compute the Hilbert series, obtaining:
\begin{equation}
H(t) = \frac{1-t^{9/2}}{(1-t)(1-t^{3/2})(1-t^{7/4})(1-t^{9/4})} = 
\frac{1 -t^{3/4}+ t^{3/2}}{\left(1-t^{3/4}\right) (1-t) \left(1-t^{7/4}\right)} \ ,
\end{equation}
where $t$ keeps track of the R-symmetry in the unit of $1$, without any rescaling.  (Note however that in Appendix \ref{LauferApp}, and in particular \eref{HSLauferApp}, we use $t$ to count the R-charge in the unit of $1/4$.) The volume can then be extracted as follows:
\begin{equation}
\Vol(B_5)= 
\left(\frac{2}{3} \pi \right)^3
\lim_{s\to 0} s^3 \,H\left(t \mapsto e^{-s}\right) = \frac{2^7}{3^4\,7}\pi^3\ ,
\end{equation}
which clearly agrees with \eqref{eq:laufvol4d}.

\subsection{Metric and volume from the Reeb vector}
\label{sub:reeb}

A simple way to prove the existence of a SE metric on the base $B_5$ of the CY$_3$ singularity is by computing the Futaki invariant of the threefold, and showing that it is strictly positive. Importantly, from the Futaki invariant one can also extract the volume of the base, thereby showing its equivalence to the one computed from the $a$ central charge of the dual field theory.

We start by noticing that the threefold hypersurface equation, namely
\begin{equation}
\text{CY}_3:\quad x^2+y^3+w z^2+w^3 y =0\ \subset\ \cc^4\ ,
\end{equation}
is homogeneous with weight $w_p$ under a $\cc^*$ action assigning weights $(w_1,\ldots,w_4)$ to the $\cc^4$ ambient coordinates $(x,y,w,z)$:
\begin{equation}\label{eq:weights}
(w_1,\ldots,w_4)=(w_1,\tfrac{2}{3}w_1,\tfrac{4}{9}w_1,\tfrac{7}{9}w_1) \equiv (9,6,4,7)\ , \quad w_p = 2w_1 \equiv 18\ ,
\end{equation}
having chosen $w_1 \equiv 9$ so that all weights are integer. (Observe that the only other option would be to have $w_p=0$; however this implies the trivial action $(w_1,\ldots,w_4)=(0,\ldots,0)$, which we disregard.)
Geometrically this $\cc^*$ action corresponds to a $\U(1)$ isometry of the base, which is generated by a vector field $\xi$, commonly known as Reeb vector. Under the AdS/CFT dictionary this vector is dual to the $\U(1)_R$ symmetry of the 4d $\None$ model. Observe that, since the Reeb vector field is generated by a single $\cc^*$ action (i.e. $\U(1)$ isometry), there is no other global $\U(1)$ the R-symmetry generator could mix with: this isometry is already the superconformal R-symmetry.

Let us fix $\xi = \kappa (9,6,4,7)$ as generator of this $\U(1)$, where $\kappa$ is a normalization to be determined shortly. To compute the Futaki invariant of the CY$_3$, we first need to find additional $\cc^*$ actions (generated by four-vector $\lambda$'s) that commute with $\xi$, and perturb the latter with a test parameter $\epsilon$ as follows: $\xi \to \xi + \epsilon \lambda$. We then need to compute the index character of Martelli--Sparks--Yau \cite{martelli-sparks-yau-volmin}, and from the latter extract the invariant. Let us see how this works concretely.

Consider e.g. 
\begin{equation}\label{eq:tests}
\lambda = (0,0,2,-1)\ ,
\end{equation}
and define the \emph{test configuration} $\xi + \epsilon \lambda$.\footnote{One might correctly wonder whether one should check positivity of the Futaki invariant against an infinite number of test configurations (thereby making this approach completely impractical). However in favorable situations (such as for toric CY$_3$'s) it is possible to show that one only needs to test against a finite number of \emph{$T$-equivariant test configurations} with normal special fiber (i.e. when $\epsilon \to 0$), and even construct them explicitly \cite{ilten-suss}. (Indeed for the 4d parents of Laufer and of the theory discussed in in Section \ref{case2} they were very recently constructed in \cite{Fazzi:2019gvt}.) Here $T\equiv \U(1)^r$ is the maximal (compact) torus of isometries of the CY$_3$ (defined by an equation $p(x,y,z,t)=0$ inside $\cc^4$); $r$ is the number of independent $\cc^*$ actions acting homogeneously on $p$ (e.g. $r=3$ in the toric case, while $r=1$ for Laufer). In the non-toric case (which is the one of interest in this paper), one is left with the task of testing against $r$ $T$-equivariant configurations $\lambda$, which can easily be guessed (see e.g. \cite[Sec. 8]{collins-szekelyhidi} for other examples).}
Now we compute the index character $I$ of the threefold with Reeb vector $\xi$ and test parameter $\epsilon$ as follows \cite{martelli-sparks-yau-volmin}:
\begin{equation}
I(s;\xi,\epsilon) \equiv \frac{1-e^{w_p s}}{\prod_{i=1}^4 1-e^{w_i s}} \sim \frac{2a_0(\xi,\epsilon)}{s^{3}}+\frac{a_1(\xi,\epsilon)}{s^{2}} + \ldots \label{eq:indchar}\ .
\end{equation}
as $s\to 0$. (This index coincides with the Hilbert series of the threefold singularity with R-symmetry fugacity $e^{-s}$ when $\epsilon=0$ \cite[Thm. 3]{collins-szekelyhidi-Ksemi}.) For the test configuration $\xi + \epsilon \lambda$ with $\lambda$ as in \eqref{eq:tests} we obtain the following coefficients:
\begin{equation}
a_0(\xi,\epsilon) = \frac{1}{12 \kappa  (7 \kappa-\epsilon ) (2 \kappa +\epsilon)}\ , \quad a_1(\xi,\epsilon) = \frac{8 \kappa +\epsilon}{12 \kappa  (7 \kappa-\epsilon ) (2 \kappa +\epsilon)}\ .
\end{equation}
When the test parameter $\epsilon$ vanishes, the coefficients $a_i(\xi,0):\mathcal{C} \to \rr $ are smooth functions of $\xi$, and for a CY$_3$ are given by \cite[Sec. 4.5]{collins-phd}
\begin{equation}\label{eq:a0a1}
a_0(\xi,0) = \frac{1}{2\pi^{3}} \Vol(B_5) \ , \quad a_1(\xi,0)=\frac{1}{6\pi^{3}}\Vol(B_5)\ ,
\end{equation}
where $\Vol(B_5)$ is the volume of the base $B_5$ of the CY$_3$. 
Notice that the ratio $a_1/a_0$ is always a number independent of $\xi$ which only depends on the dimensionality of the CY singularity \cite[Prop. 6.4]{collins-szekelyhidi}. For a threefold,
\begin{equation}\label{eq:reebnorm}
\left(\frac{a_1}{a_0}\right)(\xi,0) \equiv \frac{a_1(\xi,0)}{a_0(\xi,0)} = 3\ ,
\end{equation}
which can be used to fix the normalization of the Reeb vector that minimizes the volume. For the test configuration in \eqref{eq:tests} we obtain $\kappa = \tfrac{3}{8}$; thus:
\begin{equation}\label{eq:reeblaufer}
\xi_\text{Laufer} = \left( \frac{27}{8},\frac{9}{4},\frac{3}{2},\frac{21}{8}\right)\ .
\end{equation}
From \eqref{eq:indchar} one can then compute the Futaki invariant of the singularity, which can be defined as follows \cite[Def. 2.2]{collins-szekelyhidi}:
\begin{equation} \label{eq:futaki}
\text{Fut}(\xi,\lambda) =\frac{a_0(\xi,0)}{2} \left[\frac{d}{d\epsilon}\left(\frac{a_1}{a_0}\right)(\xi+\epsilon \lambda) \right]_{\epsilon=0}+ \frac{1}{6}\left(\frac{a_1}{a_0}\right)(\xi,0) \left[\frac{d}{d\epsilon}a_0(\xi+\epsilon \lambda) \right]_{\epsilon=0}\ .
\end{equation}
We obtain:
\begin{equation}
\text{Fut}(\xi,\lambda)  = \frac{2^5}{3^5 \, 7^2} > 0\ ; \quad a_0(\xi,0) \equiv a_0(\xi) = \frac{2^6}{3^4 \, 7}\ .
\end{equation}
This conclusion bears two important results. On the one hand, we see that Laufer's threefold singularity admits a SE metric on its base $B_5$, thereby allowing one to apply the usual AdS/CFT logic to the 4d $\None$ model. On the other, it gives us the volume of the base according to \eqref{eq:a0a1}:
\begin{equation}
\Vol(B_5) = 2\pi^3 a_0(\xi) = \frac{2^7}{3^4 \, 7} \pi^3\ ,
\end{equation}
which coincides exactly with what we computed in field theory from the $a$ central charge, i.e. \eqref{eq:laufvol4d}.

\subsection{Three-adjoint Laufer}

Using the three-adjoint generalization of 3d Laufer's theory studied in Section \ref{sub:laufer-n}, we obtain for the 4d parent:
\begin{subequations}\label{eq:aclaufer-n}
\begin{align}
a_\text{Laufer-$n$} &= \frac{27 (2 n+1) (6 n+1)}{64 (n+1)^3} N^2 -\frac{3 (4 n (49 n+8)+49)}{256 (n+1)^3}\ , \\ 
c_\text{Laufer-$n$} &= \frac{27 (2 n+1) (6 n+1)}{64 (n+1)^3} N^2 -\frac{4 n (137 n+4)+107}{256 (n+1)^3}\ ,
\end{align}
\end{subequations}
which correctly reduce to \eqref{eq:aclaufer} for $n=1$. The central charges are equal at order $N^2$, as it should be for a holographic theory. They were obtained with the R-charge assignment of Table \eqref{tab:lauf-ncharges}, which maximizes the trial $a$ central charge. The volume of the base obtained from the $a$ central charge at the superconformal fixed point reads
\begin{equation}\label{eq:lauf-nvol4d}
\Vol(B_5) = \frac{16 (n+1)^3}{27 (2 n+1) (6 n+1)} \pi^3\ ,
\end{equation}
which correctly reduces to \eqref{eq:laufvol4d} for $n=1$. The Hilbert series is given in \eqref{eq:HSlaufer-n}. By extracting its most divergent term for $t\to 1$, we reobtain the volume \eqref{eq:lauf-nvol4d}. 

Now the geometry. The hypersurface singularity equation is \cite{laufer}
\begin{equation}
\text{CY}_3:\quad x^2+y^3+w z^2+w^{2n+1} y =0\ \subset\ \cc^4\ .
\end{equation}
The single nontrivial $\cc^*$ action assigns weights
\begin{equation}\label{eq:weights-n}
(w_1,\ldots,w_4)=(w_1,\tfrac{2}{3}w_1,\tfrac{4}{3+6n}w_1,\tfrac{1+6n}{3+6n}w_1) \equiv (9,6,\tfrac{12}{1+2n},\tfrac{3+18n}{1+2n})\ , \quad w_p = 2w_1 \equiv 18\ ,
\end{equation}
choosing $w_1\equiv 9$ (and again disregarding the trivial case with $w_1\equiv 0$). It is generated by the following normalized Reeb vector,
\begin{equation}
\xi_\text{Laufer-$n$} = \left(\frac{9 (2 n+1)}{4 (n+1)},\frac{3 (2 n+1)}{2 (n+1)},\frac{3}{n+1},\frac{3 (6 n+1)}{4 (n+1)}\right)\ ,
\end{equation}
which corresponds to the superconformal $\U(1)_R$ symmetry of the 4d model. 

To compute the Futaki invariant, we can once again use the test configuration defined in \eqref{eq:tests}. Extracting the $a_0(\xi)$ coefficient from the index character, we obtain the volume \eqref{eq:lauf-nvol4d}. However the Futaki invariant is strictly positive only for $n=1$, i.e. Laufer's theory, implying once again that only this case may admit an AdS$_5$ dual with SE metric on the internal space $B_5$.\footnote{Notice that this is akin to what happens for the higher-$n$ generalizations of the cone over $V_{5,2}$, the latter being the base of the $n=2$ case in the fourfold family $X_n: z_0^n + z_1^2+z_2^2+z_3^2+z_4^2=0$ inside $\cc^5$ \cite{Fabbri:1999hw,martelli-sparks-V52}. On the field theory side, for $n>2$ certain operators violate the unitarity bound \cite{martelli-sparks-V52,Jafferis:2009th}, which is impossible for (unitary) CFT's, hence the absence of AdS$_4$ duals with SE metric on the internal space for $n>2$.}

\subsection{The 4d parent of the model in Section \ref{case2}}

In this Section we apply the above logic to the CY$_3$ singularity associated with the 4d parent of the 3d model discussed in Section \ref{case2}. It was found in Appendix \ref{appC5} that the hypersurface singularity equation in this case reads
\begin{equation}\label{eq:case2CY3top}
\text{CY}_3:\quad x^2 + wy^2 + zw^2=0\ \subset\ \cc^4\ .
\end{equation}
Any nontrivial homogeneous $\cc^*$ acting on this equation assigns weights 
\begin{equation}
(w_1, w_2 , w_3,w_4)=(w_1,w_2,2(w_1-w_2),2(-w_1+2w_2))
\end{equation}
to $(x,y,w,z)$ respectively, and $w_p = 2w_1$ to the hypersurface. There are two independent actions, given we have two free parameters, $w_1,w_2$, to play with. We can choose e.g.
\begin{equation}\label{eq:C*reeb}
\begin{array}{c|ccccc}
& w_1 & w_2 & w_3 & w_4 & w_p \\ \hline
\cc^*_1 &1 & 0& 2&-2 & 2\\
\cc^*_2 &0 & 1 & -2& 4 & 0\\
\end{array}
\end{equation}
The Reeb vector is given by a linear combination of these two actions with positive entries $\omega_i$ (since it has to act effectively on the coordinates $z_i$):
\begin{equation}
\xi = \kappa_1 \xi_1 + \kappa_2 \xi_2\ ; \quad \mathcal{L}_\xi z_i = \omega_i z_i\ , \quad \omega_i>0\ ,
\end{equation}
with $\xi_i$ generating $\cc^*_i$ of Table \eqref{eq:C*reeb}. The combination that corresponds to the superconformal $\U(1)_R$ symmetry of the 4d model is fixed by volume minimization \cite{martelli-sparks-yau-volmin} (whereas the leftover overall normalization is fixed by requiring \eqref{eq:reebnorm}, as before). 

To compute the Futaki invariant, and thus the volume, we can use the following test configurations (see \cite[Sec. 5.1]{Fazzi:2019gvt} for details)
\begin{equation}
\lambda^{(1)}=(0,1,1,-2)\ , \quad \lambda^{(2)}=(0,1,-2,5)\ .
\end{equation}
Proceeding as above, and minimizing w.r.t. both parameters $\xi_i$, we obtain
\begin{equation}
\text{Fut}(\xi,\lambda^{(1)}) = \text{Fut}(\xi,\lambda^{(2)})  = \frac{1}{24} >0\ .
\end{equation}
Therefore a SE metric exists on the base of the CY$_3$ given by \eqref{eq:case2CY3top}. Extracting the $a_0^{(j)}(\xi,0)\equiv a_0(\xi)$ coefficient (for each $j=1,2$) yields the volume
\begin{equation}\label{eq:volcase2geo}
\Vol(\text{SE}_5) = 2\pi^3 a_0(\xi) = \frac{\pi^3}{3\sqrt{3}}\ .
\end{equation}
The normalized Reeb vector that minimizes the volume is given by
\begin{equation}
\xi = \left(\frac{3}{2} (\sqrt{3}+1),\frac{1}{2} (\sqrt{3}+3),2 \sqrt{3},3-\sqrt{3}\right)\ .
\end{equation}
We are now ready to compare the result \eqref{eq:volcase2geo} with what one extracts from the $a$ central charge of the dual 4d model. Looking at Table \ref{tab:4dmodels}, we see that the trial $a$ charge is given by
\begin{equation}
a(\Delta) = \frac{27}{8}\Delta (\Delta-1)(\Delta-2) N^2 + \ldots\ ,
\end{equation}
where the ellipsis denotes the order $N^0$ contribution. The value of the R-charge $\Delta_*$ at the superconformal fixed point is obtained via $a$-maximization, which yields $\Delta_* = \tfrac{3-\sqrt{3}}{3}\approx 0.422$. Thus, at large $N$, we obtain the volume
\begin{equation}
\Vol(B_5) = \frac{\pi^3}{4} \frac{N^2}{a(\Delta_*)} = \frac{\pi^3}{3\sqrt{3}}\ ,
\end{equation}
which nicely matches the expression \eqref{eq:volcase2geo}.

\section{Further directions}
\label{sec:further}

In this paper we have studied a class
of 3d $\mathcal{N}=2$ CS-matter quivers with gauge group $\prod_{a=1}^r \U(N_a)_{k_a}$, where the unequal ranks $N_a = n_a N$ satisfy the condition $\sum_{a=1}^{r} N_a k_a = 0$.
We have shown that the free energy of these models exhibits 
the expected $N^{3/2}$ scaling of the degrees of freedom at large $N$, and
we have performed the $F$-maximization procedure when necessary.
We have compared these results to the ones obtained by extracting the most 
divergent contribution of the Hilbert series, by sending the R-symmetry fugacity to one.
If the models admit an AdS$_4$ dual description, this procedure is equivalent to the computation of the volume of the (seven-dimensional base of the) Calabi--Yau fourfold.
We have shown that the candidate volume computed from the Hilbert series matches the one obtained from the free energy.
This is a first step towards the holographic interpretation of the models discussed here.

We have tackled an analogous problem for the 4d $\mathcal{N}=1$ parent 
quivers. In this case we have matched the volumes obtained from the $a$ central charge 
to the ones obtained from the Hilbert series. Moreover, for the models in which the Calabi--Yau threefold is defined by a single hypersurface equation, we have proven the existence of a Sasaki--Einstein metric on its five-dimensional base by showing strict positivity of its Futaki invariant, which is equivalent to K-stability \cite{collins-szekelyhidi-Ksemi}. We have matched the volume extracted from this invariant with the one computed from the $a$ central charge. Actually, since all our 4d models are associated to complete intersection Calabi--Yau threefolds, we may repeat this calculation (as explained in footnote \ref{foot:ind}) and show the existence of a Sasaki--Einstein metric for all models.

A first generalization of our analysis consists of considering the quivers with chiral flavors discussed in \cite{benini-closset-cremonesi1} and generalizing them by allowing different-rank gauge groups, such that $\sum_a n_a(k_a +\tfrac{1}{2}(n_{\mathbf{f},a}-n_{\overline{\mathbf{f}},a}))=0$ (see \eqref{level:condition}, \eqref{fundamental:condition}). Furthermore we believe that our work raises many questions that we leave for future investigation. A first problem regards the existence of a holographically dual description of the 3d models.
We have not been able to identity from the symmetries or from the free energies 
computed here any relation with known supergravity solutions, neither the ones 
already associated to other field theory duals (e.g. the ABJM model) 
nor the ones that have not been so far given any field theory interpretation (e.g. 
the field theory dual description of M-theory on AdS$_4 \times M^{3,2}$ \cite{Castellani:1983mf,Fabbri:1999hw,Martelli:2008rt}). 
This search requires an extension of our analysis to other  models
sharing the same features of the ones discussed here. For this reason in Appendix \ref{families} we have provided some infinite families of 3d $\mathcal{N}=2$ CS-matter quivers with varying ranks satisfying the condition $\sum_a N_a k_a = 0$.
While we have not provided a classification scheme for these families of models, it is possible that it can be furnished by the holographic picture.
Even in absence of such a result, it should be possible to extend the techniques 
discussed in Section \ref{4dholo} in order to prove the existence of a Sasaki--Einstein metric on the base of the Calabi--Yau fourfold, corroborating the holographic interpretation of the 3d theories studied here.\footnote{For instance, the CY$_4$'s described by the last two Hilbert series in \eqref{HSkmkk} are amenable to being studied in the way explained in footnote \ref{foot:ind}, since they are complete intersections defined by two equations. One could thus check whether a SE metric on their seven-dimensional base exists.} Along these lines, it should be interesting to look for connections between our results
and those of \cite{Crichigno:2017rqg} regarding the volumes of seven-manifolds associated to non-toric cones.  For instance, with our techniques it is easy to show that in the $D_k$ fourfold family (defined by $z_0^2 + z_1^2 +z_2^2 +z_3^{k-1} + z_3 z_4^2=0$ inside $\cc^5$) only $D_4$ admits a SE metric (confirming the statement given towards the end of \cite[Sec. 5.1]{Gauntlett:2006vf}), with volume given by $\text{Vol}(\text{SE}_7)/\text{Vol}(S^7) = 2401/4608$. It would be interesting to identify the dual 3d model (including the CS assignment). 

We conclude by mentioning that another interesting possibility consists in topologically twisting the 3d models, and computing their topologically twisted index \cite{Benini:2015noa}.
At large $N$, the analysis is very similar to the one we performed for the 
free energy \cite{Hosseini:2016tor,Hosseini:2016ume}, and this index has indeed been recently computed for some of the ADE quivers in \cite{Jain:2019lqb}. 

\section*{Acknowledgments}

We would like to thank O.~Bergman, S.~Cremonesi, A.~Hanany, S.~M.~Hosseini, D.~Rodr\'iguez-G\'omez, A.~Tomasiello, and A.~Zaffaroni for valuable comments. M.F.~is indebted to A.~Collinucci, S.~S.~Razamat, O.~Sela, A.~Tomasiello, and R.~Valandro for collaboration on related topics. The work of M.F. was performed in part at the Simons Center for Geometry and Physics, Stony Brook University during the XVI Simons Summer Workshop, and in part at the Universities of Milan and Milan-Bicocca. The work of M.F.~and N.M.~was performed in part at the Aspen Center for Physics, which is supported by the National Science Foundation grant PHY-1607611. M.F.~acknowledges financial support from the Aspen Center for Physics through a Jacob Shaham Fellowship Fund gift. The work of M.F.~and A.N.~is partially supported by the Israel Science Foundation under grant No.~504/13, 1696/15, 1390/17, 2289/18 and by the I-CORE Program of the Planning and Budgeting Committee. 

\appendix

\newpage
\section{Infinite families}
\label{families}

So far we have discussed models with a fixed, small number of gauge groups.
In principle one may expect the existence of models with a large amount of gauge groups.
In this Section we list a set of one-parameter families of this type, i.e. chains of gauge groups 
where the long-range force cancellation is at work, allowing the $N^{3/2}$ scaling of the free energy at large $N$.
Below we list the quivers, the ranks, and the total number of gauge groups as a function of the parameter $r \in \mathbb{N}$. (The constraint $\sum_a N_a k_a =0$ is always enforced. Also, for each family there is an obvious lower bound on $r$ that follows from the structure of the quiver.)
We also provide a superpotential term, even if in most of the cases exactly marginal deformations can be added. For this reason we do not specify the superpotential coupling.
Observe that traces are understood and that
some of the interactions can also be related to double-trace deformations.
Indeed at this level of the discussion the superpotential is only necessary in order to provide the constraints on the R-charges that 
enforce the cancellation of the long-range forces in the free energy.

In each quiver we label the matter fields (bifundamentals $Q_{ij}$ and adjoints $\Phi_{ii}, \Psi_{ii}$) with the R-charges that follow from the superpotential, upon turning off the mixing with the baryonic symmetries. Dashed arrows indicate the ``region'' where one can extend the quiver indefinitely, adding in the obvious way gauge groups with the specified rank and bifundamentals with the specified R-charge.
\begin{enumerate}
\item \label{fam1}
The gauge groups is $\U(N)_{k_1} \times \U(2N)_{k_2}
\times 
\dots \times  \U(2N)_{k_{r-1}} \times \U(N)_{k_r}$
with the constraint $\sum_{a=1}^{r} N_a k_a = 0$.
The superpotential reads
\begin{align}
W_{1}  =&\  (Q_{1,2}Q_{2,1})^2
+ Q_{1,2} \Phi_{2,2}^2 Q_{2,1} 
+
\sum_{i=2}^{r-2} \Big(
Q_{i+1,i} \Phi_{i,i} Q_{i,i+1}\ +
\nonumber \\
&+ 
Q_{i,i+1} \Phi_{i+1,i+1} Q_{i+1,i}
\Big)
+
 Q_{r,r-1} \Phi_{r-1,r-1}^2 Q_{r-1,r} 
 +
 (Q_{r-1,r}Q_{r,r-1})^2\ .
\end{align}
The quiver is
\begin{equation}
\includegraphics[scale=.85]{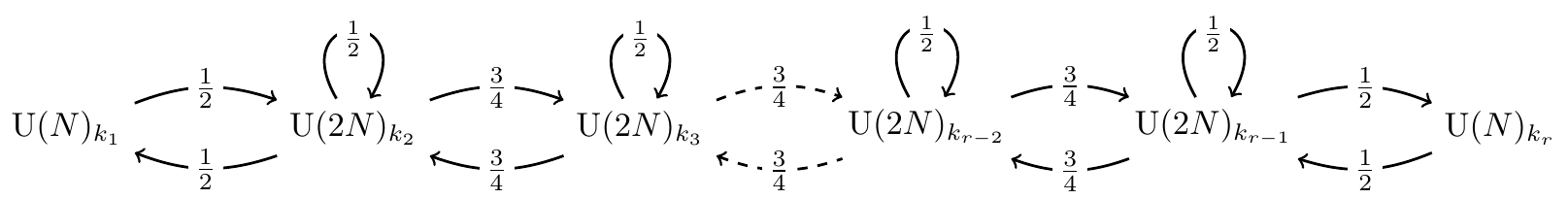} 
\end{equation}
\item \label{fam2}
The gauge groups is 
$\U(N)_{k_1} \times \U(2N)_{k_2}
\times 
\dots \times  \U(2N)_{k_{2r}}$
with the constraint $\sum_{a=1}^{2r} N_a k_a = 0$. The superpotential reads
\begin{align}
W_{2}  =&\  (Q_{1,2}Q_{2,1})^2
+ Q_{1,2} \Phi_{2,2}^2 Q_{2,1} 
+
\sum_{i=2}^{r-1} \Big(
Q_{i+1,i} \Phi_{i,i} Q_{i,i+1}\ +
\nonumber \\
&+
Q_{i,i+1} \Phi_{i+1,i+1} Q_{i+1,i}
\Big)
+
\Phi_{r,r} \Psi_{r,r}^2\ .
\end{align}
The quiver is
\begin{equation}
\includegraphics[scale=1]{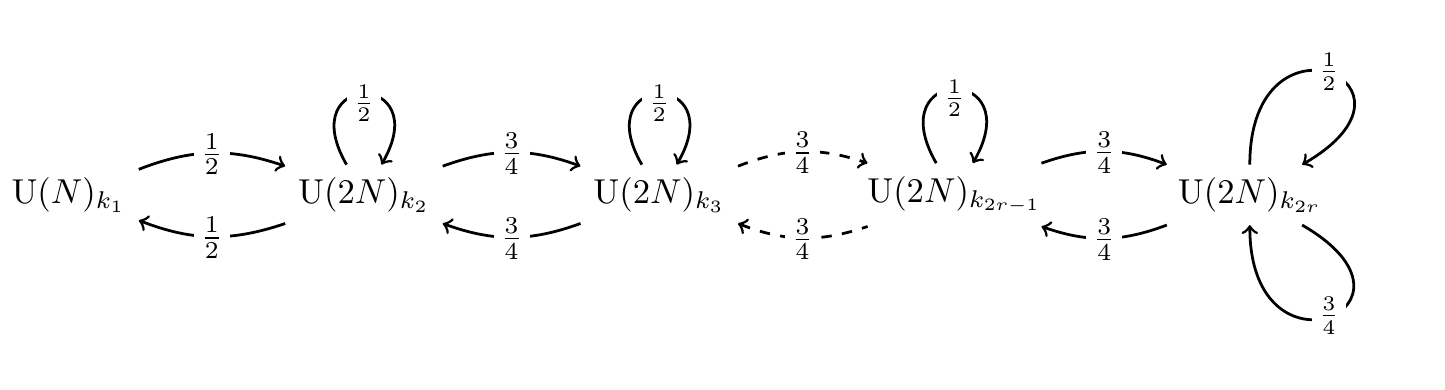}
\end{equation}
\item \label{fam3}
The gauge groups is 
$\U(N)_{k_1} \times \U(2N)_{k_2} \times 
\dots \times  \U(2N)_{k_{2r}}$
with the constraint $\sum_{a=1}^{2r} N_a k_a = 0$. The superpotential reads
\begin{align}
W_{3} =&\
(Q_{1,2}Q_{2,1})^2
+
Q_{2,1} Q_{1,2} (Q_{2,3} Q_{3,2})^2
+
\sum_{i=2}^{2(r-1)} Q_{i,i+1} Q_{i+1,i+2} Q_{i+2,i+1} Q_{i+1,i}\ +
\nonumber \\
&+
Q_{2r-1,2r} \Phi_{2r,2r}^2 Q_{2r,2r-1}\ .
\end{align}
The quiver is
\begin{equation}
\includegraphics[scale=.85]{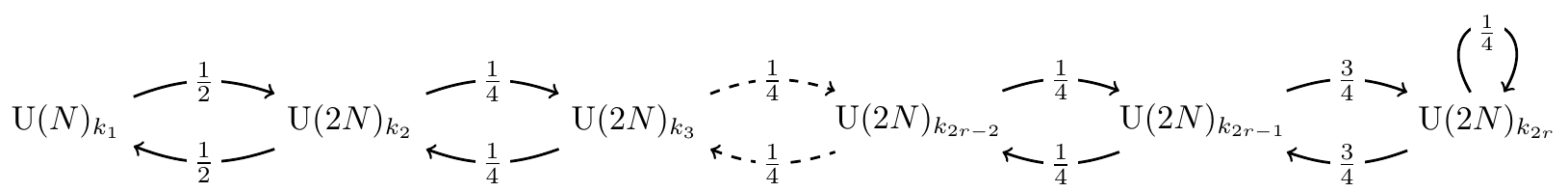}
\end{equation}
\item \label{fam4}
The gauge groups is 
$\U(N)_{k_1} \times \U(2N)_{k_2}
\times 
\dots \times  \U(2N)_{k_{2r+1}}$
with the constraint $\sum_{a=1}^{2r+1} N_a k_a = 0$. The superpotential reads
\begin{align}
W_{4} =&\
(Q_{1,2}Q_{2,1})^2
+
Q_{2,1} Q_{1,2} (Q_{2,3} Q_{3,2})^2
+
\sum_{i=2}^{2 r} Q_{i,i+1} Q_{i+1,i+2} Q_{i+2,i+1} Q_{i+1,i}\ +
\nonumber \\
&+
Q_{2r-1,2r} \Phi_{2r,2r}^2 Q_{2r,2r-1}\ .
\end{align}
The quiver is
\begin{equation}\label{fig:infIV}
\includegraphics[scale=.85]{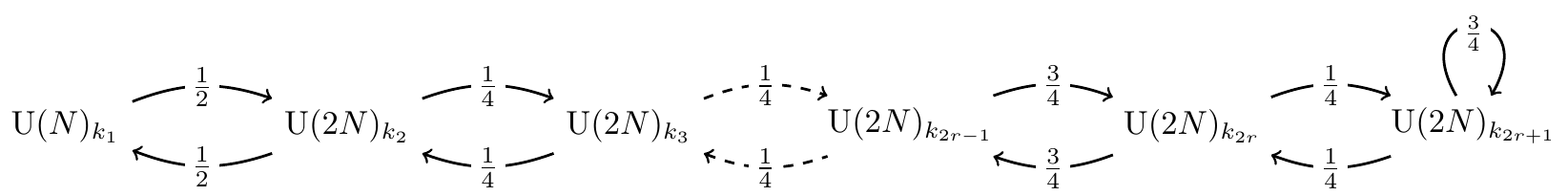}
\end{equation}
\item \label{fam5}
The gauge groups is 
$\U(N)_{k_1} \times \U(2N)_{k_2}
\times 
\dots \times  \U(2N)_{k_{2r-1}} \times \U(N)_{k_{2r}}$
with the constraint $\sum_{a=1}^{2r} N_a k_a = 0$. The superpotential reads
\begin{align}
W_{5} =&\
(Q_{1,2}Q_{2,1})^2
+
Q_{2,1} Q_{1,2} (Q_{2,3} Q_{3,2})^2
+
\sum_{i=2}^{2 r-3} Q_{i,i+1} Q_{i+1,i+2} Q_{i+2,i+1} Q_{i+1,i}\ +
\nonumber \\
&+
(Q_{2r-1,2r-2} Q_{2r-2,2r-1})^2 Q_{2r-1,2r} Q_{2r,2r-1} 
+
(Q_{2r-1,2r} Q_{2r,2r-1} )^2
\end{align}
The quiver is
\begin{equation}\label{fig:infV}
\includegraphics[scale=.85]{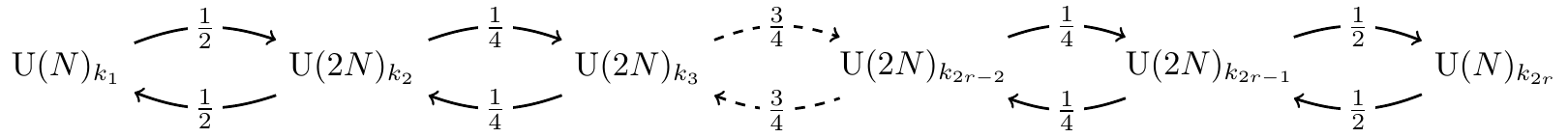}
\end{equation}
\item \label{fam6}
The gauge groups is 
$\U(N)_{k_1} \times \U(2N)_{k_2}
\times 
\dots \times  \U(2N)_{k_{4r+2}} \times \U(N)_{k_{4r+3}}$
with the constraint $\sum_{a=1}^{4r+3} N_a k_a = 0$. The superpotential reads
\begin{align}
W_{6} =&\ 
(Q_{1,2}Q_{2,1})^2
+
Q_{2,1} Q_{1,2} (Q_{2,3} Q_{3,2})^2
+
\sum_{i=2}^{2r} Q_{i,i+1} Q_{i+1,i+2} Q_{i+2,i+1} Q_{i+1,i}\ +
\nonumber 
\\
&\ +
Q_{2r+1,2r+2} {\Phi_{2r+2,2r+2}} Q_{2r+2,2r+1}
+
Q_{2r+3,2r+2} {\Phi_{2r+2,2r+2}} Q_{2r+2,2r+3}\ +
\nonumber 
\\
&\ +
\sum_{i=2r+2}^{4r} Q_{i,i+1} Q_{i+1,i+2} Q_{i+2,i+1} Q_{i+1,i}
+
(Q_{4r+2,4r+3} Q_{4r+2,4r+3}  )^2\ +
\nonumber 
\\
&\ +
(Q_{4r+2,4r+1} Q_{4r+1,4r+2})^2 Q_{4r+2,4r+3} Q_{4r+3,4r+2} \ .
\end{align}
The quiver is
\begin{equation}\label{fig:infVI}
\includegraphics[scale=.8]{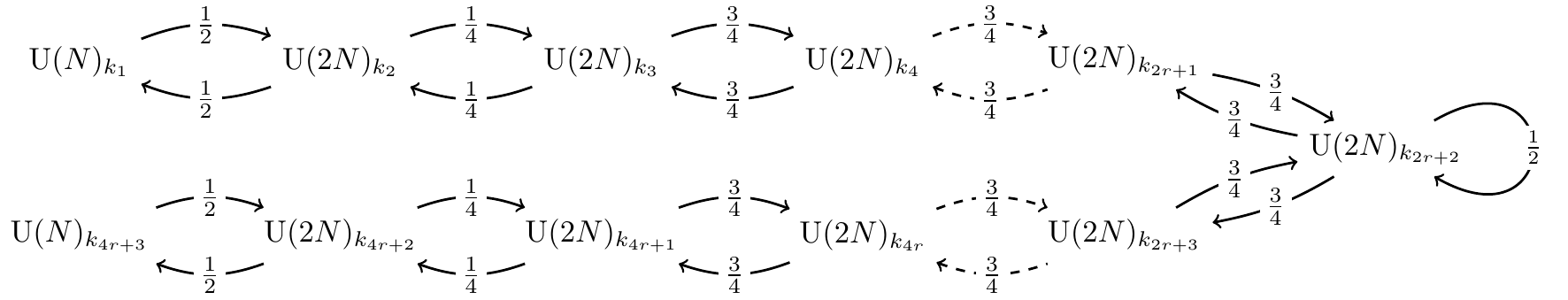}
\end{equation}
\item \label{fam7}
The gauge groups is 
$\U(N)_{k_1} \times \U(2N)_{k_2}
\times 
\dots \times  \U(2N)_{k_{4r}} \times \U(N)_{k_{4r+1}}$
with the constraint $\sum_{a=1}^{4r+1} N_a k_a = 0$. The superpotential reads
\begin{align}
W_{7} =&\
(Q_{1,2}Q_{2,1})^2
+
Q_{2,1} Q_{1,2} (Q_{2,3} Q_{3,2})^2
+
\sum_{i=2}^{2r-1} Q_{i,i+1} Q_{i+1,i+2} Q_{i+2,i+1} Q_{i+1,i}\ +
\nonumber 
\\
&\ +
Q_{2r,2r+1} {\Phi_{2r+1,2r+1}} Q_{2r+1,2r}
+
Q_{2r+2,2r+1} {\Phi_{2r+1,2r+1}} Q_{2r+1,2r+2}\ +
\nonumber 
\\
&\ +
\sum_{i=2r+1}^{4r-1} Q_{i,i+1} Q_{i+1,i+2} Q_{i+2,i+1} Q_{i+1,i}
+
(Q_{4r,4r+1} Q_{4r,4r+1}  )^2\ +
\nonumber 
\\
&\ +
(Q_{4r,4r-1} Q_{4r-1,4r})^2 Q_{4r,4r+1} Q_{4r+1,4r} \ .
\end{align}
The quiver is
\begin{equation}\label{fig:infVII}
\includegraphics[scale=.8]{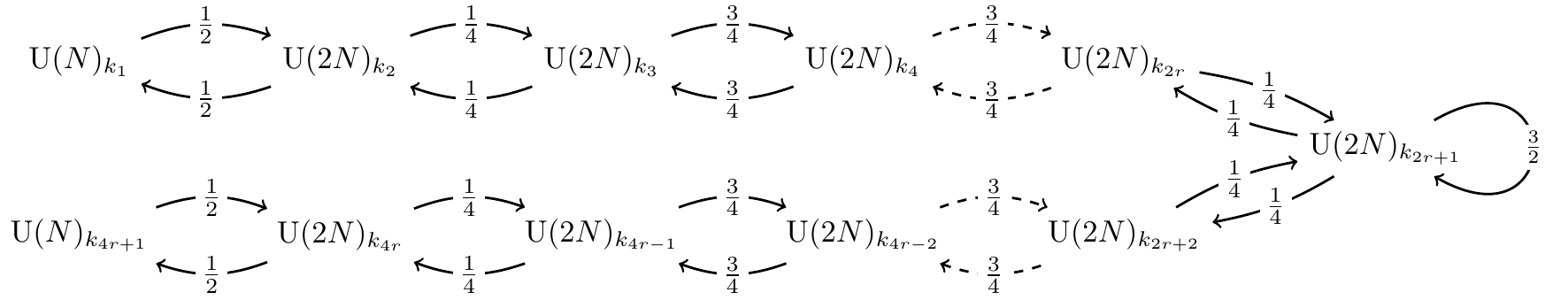}
\end{equation}
\end{enumerate}
Each of the above families admits a 4d parent, which is obtained as usual by forgetting the CS interactions and replacing the unitary groups with $\SU$ ones. We can then compute the $a$ and $c$ central charges, which turn out to agree at large $N$. (In other words, the gravitational anomaly $c-a=-\tfrac{1}{16}\Tr R$ is always an order $N^0$ number.) We can also extract the volume of the candidate gravity dual $B_5$ from the large-$N$ Hilbert series, and check that it matches with the one obtained from the $a$ central charge. (To prove the existence of a SE metric on $B_5$ one can follow the steps explained in Section \ref{4dholo}; the hypersurface equation(s) defining the CY$_3$ can be extracted from the large-$N$ Hilbert series whenever the former is a complete intersection.)
We summarize the results for a single representative at fixed $r$ of each infinite family in Table \ref{tab:inf}. 
\begin{table}[ht!]
\centering
\begin{tabular}{c|cccccc}
family &
$r$&&
$H(t)$  &
$\Vol(\text{SE}_5)/\pi^3$&
$a(N)/N^2$&
$c-a$\\
\hline
\rule{0pt}{20pt} \ref{fam1} &
6&&
$\ddfrac{1+t^7}{(1-t) (1-t^{3/2})(1-t^{13/2})}$ &
$\ddfrac{64}{1053}  $ &
$\ddfrac{1053}{256} $ &
$\ddfrac{1}{4}$\\[5pt]
\rule{0pt}{20pt} \ref{fam2} &
6&&
$\ddfrac{1+t^{27/4}}{(1-t) (1-t^{3/2}) (1-t^{25/4})}$ 
&
$\ddfrac{128 }{2025} $
&
$\ddfrac{2025}{512}$
&
$\ddfrac{11}{64}$
\\[5pt]
\rule{0pt}{20pt} \ref{fam3} &
4&&
$\ddfrac{1+t^{15/4}}{(1-t) \left(1-t^{3/2}\right) \left(1-t^{13/4}\right)}$ 
&
$\ddfrac{128 }{1053}  $
&
$\ddfrac{1053}{512}$
&
$\ddfrac{13}{64}$
\\[5pt]
\rule{0pt}{20pt} \ref{fam4} &
5&&
$ \ddfrac{1+t^{19/4}}{(1-t) \left(1-t^{3/2}\right) \left(1-t^{17/4}\right)} $ &
$\ddfrac{128 }{1377}  $ &
$\ddfrac{1377}{512}$ &
$\ddfrac{19}{64}$ 
\\[5pt]
\rule{0pt}{20pt} \ref{fam5} &
6&&
$ \ddfrac{1+t^5}{(1-t) (1-t^{3/2})(1-t^{9/2})} $ &
$\ddfrac{64 }{729}$ &
$\ddfrac{729}{256}$ &
$\ddfrac{3}{8}$ 
\\[5pt]
\rule{0pt}{20pt} \ref{fam6} &
7&&
$\ddfrac{1+t^{13/2}}{(1-t) \left(1-t^{3/2}\right) \left(1-t^6\right)}$ &
$\ddfrac{16 }{243}  $ &
$ \ddfrac{243}{64} $ &
$\ddfrac{13}{32}$
\\[5pt]
\rule{0pt}{20pt} \ref{fam7} &
5&&
$
\ddfrac{1+t^{7/2}}{(1-t) \left(1-t^{3/2}\right) \left(1-t^3\right)}
$
&
$\ddfrac{32 }{243}  $ &
$\ddfrac{243}{128}$ &
$\ddfrac{11}{32}$
\end{tabular}
\caption{We collect the large-$N$ Hilbert series, the volume of the (five-dimensional internal space of the) putative gravity dual $\text{AdS}_5 \times B_5$, the $a$ central charge, and the gravitational anomaly for the seven infinite families of 4d models obtained from the 3d ones (by forgetting the CS interactions and replacing U with SU). The total number of gauge groups is related to the chosen value of $r$, as explained in Appendix \ref{families}.}
\label{tab:inf}
\end{table}

\section{\texorpdfstring{Review: the free energy at large $N$}{Review: the free energy at large N}}
\label{AppAnton}

In this Section we collect some results for the evaluation of the large-$N$ free energy
that are useful for our analysis.
We start by summarizing the various contributions to the free energy functional 
$F\left[ \{\lambda\} \right]$ in (\ref{partition:function:gen}):
\begin{itemize}
  \item Chern--Simons terms:
    \begin{equation}
    F^{(a)}_{\mathrm{CS} }\left[ \{\lambda\} \right]=-\frac{i}{4\pi}\sum\limits_{i=1}^{n_a N}k_a\lambda_{a,i}^2\,,
    \label{CS:free:energy}
    \end{equation}
    where $k_a$ is the CS level at node $a$. 
   \item Vector multiplet:
     \begin{equation}
     F^{(a)}_{\mathrm{vec.} }\left[ \{\lambda\} \right]=-\sum\limits_{i<j}\log\left[4\sinh^2\left( \frac{\lambda_{a,i}-\lambda_{a,j}}{2} \right) \right]\,.
     \label{vector:free:energy}
     \end{equation}
   \item Adjoint chiral multiplet:
     \begin{align}
     F^{(a)}_{\mathrm{adj.} }\left[ \{\lambda\},\,\Delta \right] =& -\frac{1}{2}\sum\limits_{i,j=1}^{n_a N}\left[l\left(1-\Delta+\frac{i}{2\pi}
       \left(\lambda_{a,j}-\lambda_{a,i}\right)\right)\nonumber
       \right.\\ 
        &\left. 
       +\ l
       \left(1-\Delta+\frac{i}{2\pi}\left(\lambda_{a,i}-\lambda_{a,j}  \right)  \right)\right]\, ,
     \label{adjoint:free:energy}
     \end{align}
     where $\Delta$ is the R-charge of the multiplet and we have symmetrized all the expressions. The $l$-function used above was defined by Jafferis in \cite{Jafferis:2010un}:
     \begin{align}
     l(z)\equiv -z\log\left( 1-e^{2\pi i z} \right)+\frac{i}{2}\left( \pi z^2 +\frac{1}{\pi}\mathrm{Li}_2\left( e^{2\pi i z} \right)\right)-\frac{i \pi}{12}\ .
     \label{l:function}
     \end{align}
   \item Bifundamental chiral multiplet connecting nodes $a$ and $b$:
     \begin{align}
     F^{(a,b)}_{\mathrm{bif.} }\left[ \{\lambda\},\,\Delta_{ab} \right]=-\sum\limits_{i=1}^{n_a N}\sum\limits_{j=1}^{n_b N}l
     \left( 1-\Delta_{ab}+\frac{i}{2\pi}\left(\lambda_{a,i}-\lambda_{b,j} \right)\right)\ .
     \label{bifund:free:energy}
     \end{align}
    \item Fundamental and anti-fundamental multiplets:
     \begin{align}
     F^{(a)}_{\mathrm{fund.} }\left[ \{\lambda\},\,\Delta \right]=-\sum\limits_{i=1}^{n_a N} l\left( 1-\Delta +\frac{i}{2\pi}\lambda_{a,i}\right)\,,\nonumber\\
     F^{(a)}_{\mathrm{afund.} }\left[ \{\lambda\},\,\Delta \right]=-\sum\limits_{i=1}^{n_a N} l\left( 1-\Delta -\frac{i}{2\pi}\lambda_{a,i}\right)\,.
     \label{fund:free:energy}
     \end{align}
 \end{itemize}
In order to find the large-$N$ free energy from these expressions we study the solutions of the saddle-point equations:
     \begin{align}
     \frac{d F\left[ \{\lambda\} \right]}{d\lambda_{a,i}}=0\, .
     \label{saddle:point:general}
     \end{align}
     Then after finding solutions for these equations we can substitute them back into the partition function and get the final result for the free energy. 
     The following expressions  are very useful
     to find 
     the saddle-point equations: 
\begin{subequations}\label{eigenvalue:forces}
\begin{align}
\frac{\partial F^{(a)}_{\mathrm{CS} }}{\partial \lambda_{b,i}} =& -\delta_{a,b}\frac{i k_a}{2\pi}\lambda_{a,i}\ , \\
 \frac{\partial F^{(a)}_{\mathrm{vec.} }}{\partial \lambda_{b,i}}= &-\delta_{a,b}\sum\limits_{i\neq j}^{n_a N} 
     \coth\left( \frac{\lambda_{a,i}-\lambda_{a,j}}{2}\right)\ ,\\
     \frac{\partial F^{(a)}_{\mathrm{adj.} }}{\partial \lambda_{b,i}}=&-\delta_{a,b}\sum\limits_{j=1}^{n_a N}\left(\left[ \frac{\Delta-1}{2}
    -i\frac{\lambda_{a,i}-\lambda_{a,j}}{4\pi}\right]\coth\left[\frac{\lambda_{a,i}-\lambda_{a,j}}{2}-i \pi \left( 1-\Delta \right) \right]+\right.\nonumber\\
     &+ \left.\left[ \frac{\Delta-1}{2}
     +i\frac{\lambda_{a,i}-\lambda_{a,j}}{4\pi}\right]\coth\left[\frac{\lambda_{a,i}-\lambda_{a,j}}{2}+i \pi \left( 1-\Delta \right) \right]\right)\ , \\
 \frac{\partial F^{(a,b)}_{\mathrm{bif.} }}{\partial \lambda_{c,i}} =&-\delta_{a,c}\sum\limits_{j=1}^{n_b N}
     \left[ \frac{\Delta_{ab}-1}{2} -i\frac{\lambda_{a,i}-\lambda_{b,j}}{4\pi}\right]\coth\left[\frac{\lambda_{a,i}-\lambda_{b,j}}{2}-i \pi \left( 1-\Delta_{ab} \right) \right] +\nonumber\\
&-\delta_{b,c}\sum\limits_{j=1}^{n_a N}\left[ \frac{\Delta_{ab}-1}{2}
     +i\frac{\lambda_{b,i}-\lambda_{a,j}}{4\pi}\right]\coth\left[\frac{\lambda_{b,i}-\lambda_{a,j}}{2}+i \pi \left( 1-\Delta_{ab} \right) \right]\ , \\
      \frac{\partial F^{(a)}_{\mathrm{(a)fund.} }}{\partial \lambda_{b,i}}=& -\delta_{a,b}\left[ \frac{\Delta-1}{2}
   \mp i\frac{\lambda_{a,i}}{4\pi}\right]\coth\left[\frac{\lambda_{a,i}}{2}+i \pi \left( 1-\Delta \right) \right]\ .
     \end{align}
\end{subequations}
%
     As usual, if we interpret the matrix model (\ref{partition:function:gen}) as a 1d system of interacting particles at positions $\lambda_{a,i}$,
     then the expressions above play the role of forces. In particular the CS term together with the contribution (anti)fundamental multiplets
     in this picture plays the role of the central potential force, while all other terms describe interactions between the various groups of particles.
     Equivalently we can refer to them as interactions between eigenvalues.


     We are interested in finding solutions to the saddle-point equations. In general this is a complicated problem. Hence we will always work in the large-$N$ approximation which greatly
     simplifies the problem of finding solutions. 
      Nevertheless, in order to develop intuition and check the 
      large-$N$ results, it is very useful to find explicit numerical solutions 
      for various values of the parameters that we have in theory. To find these numerical solutions we use the standard technique of the heat equation. In particular, instead of 
      solving equations (\ref{saddle:point:general}), we solve the following system of heat equations:
      \begin{equation}
      \tau \frac{d \lambda_{a,i}(t)}{dt}=\frac{dF\left[ \{\lambda\} \right]}{d\lambda_{a,i}}\,,
      \label{heat:equations}
      \end{equation}
      where we artificially introduce a ``time" variable $t$, which our eigenvalues $\lambda_{a,i}$ depend on. The parameter $\tau$ in the equations above plays 
      the role of the heat capacity. With the appropriate choice of this parameter and initial conditions at $t=0$, the solution of (\ref{heat:equations}) converges to the solutions 
      of our original system of equations (\ref{saddle:point:general}) asymptotically at large times $t\to\infty$. 


An important class of theories, many of which have holographic duals, comprises those with no long-range forces. This means that the forces 
     between the eigenvalues cancel at large separations $|\lambda_{a,i}-\lambda_{b,j}|\gg 1$. Using the expressions (\ref{eigenvalue:forces}) we can derive corresponding expressions for long-range forces considering the large separation limit:
\begin{subequations}\label{long:range:forces}
     \begin{align}
     \left.\frac{\partial F^{(a)}_{\mathrm{vec.} }}{\partial \lambda_{b,i}}\right|_\text{LR} &= -\delta_{a,b}\sum\limits_{i\neq j}^{n_a N} \mathrm{sign}\,\mathrm{Re}\left(\lambda_{a,i}-\lambda_{a,j}\right)\ , \\
  \left. \frac{\partial F^{(a,b)}_{\mathrm{bif.} }}{\partial \lambda_{c,i}}\right|_\text{LR} &=-\delta_{a,c}\sum\limits_{j=1}^{n_b N}\left[ \frac{\Delta_{ab}-1}{2}
	 -i\frac{\lambda_{a,i}-\lambda_{b,j}}{4\pi}\right]\mathrm{sign}\,\left[\mathrm{Re}\left(\lambda_{a,i}-\lambda_{b,j}\right)\right]+\nonumber\\
  &\quad-\delta_{b,c}\sum\limits_{j=1}^{n_a N}\left[ \frac{\Delta_{ab}-1}{2}
     +i\frac{\lambda_{b,i}-\lambda_{a,j}}{4\pi}\right]\mathrm{sign}\,\left[\mathrm{Re}\left(\lambda_{b,i}-\lambda_{a,j}\right)\right]\ ,\\
     \left. \frac{\partial F^{(a)}_{\mathrm{adj.} }}{\partial \lambda_{b,i}}\right|_\text{LR} &= -\delta_{a,b}\left( \Delta-1 \right)
     \sum\limits_{j=1}^{n_a N}\mathrm{sign}\,\left[\mathrm{Re}\left(\lambda_{a,i}-\lambda_{a,j}\right)\right]\ .
       \end{align}
       \end{subequations}    
First of all, let us split the eigenvalues of all nodes in groups of equal sizes:
     \begin{eqnarray}
        \lambda_{a,i}\rightarrow \lambda_j^{(a,I)}\,,
     \label{eigenvalue:splitting}
     \end{eqnarray}
     where indices on the l.h.s. run through $a=1,\dots,r,\,i=1,\dots,n_aN$ and on the r.h.s. $a=1,\dots,r,\, I=1,\dots,n_a,\, j=1,\dots,N$. There is a one-to-one 
     correspondence $i\leftrightarrow (I,j)$ between the indices of eigenvalues on two sides of (\ref{eigenvalue:splitting}).
     
     Now using the expressions in (\ref{long:range:forces}) let us write down the long-range forces acting on the eigenvalue $\lambda_{a,i}$
     for the general quiver theory described in Section \ref{subsec:setting}:
     \begin{align}
       \left.\frac{\partial F\left[ \{\lambda\},\{\Delta\} \right]}{\partial \lambda_{i}^{(a,I)}}\right|_{\mathrm{LR}}=&
       \left( 1+\sum\limits_{i \in \mathrm{adj}~ a}\left( \Delta_i^{(a)}-1 \right) \right)\sum\limits_{J=1}^{n_a}\sum\limits_{j=1}^{n_aN}\sign\,
       \left[\Re \left( \lambda_i^{(a,I)}-\lambda_j^{(a,J)} \right)\right]+\nonumber\\
       &+
       \sum\limits_{b\in (a,b)}\sum\limits_{J=1}^{n_b}\sum\limits_{j=1}^N\left( \frac{\Delta_{ab}+\Delta_{ba}}{2}-1 \right)\sign\,
       \left[\Re \left( \lambda_i^{(a,I)}-\lambda_j^{(b,J)} \right)\right]\nonumber \\
       =&\ 0\ , 
     \end{align}
      where in the first term on the r.h.s the sum runs over adjoints based at node $(a)$ and in the second term the sum runs over all nodes $(b)$ connected to node $(a)$ by bifundamentals. The two terms in the equation above in general have different functional dependence, so that it is difficult to simultaneously satisfy the condition of long-range force cancellation for all of the eigenvalues. In order to do this we should assume that the distributions of eigenvalues inside each of the groups 
      are the same along the real axis, and only the imaginary part of the eigenvalues differs for different groups and nodes. Namely, we assume: 
      \begin{align}
      \lambda_i^{(a,I)}=N^\alpha\,x_i+iy_i^{(a,I)}\,,\quad \forall ~ ~ a,I\,.
       \label{eigenvalue:ansatz}
      \end{align}
      This Ansatz imposed by the long-range force cancellation was used before in many works \cite{Herzog:2010hf,Jafferis:2011zi} and in particular 
      in  \cite{Crichigno:2012sk,Gulotta:2011vp} to consider $D$-type quivers, which, like our quivers, feature gauge groups of different ranks.
Using this Ansatz the long-range force cancellation condition reduces to the following simple algebraic equation for the parameters of the gauge theory:
      \begin{equation}
      \sum\limits_{b\in (a,b)}n_b\left( \Delta_{ab}+\Delta_{ba} \right)+2n_a\sum\limits_{i\in \mathrm{adj.}~a}\Delta_i^{(a)}=2\left[ n_a\,\left(n^{(a)}_\text{adj.}-1\right)
      +\sum\limits_{b\in (a,b)}n_b\right]\ .
      \label{long:range:force:condition}
      \end{equation}
         In order to compare our results with the holographic dual computations we should consider the large-$N$ limit on field theory side. In this limit the matrix models greatly simplify. We will always be interested in quiver theories which satisfy the long-range force cancellation condition (\ref{long:range:force:condition}), as well as the following relation between levels and ranks of the gauge groups,
       \begin{equation}
       \sum\limits_{a=1}^{r}k_an_a=0\ ,
       \label{level:condition}
       \end{equation}
and numbers of fundamental and anti-fundamental multiplets of node $a$ (which \emph{need not} be equal for chiral-like fundamental flavors):
      \begin{equation}
      \sum\limits_{a=1}^r n_a\left(n_{\mathbf{f},a}-n_{\overline{\mathbf{f}},a}  \right)=0\ .
       \label{fundamental:condition}
      \end{equation}
Provided all these conditions are satisfied, the theory exhibits the $N^{3/2}$ scaling of the (planar) free energy, that is expected from the holographic dual theory.

Let us take the large-$N$ limit of all expressions (\ref{CS:free:energy})-(\ref{bifund:free:energy}) contributing to the free energy functional, and in the corresponding expressions 
       (\ref{eigenvalue:forces}) for the forces acting between eigenvalues. Thanks to this limit it makes sense to pass to continuous 
       distributions in our Ansatz (\ref{eigenvalue:ansatz}), by replacing $x_i$ and $y_i^{(a,I)}$ with the continuous functions $x(s)$ and $y^{(a,I)}(s)$, such that $x_i=x\left( i/N \right)$ and 
       $y_i^{(a,I)}=y^{(a,I)}\left( i/N \right)$. We also introduce an eigenvalue density along the real axis: 
       \begin{equation}
       \rho(x)=\frac{ds}{dx}\,,
       \label{eigenvalue:density:def}
       \end{equation}
       satisfying the normalization condition
       \begin{equation}
       \int_{x_1}^{x_2} dx \rho(x)=1\,,
	\label{eig:dens:norm}
       \end{equation}
       where $x_{1,2}$ denote the endpoints of the distribution's support.

       The imaginary parts of the eigenvalues are then expressed as functions $y_{(a,I)}(x)$ of $x$. All sums over eigenvalues in the expressions for the free energy functional and forces turn into integrals over $x$:
       \begin{equation}
       \sum_{i=1}^N\to N\int dx \rho(x)\,.
       \end{equation}
       Also, the leading order contribution in $N$ cancels due to the long-range force cancellation (\ref{long:range:force:condition}), and 
       one needs to expand all the expressions to the sub-leading order. This has been done for quiver theories with equal ranks in \cite{Jafferis:2011zi}, and 
       for $D$-type quivers in \cite{Gulotta:2011vp}. Below we present a generalization of these rules to the quivers considered in this paper (i.e. with varying ranks and adjoint fields).

        Let us write down the large-$N$ contributions of various multiplets to the free energy functional:
	\begin{itemize}
	  \item Chern--Simons term of node $a$:
	     \begin{equation}
	        F^{(a)}_{\mathrm{CS}}=\frac{k_a}{2\pi}N^{1+\alpha}\int dx \rho(x)\,x\,y^{(a,I)}(x)-\frac{i k_a}{4\pi}N^{1+2\alpha}\int dx \rho(x)\,x^2\,,
	     \label{free:energy:CS:largeN}
	     \end{equation}
	     where we have omitted subleading term of order one. 
	   \item For the bifundamental chiral  multiplet contribution combined with that of the  vector multiplet
	     we use the following expression (that can be found in \cite{Jafferis:2011zi}):
	     \begin{align}
	     F_\text{bif.}^{(a,b)} =&\ \frac{N^{2-\alpha}}{12\pi}\sum\limits_{I=1}^{n_a}\sum\limits_{J=1}^{n_b}
	     \int dx\,\rho(x)^2\,\left( \pi^2-f_{(aI,bJ)}\left(x,\Delta_{ab}\right)^2 \right) \cdot
	     \nonumber \\
	     & \cdot 
	     \left[ 2f_{(aI,bJ)}\left( x,\Delta_{ab} \right)-3\delta y_{(aI,bJ)}(x)-6\pi\left( \Delta_{ab}-1 \right) \right]\,,
	     \label{free:energy:bifund:largeN}
	     \end{align}
	     where the function $f_{(aI,bJ)}(x,\Delta)$ is defined as follows:
	     \begin{equation}
	        f_{(aI,bJ)}(x,\Delta) \equiv\Arg \left[\exp \left({i\delta y_{aI,bJ}(x)
	        +2\pi i \left( \Delta-1/2 \right) } \right)\right]\ ,
	     \label{f:function}
	     \end{equation}
	     and 
	     \begin{align}
	     \delta y_{(aI,bJ)}(x)\equiv y^{(a,I)}(x)-y^{(b,J)}(x)\,.
	     \end{align}

             Notice that when solving the matrix models, the authors in \cite{Jafferis:2011zi} assume that 
	     $\left|\delta y_{aI,bJ}+\pi \Delta_{ab}-\pi \Delta_{ba}\right|\leq \pi\left(\Delta_{ab}+\Delta_{ba}  \right)$, i.e. arguments of $\Arg$ belong to the principal 
	     value. This simplifies considerations a lot. As we will see later on, for us it is crucial to keep the most general form of expression (\ref{free:energy:bifund:largeN}).

	   \item The contribution of an adjoint chiral of R-charge $\Delta$  is given by half the contribution of a pair of 
	     bifundamental multiplets (\ref{free:energy:bifund:largeN}) with $a=b$ and $\Delta_{ab}=\Delta_{ba}=\Delta$:
	     \begin{align}
	     F_\text{adj.}^{(a)}=&\ \frac{N^{2-\alpha}}{24\pi}\sum_{I,J=1}^{n_a}
	     \int dx\,\rho(x)^2\,\left( \pi^2-f_{(aI,aJ)}\left(x,\Delta\right)^2 \right) \cdot 
	     \nonumber \\
&\cdot
	     \left[ 2f_{(aI,aJ)}\left( x,\Delta \right)-3\delta y_{(aI,aJ)}(x)-6\pi\left( \Delta-1 \right) \right]+\left(I\leftrightarrow J  \right)\, .
	      \label{free:energy:adjoint:largeN}
	     \end{align}

          \item The contribution of fundamental and anti-fundamental chirals of R-charge $\Delta$ is given by 
	  \begin{align}
           F_\text{fund.}^{(a)\!}&=\frac{i n_a}{8\pi}N^{1+2\alpha}\!\!\int \!\! dx \rho(x)\,x|x|\! -\! N^{1+\alpha}\sum\limits_{I=1}^{n_a}\int\!\! dx \rho(x)\,|x|
	   \left(\!\frac{1}{4\pi}y^{(a,I)}(x)+\frac{\Delta-1}{2}\! \right)\, ,\nonumber\\
	   F_\text{afund.}^{(a)}\! &=-\frac{i n_a}{8\pi}N^{1+2\alpha}\!\! \int \!\! dx \rho(x)\,x|x|\! +\! N^{1+\alpha}\sum\limits_{I=1}^{n_a}\int\!\! dx \rho(x)\,|x|
	   \left(\!\frac{1}{4\pi}y^{(a,I)}(x)-\frac{\Delta-1}{2} \!\right)\, .
	   \label{free:energy:fund:largeN}
	  \end{align}

	   \item Finally there is a contribution to the free energy functional that comes solely from the vector multiplet corresponding to the interaction between 
	     different groups of eigenvalues labeled by $I$ and $J$, belonging to the same node $a$. To write down this expression we notice the following functional identity  
	     \begin{equation}
	     \log\left[ 4\sinh(\pi x)^2 \right]=l(-1-i x)+l(-1+ix)\, .
	     \end{equation}
Then the vector contribution in (\ref{vector:free:energy}) can be written in the following form:
\begin{align}
	     F^{(a)}_\text{vec.} [ \{ \lambda \}] =& -\frac{1}{2}\sum_{I\neq J}^{n_a} \sum_{i,j}^N \left[l\left( -1-\frac{i}{2\pi}\left(\lambda_i^{(a,I)}-\lambda_j^{(a,J)}  \right) \right)+ \right. \nonumber \\
	     & + \left. l\left( -1+\frac{i}{2\pi}\left(\lambda_i^{(a,I)}-\lambda_j^{(a,J)}  \right) \right) \right]\ ,
\end{align}
which can be treated as the contribution of an adjoint multiplet (\ref{adjoint:free:energy}) of R-charge $\Delta=2$. Then we can directly use 
the	     expression (\ref{free:energy:adjoint:largeN}) with $\Delta=2$. In particular using the symmetry $f_{(aI,aJ)}\left( x,\Delta=2 \right)=f_{(aJ,aI)}\left( x,\Delta=2 \right)$
	     we get:
	     \begin{equation}
	     F_{\mathrm{vec.}}^{(a)}=-\frac{1}{2}\sum_{I\neq J}^{n_a}N^{2-\alpha}\int dx \, \rho(x)^2 \left[ \pi^2- f_{(aI,aJ)}^2\left( x,\Delta=2 \right)\right]\, .
	     \label{free:energy:vec:largeN}
	     \end{equation}
	\end{itemize}
This is the full list of expressions we will need to work out the details of the matrix models and find the free energy of the quiver theories we are interested in. Let us comment on the comparison of the rules listed above with those presented in \cite{Gulotta:2011vp,Crichigno:2012sk}, were $D$-type quivers where 
   considered. Unfortunately the authors of these papers do not summarize the contributions of different multiplets, but instead directly write the full expression for the free energy.  By accurately comparing \cite[Eq. (2.4)]{Crichigno:2012sk} with the various contributions listed above we can see that expressions for the free energy 
   functionals match, but only for theories containing vector multiplets and pairs of bifundamentals or R-charge $\Delta_{ab}=\Delta_{ba}=1/2$ connecting them.
   On top of this, to see the matching explicitly it is important to use the long-range force cancellation condition (\ref{long:range:force:condition}).  Finally, while the draft of our paper was in preparation, \cite{Jain:2019lqb} appeared on the arXiv, discussing the free energies and twisted indices of ${\cal N}=2$ $A$ and $D$-type theories. 
The expressions in this paper are also grouped in a slightly different way but can be compared against ours, and shown to be the same after the long-range force cancellation is imposed.

Finally let us discuss the $N^{3/2}$ scaling of the free energy. In order to achieve an equilibrium distribution of eigenvalues, the central forces coming from the
CS terms and fundamental multiplets should balance out all other forces. If we want our free energy to scale as $N^{3/2}$ the leading $N^{1+2\alpha}$ terms 
in CS (\ref{free:energy:CS:largeN}) and fundamental multiplet (\ref{free:energy:fund:largeN}) should not contribute. This can be achieved provided both 
conditions (\ref{level:condition}) and (\ref{fundamental:condition}) are satisfied simultaneously. Then the remaining terms of order $N^{1+\alpha}$ should 
be balanced with the terms of order $N^{2-\alpha}$ coming from the contributions of bifundamental (\ref{free:energy:bifund:largeN}) and adjoint 
(\ref{free:energy:adjoint:largeN}) multiplets. This obviously leads to $\alpha=1/2$, and to desired $N^{3/2}$ scaling of the free energy.

\section{The toric phase}
\label{apptor}

In this Appendix we study the volumes of the toric phase
of the model discussed in Section \ref{C3Z2Z2}.
 We can study the volumes by using the tiling and the toric diagram
 using the algorithms introduced in \cite{Ueda:2008hx,Imamura:2008qs,Hanany:2008fj}.
 For completeness let us summarize the construction, referring the interested reader
to the original references for more complete explanations.
 
The quiver of a toric gauge theory can be equivalently represented on a two-torus.
Such a periodic quiver is named planar quiver and it encodes all the information
about the superpotential.
This  structure can be further dualized to obtain the dimer, also known as brane tiling.
In the brane tiling the edges correspond to the fields, the faces to the gauge groups and 
there is a bipartite structure of nodes corresponding to the signed superpotential terms.
A useful notion is the one of dimer covers, or perfect matchings (PM): they correspond to
all the sets of edges in which each node is incident to exactly one edge.
The toric diagram that encodes the singularity 
probed by the D3-brane in the holographic correspondence
 can be obtained from the homologies of the PM
w.r.t. the two cycles of the two-torus on which the 
 brane tiling lives.
\begin{figure}
\includegraphics[width=14cm]{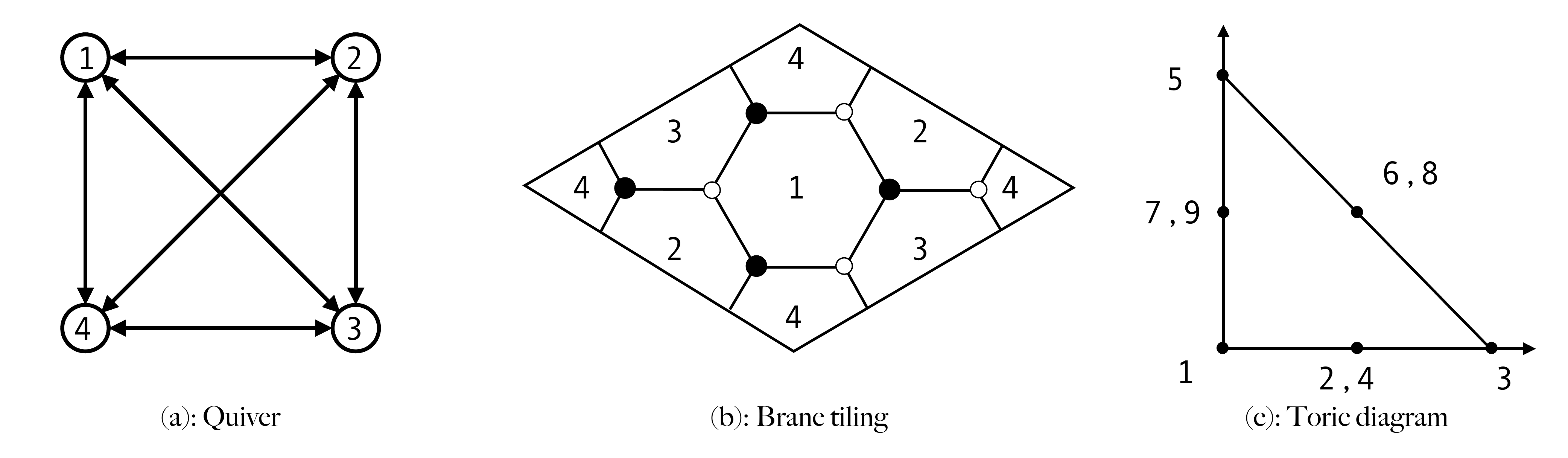}
\caption{Quiver, tiling and toric diagram of the 4d parent of the 3d model studied in Section \ref{C3Z2Z2}.}
\label{fig:toricC3Z2Z2}
\end{figure}
The lattice points associated to the PM in the toric diagram of Figure \ref{fig:toricC3Z2Z2} are
\begin{equation}
\begin{array}{lcccr}
\Pi_1 = Q_{41} Q_{14} Q_{23} Q_{32}\ ,
& &
\Pi_2 = Q_{12} Q_{34} Q_{32} Q_{14}\ ,
& &
\Pi_3 = Q_{12} Q_{34} Q_{43} Q_{21}\ ,
\\[10pt]
\Pi_4 = Q_{41} Q_{21} Q_{23} Q_{43}\ ,
& &
\Pi_5 = Q_{24} Q_{42} Q_{13} Q_{31}\ ,
& &
\Pi_6 =  Q_{12} Q_{13} Q_{43} Q_{42}\ ,
\\[10pt]
\Pi_7 = Q_{24} Q_{13} Q_{14} Q_{23}\ ,
& &
\Pi_8 = Q_{24} Q_{21} Q_{31} Q_{34}\ ,
& &
\Pi_9 = Q_{41} Q_{31} Q_{32} Q_{42}\ ,
\\
\end{array}
\end{equation}
where the enumeration refers to the one in the toric diagram.

This description can be used to study the 3d version of the model, where the gauge nodes are decorated by CS levels. 
Observe that this model is somehow unique.
At the geometric level  it is the only 4d 
vector-like model that has points lying on each side of the perimeter 
of the toric diagram.
It is also the only vector-like model that admits a non-toric phase
that can be studied using the 3d version of Seiberg duality.
Moreover the 3d version  of this theory has not been studied yet in the literature. In the 3d picture we treat the nodes as $\U(N)$ and the CS levels are chosen such that $k_1+k_2+k_3+k_4=0$.
Following the analysis of Section \ref{C3Z2Z2}, we can and will make two distinct choices:
\begin{enumerate}
\item[$i)$]
$k_1 = -k_2 = k_3=-k_4 = k$;
\item[$ii)$]
$k_1 = -k_2 =k$ and $k_3=-k_4 = 0$.
\end{enumerate}
The volume of the SE internal space can be computed from the toric diagram using the Reeb vector 
$\mathbf{b} = (b_1,b_2,b_3,4)$ \cite{Martelli:2005tp},
a constant-norm Killing vector field that commutes 
with the isometries of the SE$_7$ base of the conical CY$_4$.
We refer to the volume of the the base as $\Vol(\text{SE}_7)$,
and to the volumes of the five-cycles 
over which M5-branes can be wrapped
as $\Vol(\Sigma_i)$.
The R-charge of an M5 wrapped on a cycle $\Sigma_i$  is given by:
\begin{equation}
\Delta_i^\text{M5} = \frac{\pi}{6} \frac{\Vol(\Sigma_i)}{\Vol(\text{SE}_7)}\ .
\end{equation}
It identifies the R-charge of a PM associated to an 
external corner of the toric diagram as a function of the components of the Reeb 
vector.
The exact R-charge is obtained after minimization of the 
volume
\begin{equation}
\Vol(\text{SE}_7) = \frac{\pi^4}{12}  \sum_{i} \Vol(\Sigma_i)\ .
\end{equation}
The volumes $\Vol(\Sigma_i)$ can be calculated from the toric diagram thanks to the algorithm of 
\cite{Martelli:2005tp}, which was extended to three dimensions in \cite{Hanany:2008fj}.
By defining the four-vectors on the three-dimensional  lattice describing the toric diagram 
as $v_i=(w_i,1)$, and by considering the counterclockwise 
sequence $w_{k}$, $k=1,\dots,n_i$ of vectors adjacent to a given vector $v_i$, one has
\begin{equation}\label{MSvol}
\Vol(\Sigma_i) = \sum_{k=2}^{n_{i}-1}  \frac{
\langle v_i,w_{k-1},w_k,w_{k+1}\rangle \langle v_i,w_k,w_1,w_{n_i}\rangle}{\langle v_i,b,w_k,w_{k+1}\rangle
\langle v_i,b,w_{k-1},w_k\rangle \langle v_i,b,w_1,w_{n_i}\rangle }\ ,
\end{equation}
where  $\langle \cdot ,\cdot ,\cdot ,\cdot \rangle$ represents the determinant and the $\cdot $ are column vectors.

In the following we will compute the volumes by using this algorithm for the two choices of CS levels presented above.

\subsection{First Chern--Simons assignment}
We chose the levels as $k_1 = -k_2 = k_3=-k_4 = k$ and in this way the only PM that are lifted in the toric diagram of the 3d theory are $\Pi_2$ and $\Pi_4$, by $k$ and $-k$ units respectively. In the figure below we represent the toric diagram of the 3d theory for $k=1$.
\begin{equation}
\includegraphics[scale=.2]{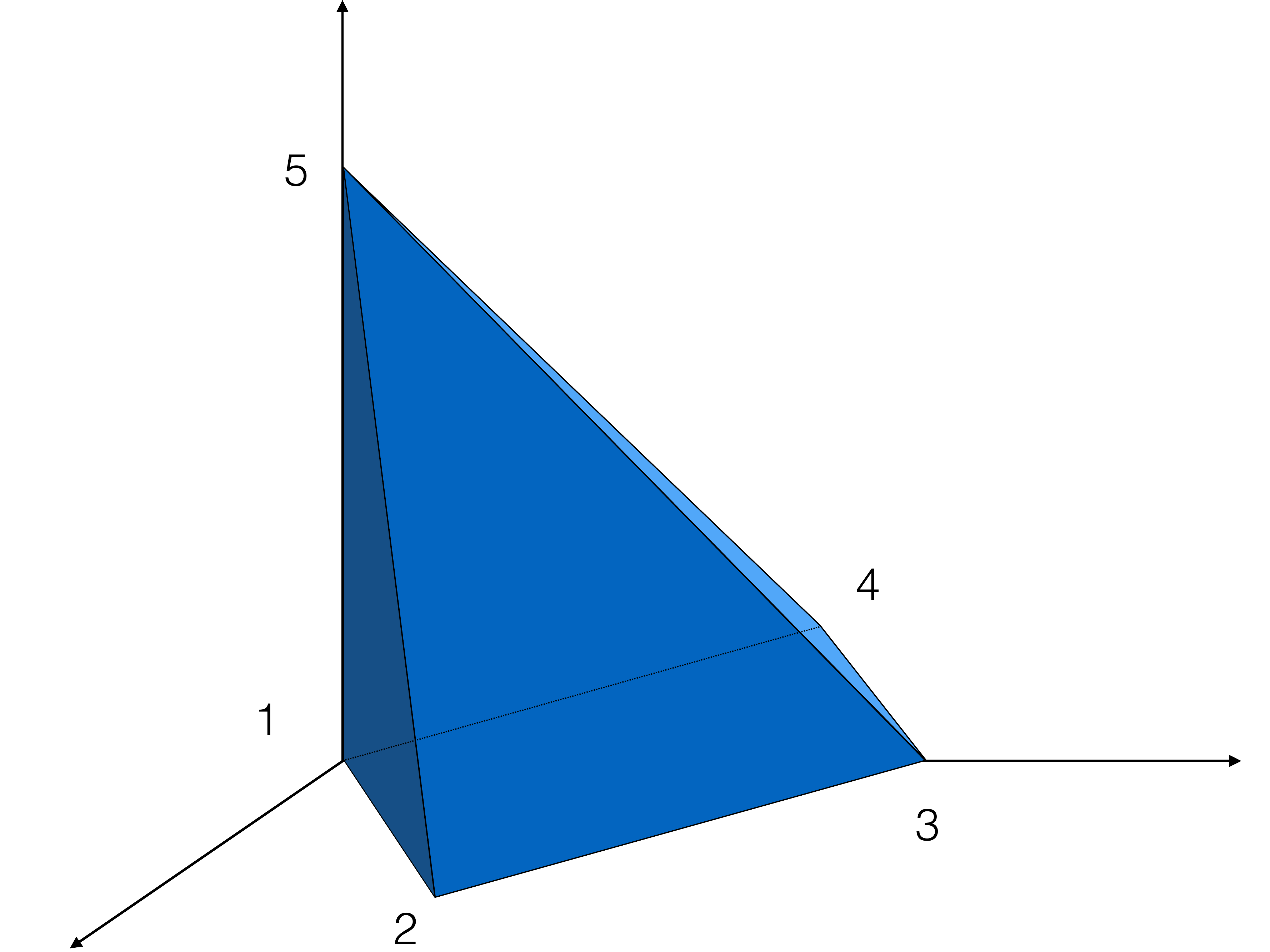} \nonumber
\end{equation}
From here we can compute the volumes, in terms of the components 
of the Reeb vector $\mathbf{b}= (b_1,b_2,b_3,4)$.
We have
\begin{eqnarray}
\Vol(\Sigma_1) &=& \frac{\pi^4}{12}
 \frac{2}{b_3(b_2^2-b_1^2)}\ , \nonumber \\
\Vol(\Sigma_2) &=&  \frac{\pi^4}{12} \frac{2}{b_3(b_2-b_1)(8-b_1-b_2-b_3)}\ , \nonumber \\
\Vol(\Sigma_3) &=&  \frac{\pi^4}{12} \frac{2}{b_3((b_2+b_3-8)^2-b_1^2)}\ ,  \\
\Vol(\Sigma_4) &=&  \frac{\pi^4}{12} \frac{2}{b_3(b_2+b_1)(8+b_1-b_2-b_3)}\ , \nonumber \\
\Vol(\Sigma_5) &=& \frac{\pi^4}{12} \frac{2(8-b_3)}{(b_2^2-b_1^2)((b_2+b_3-8)^2-b_1^2)}\ . \nonumber 
\end{eqnarray}
The volume of the internal space reads 
\begin{eqnarray}
\Vol(\text{SE}_7)  = \sum_{i=1}^5 \Vol(\Sigma_i) = 
\frac{4 \, \pi^4 (b_3-8)}{3 \, b_3(b_2^2-b_1^2)(b_1^2-(b_2+b_3-8)^2)}\ .
\label{vol:SE7:1}
\end{eqnarray}
This volume is minimized by $b_1=0$, $b_2 = 3$ and $b_3=2$, 
such that $\Vol(\text{SE}_7)  = \tfrac{4 \pi^4}{81}$.
We can also read off the R-charges from the geometry as follows.
The R-charges of the PM are given by 
\begin{equation}
\Delta_{\Pi_i} = \frac{2 \, \Vol(\Sigma_i)}{\Vol(\text{SE}_7)}\ ,
\label{R:charge:PM}
\end{equation}
and from here we can read the charges of the fields in terms of the 
charges of the PM.
We have
\begin{eqnarray}
&&
\Delta_{Q_{14}} = \Delta_{Q_{32}}
=
\Delta_{\Pi_1}+\Delta_{\Pi_2}
=
\frac{(8+b_1-b_2-b_3)}{4} 
\equiv 
R_1\ ,
\nonumber 
\\
&&
\Delta_{Q_{41}} = \Delta_{Q_{23}}
=
\Delta_{\Pi_1}+\Delta_{\Pi_4}
=
\frac{(8-b_1-b_2-b_3)}{4} 
\equiv 
R_2\ ,
\nonumber \\
&&
\Delta_{Q_{24}} = \Delta_{Q_{42}} = \Delta_{Q_{13}} = \Delta_{Q_{31}} 
= 
\Delta_{\Pi_5}
=
\frac{b_3}{4}
\equiv 
R_3\ ,
\\
&&
\Delta_{Q_{12}} = \Delta_{Q_{34}}
=
\Delta_{\Pi_2} + \Delta_{\Pi_3}
=
\frac{b_2+b_1}{4}
\equiv
2-R_2-R_3\ ,
\nonumber \\
&&
\Delta_{Q_{21}} = \Delta_{Q_{43}}
=
\Delta_{\Pi_3} + \Delta_{\Pi_4}
=
\frac{b_2-b_1}{4}
\equiv
2-R_1-R_3\ .
\nonumber 
\end{eqnarray}
At the fixed point the charges are $R_1 = R_2 = 3/4$ and $R_3 =1/2$.

We can now match the internal space volume (\ref{vol:SE7:1}) with the volume computed from the 
	free energy (\ref{FS3:nontoric:2})  of the dual phase. We can easily see that the values at the fixed point indeed match. We can also match the volumes away from the fixed point. For this we need to identify the R-charges $\Delta_i$ of the adjoint fields in the dual theory with the R-charges of the toric phase, and thus with the components $b_i$ of the Reeb vector.
The correct identification is:
\begin{equation}
\Delta_i=2-Y_{1i}-Y_{i1}\ ,\quad  Y_{j1}+Y_{1i}=\Delta_{Q_{ji}}\,.
\label{toric:nontoric:charges}
\end{equation}
To precisely match the volumes we also need to take $b_1=0$. In the non-toric phase this operation corresponds 
to setting the R-charges of all bifundamentals to be equal in pairs (i.e. $\Delta_{\tilde{Q}_{ij}}=\Delta_{\tilde{Q}_{ji}}$ in the dual phase).
Equivalently, this means that we turn off possible baryonic symmetries. Then, using the identification \eqref{toric:nontoric:charges}, we can indeed check that the volumes match even away from the fixed point.

\subsection{Second Chern--Simons assignment}

We choose levels $k_1=-k_2=k$, $k_3=k_4=0$. The toric diagram of the 3d theory for $k=1$ is shown in the figure below.
\begin{equation}
\includegraphics[width=6cm]{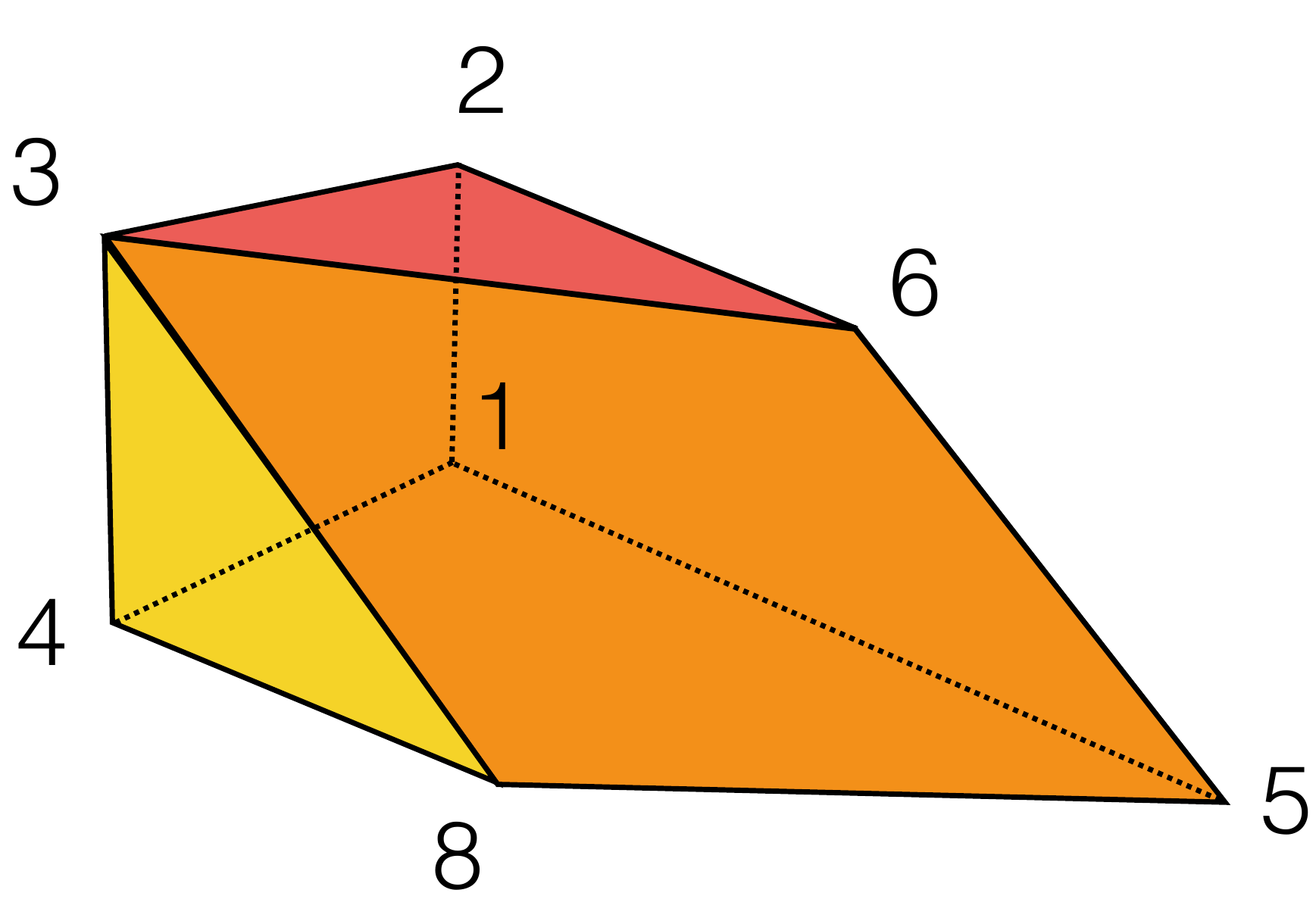} \nonumber
\end{equation}  
The volumes read:
\begin{eqnarray}
\Vol(\Sigma_1) &=&  \frac{\pi^4}{12}\,  \frac{1}{b_1 b_2 b_3}\ ,
\nonumber \\
\Vol(\Sigma_2) &=&\frac{\pi^4}{12} \, 
\frac{8-b_1-b_3}{ b_2 \left(4-b_1\right) \left(4-b_3\right) \left(8-b_1-b_2-b_3\right)}\ ,
\nonumber \\
\Vol(\Sigma_3) &=&  \frac{\pi^4}{12}\,  \frac{1}{b_2 b_3(4-b_1)}\ ,
\nonumber \\
\Vol(\Sigma_4) &=&  \frac{\pi^4}{12} \, \frac{1}{b_1 b_2 (4-b_3)}\ ,
\\
\Vol(\Sigma_5) &=& \frac{\pi^4}{12} \, \frac{1}{b_1 b_3 \left(8-b_1-b_2-b_3\right)}\ ,
\nonumber \\
\Vol(\Sigma_6) &=& \frac{\pi^4}{12} \, \frac{1}{b_3\left(4-b_1\right)  \left(8-b_1-b_2-b_3\right)}\ ,
\nonumber  \\
\Vol(\Sigma_8) &=&  \frac{\pi^4}{12} \, \frac{1}{b_1 \left(4-b_3\right) \left(8-b_1-b_2-b_3\right)}\ ,
\nonumber
\end{eqnarray}
  and
\begin{equation}
  \Vol(\text{SE}_7) =  \frac{4 \pi^4}{3}
  \frac{ \left(8-b_1-b_3\right)}{  b_1 b_2 b_3  \left(4-b_1\right)\left(4-b_3\right) \left(8-b_1-b_2-b_3\right)}\ .
  \label{volS7:2}
\end{equation}
  At the fixed point, $b_1 = b_3 = 8/5$, $b_2 =12/5$,
  and $\Vol(\text{SE}_7) = \tfrac{3125}{41472} \pi^4$.
  The R-charges are
\begin{eqnarray}
&&
 \Delta_{Q_{23}} = \Delta_{Q_{41}} =     
\Delta_{Q_{24}} =   \Delta_{Q_{42}} =  
\Delta_{Q_{13}} =  \Delta_{Q_{31}} = 
\Delta_{Q_{32}} =  \Delta_{Q_{14}} 
 = \frac{3}{5}  \ ,
 \nonumber \\
&&
\Delta_{Q_{12}} =   \Delta_{Q_{34}}  =  
\Delta_{Q_{21}} =   \Delta_{Q_{43}} = 
\frac{4}{5}\ .
\end{eqnarray}  
Just as for the first CS assignment, we can match the volumes (\ref{volS7:2}) obtained from the calculation above and from the free energy (\ref{FS3:nontoric:1}) of the dual phase. At the fixed point the match can be seen directly. Away from the fixed point we should once again use the identification 
(\ref{toric:nontoric:charges}) between the R-charges of the two dual theories, on top of the following relations for the R-charges of the toric phase:
\begin{equation}
\begin{array}{ll}
\Delta_{Q_{14}}=\Delta_{Q_{32}}=\Delta_{\Pi_{1}}+\Delta_{\Pi_{2}}\ ,
&
\hspace{.5cm}
\Delta_{Q_{23}}=\Delta_{Q_{41}}=\Delta_{\Pi_{1}}+\Delta_{\Pi_{4}}\ , 
\\
\Delta_{Q_{13}}=\Delta_{Q_{42}}=\Delta_{\Pi_{5}}+\Delta_{\Pi_{6}}\ ,
&
\hspace{.5cm}
\Delta_{Q_{24}}=\Delta_{Q_{31}}=\Delta_{\Pi_{5}}+\Delta_{\Pi_{8}}\ ,  
\\
\Delta_{Q_{12}}=\Delta_{\Pi_{2}}+\Delta_{\Pi_{3}}+\Delta_{\Pi_{6}}\ ,
&
\hspace{.5 cm}
\Delta_{Q_{21}}=\Delta_{\Pi_{3}}+\Delta_{\Pi_{4}}+\Delta_{\Pi_{8}}\ , 
\\
\Delta_{Q_{34}}=\Delta_{\Pi_{2}}+\Delta_{\Pi_{3}}+\Delta_{\Pi_{8}}\ ,
&
\hspace{.5 cm}
\Delta_{Q_{43}}=\Delta_{\Pi_{3}}+\Delta_{\Pi_{4}}+\Delta_{\Pi_{6}}\ ,
\end{array}
\end{equation}
where $\Delta_{\Pi_i}$ stands for the R-charges of the PM (that can be evaluated from the volumes $\Vol\left( \Sigma_i \right)$ using 
(\ref{R:charge:PM}).) Finally, in order to switch off possible baryonic symmetries (i.e. in order to set the R-charges of all bifundamentals to be equal in each pair) we should assume $b_3=b_1$ for the components of the Reeb vector. Under these assumptions it is straightforward to check that the volumes indeed match.

\section{The Hilbert series}
\label{appNoppie}
In this Appendix, we describe in detail and with examples how to compute the Hilbert series for the Calabi--Yau singularities associated with the quiver theories in this paper.  The two main purposes of the Hilbert series are to \emph{i)} understand the generators of the moduli space and the relations that they satisfy, and \emph{ii)} compute the volume of the base of the corresponding Calabi--Yau singularity. Under the assumption of existence of a SE metric on the base, computing the volume is equivalent to computing the free energy of the 3d field theory, which upon extremization allows one to extract the superconformal R-charge of the quiver fields.

The procedure to compute the Hilbert series can be described as follows.  First, we consider the 4d version of the quiver with $N=1$, so that the gauge groups are only $\U(1)$ and $\U(2)$.  From the F-terms, we then compute the Hilbert series of the space of the F-term solutions. The latter is also known as the Master space, and was studied extensively in \cite{Forcella:2008bb, Forcella:2008eh}. By integrating the Master space Hilbert series over the Haar measure of the gauge group, we obtain the Hilbert series for the Calabi--Yau \emph{threefold} associated with the quiver in question.  The latter can also be compared and is in agreement with the result obtained by the procedure of \cite{Eager:2010yu} in the large $N$ limit. 

After assigning appropriate Chern--Simons levels to the gauge group in the quiver, one can now follow the procedure describing in \cite{Cremonesi:2016nbo} and compute the Hilbert series for the corresponding Calabi--Yau \emph{fourfold}, a component of the moduli space of the 3d theory.  The procedure can be summarized as follows.  We multiply the Master space Hilbert series by the contribution of the monopole operators, then integrate over the Haar measure of the gauge group and sum over the magnetic flux sectors. The result is the Hilbert series of the required CY$_4$, which we shall denote by $H(t; \Delta)$, where  $\Delta$ denotes the trial R-charge and $t$ is the corresponding fugacity. We can obtain the volume function by considering the following limit \cite{Hanany:2008fj}:\footnote{The pre-factor $\frac{\pi^4}{48}$ can be determined, for example, by considering a cone over $S^7$, which is isomorphic to $\BC^4$. The Hilbert series of $\BC^4$ is $(1-t^{1/2})^{-4}$, where the power $1/2$ corresponds to the R-charge of the four free chiral fields in 3d. Setting $t = e^{-s}$ and taking the limit $s \rightarrow0$, we see that the coefficient of the leading pole $s^{-4}$ is $16$.  Since the volume of $S^7$ is $\frac{\pi ^4}{3}$, this fixes the pre-factor to be $\frac{\pi ^4}{3 \cdot 16} = \frac{\pi ^4}{48}$.}
\begin{equation}
\label{eq:VolSE7HS}
V(\Delta)=  \frac{\pi^4}{48} \lim_{s \rightarrow 0} s^4\,H(t \mapsto e^{-s}; \Delta)\ .
\end{equation}
This function can then be extremized, and let us denote the value of $\Delta$ at its extremum by $\Delta_*$.  The latter is the superconformal R-charge, and the volume of the seven-dimensional Sasaki--Einstein base of the Calabi--Yau cone is given by
\begin{equation}
\Vol (\text{SE}_7) = V(\Delta_*)~.
\end{equation}
In the following, we demonstrate each of the aforementioned steps in detail for Laufer's threefold and its Calabi--Yau fourfold counterpart.  For the other models, we simply state the expressions and the results.

\subsection{Laufer's threefold and its fourfold counterpart}
\label{LauferApp}
Let us consider the theory in Figure~\ref{fig:laufer3d} for $N=1$ and no CS interactions, i.e.~ the gauge group is taken to be $\U(1) \times \U(2)$. For the sake of brevity, let us define
\begin{equation} \label{defP}
P_{\alpha}^\beta = A_\alpha B^\beta~.
\end{equation}
We also redefine the adjoint fields for ease of notation, and denote their traceless part by a hat, as follows:
\begin{equation}\label{eq:lauferphipsiP}
\Phi_{22}\equiv \phi = \hat{\phi} +\frac{1}{2} (\Tr \phi) \mathbf{1}_{2}~, \quad \Psi_{22}\equiv \psi = \hat{\psi} +\frac{1}{2} (\Tr \psi) \mathbf{1}_{2}~, \quad P = \hat{P} +\frac{1}{2} (\Tr P) \mathbf{1}_{2}~.
\end{equation}
The superpotential in \eqref{superpotential:laufer} is taken to be
\begin{equation} \label{supLaufer}
\begin{split}
W &= \Tr( \hat{\phi} \hat{P}) \Tr(\phi) + (\Tr P) \Tr(\hat{\phi}^2)+ (\Tr P)^2 + (\Tr \phi)^2 \Tr( \hat{\phi}^2)\ + \\
& \quad + \Tr(\hat{\phi} \hat{\psi}) (\Tr \psi) + \Tr(\hat{\psi}^2) (\Tr \phi)~,
\end{split}
\end{equation}
where $\alpha, \beta=1,2$ are $\SU(2) < \U(2)$ gauge indices.
We focus on the branch of the moduli space in which
\begin{equation} \label{branchLaufer}
\Tr(\phi) =0~, \quad  \Tr(\psi) =0~, \quad \mathcal{F}_1 \equiv \Tr P+ \Tr(\hat{\phi}^2)=0~.
\end{equation}
The relevant $F$-terms are
\begin{equation} \label{FtermsLaufer}
\mathcal{F}_2 \equiv \Tr(\hat{\phi} \hat{P})+\Tr(\hat{\psi}^2)=0~, \quad \mathcal{F}_3 \equiv \Tr(\hat{\phi} \hat{\psi}) =0~.
\end{equation}
The moduli space is generated by the following gauge-invariant combinations:\footnote{Notice that these differ from those defined in \cite[Eq. (4.10)]{collinucci-fazzi-valandro} by F-terms.}
\begin{equation} \label{genLaufer}
W\equiv\Tr(\hat{\phi}^2)~, \quad  Y\equiv \Tr(\hat{\psi}^2)~, \quad Z \equiv \Tr(\hat{\psi} \hat{P})~, \quad X\equiv \Tr(\hat{\psi} \hat{\phi} \hat{P})~.
\end{equation}
E.g. it can be checked by expanding in components that
\begin{equation}
\begin{split}
&-\CF_1^2 \CF_3^2 + 2 \CF_1 \CF_3^2 W - \CF_3^2 W^2 + 4 X^2 - 2 \CF_2^2 Y + \CF_1^2 W Y - 
 2 \CF_1 W^2 Y + W^3 Y \\
& \quad + 4 \CF_2 Y^2 - 2 Y^3 + 4 \CF_2 \CF_3 Z - 4 \CF_3 Y Z - 
 2 W Z^2=0\ .
\end{split}
\end{equation}
Setting $\CF_1=\CF_2=\CF_3=0$, we obtain
\begin{equation}
4 X^2 -2 Y^3 -2 W Z^2+ W^3 Y =0~.
\end{equation}
Redefining as follows
\begin{equation} \label{redefgenLaufer}
X = 2^{-1}x~, \quad Y= -2^{1/3} y~, \quad W = -2^{1/9} w~, \quad Z= 2^{-5/9} z~,
\end{equation}
we obtain the hypersurface equation \eqref{eq:laufer} of Laufer's threefold:
\begin{equation} \label{defeqLauer}
x^2+y^3+w z^2+w^3 y =0~.
\end{equation}
Since each superpotential term has R-charge two, we may assign the following R-charges to the chiral fields:
\begin{equation}\label{eq:Rcharges}
R[A] =\frac{1}{4}r~, \quad R[B]= 1-\frac{1}{4}r~, \quad R[\phi] = \frac{1}{2}~, \quad R[\psi]=\frac{3}{4}~,
\end{equation}
Note that we leave $r$ undetermined for now.  As we have seen above, this unknown does not affect the defining relation of the Laufer variety \eref{defeqLauer}; indeed $A$ and $B$ always appear together as a gauge-invariant combination, which we called $P$.
The R-charges of $(x,y,w,z)$ are
\begin{equation}
R[w]= 1~, \quad R[y] = 3/2~, \quad R[z] = 7/4~, \quad R[x] =9/4~.
\end{equation}
The Hilbert series for the space of the F-term solutions (i.e.~the so-called Master space) is given by
\begin{equation}
\begin{split}
H[\CF^\flat](t; r; u_1,u_2, z) &= \PE \Big[ t^r u_1 u^{-1}_2  (z+z^{-1}) + t^{4-r} u^{-1}_1 u_2  (z+z^{-1}) \\
& \quad + t^2 (z^2+1+z^{-2}) + t^3(z^2+1+z^{-2}) - t^4 - t^5 -t^6 \Big]\ ,
\end{split}
\end{equation}
where $t$ is the fugacity of the R-symmetry such that its power counts the R-charge in the unit of $1/4$; $z$ is the $\SU(2)$ gauge fugacity; $u_1$ and $u_2$ are the $\U(1)$ gauge fugacities.\footnote{The plethystic exponential (PE) of a multi-variate function $f(x_1,\ldots,x_n)$ vanishing at the origin is defined as follows:
\begin{equation}
\text{PE}[f] \equiv \sum_{k=1}^\infty \frac{f(x_1^k,\ldots,x_n^k)}{k}\ .\nonumber
\end{equation}
}  Observe that there is only one combination of such $\U(1)$ gauge symmetries acting on the matter fields.  In particular, we may define $u \equiv u_1 u_2^{-1}$ and see that only this combination of $u_1$ and $u_2$ appears in the Hilbert series:
\begin{equation}
\begin{split}
H[\CF^\flat](t; r; u, z) &= \PE \Big[ t^r u  (z+z^{-1}) + t^{4-r} u^{-1}  (z+z^{-1}) \\
& \quad + t^2 (z^2+1+z^{-2}) + t^3(z^2+1+z^{-2}) - t^4 - t^5 -t^6 \Big]~.
\end{split}
\end{equation}
The Hilbert series of Laufer's threefold can be obtained by evaluating the following Molien integral:
\begin{equation}
H[\text{Laufer}](t) = \oint_{|u|=1} \frac{du}{2 \pi i u} \oint_{|z|=1} \frac{dz}{2 \pi i z} (1-z^2) H[\CF^\flat](t;r ; u, z) ~.
\end{equation}
Computing the integrals, we obtain
\begin{equation} \label{HSLauferApp}
H[\text{Laufer}](t) = \frac{1 -t^{18}}{(1-t^4)(1-t^6)(1-t^7)(1-t^9)} = \PE \left[ t^4 + t^6 + t^7 + t^9 - t^{18} \right]~.
\end{equation}
Recall that, here, $t$ counts the R-charge in the unit of $1/4$. This is indeed a three (complex) dimensional complete intersection with four generators carrying R-charges $1$, $3/2$, $7/4$ and $9/4$, subject to a constraint with R-charge $9/2$. Notice that these are nothing but the weights \eqref{eq:weights} under the $\cc^*$ action generated by the Reeb vector field \eqref{eq:reeblaufer} on Laufer's threefold (normalized so that $R[w]=1$, i.e.~ dividing the components by $3/2$).

\subsubsection*{The fourfold}
We may construct a CY$_4$ from Laufer's threefold as follows. Let us consider the quiver in Figure~\ref{fig:laufer3d} for $N=1$ (and with CS levels $(2k,-k)$); we use notation as in \eqref{eq:lauferphipsiP}. 
We take the superpotential to be as \eref{supLaufer} and focus on the branch of the moduli space described by \eref{branchLaufer}.  The moduli space is now generated not only by gauge-invariant combinations of the quiver fields, but also by dressed monopole operators.

The Hilbert series of this branch is given by
\begin{equation}
H[\text{Laufer}_k] (t;r;x)= \sum_{m=-\infty}^\infty \oint_{|u|=1} \frac{du}{2 \pi i u} \oint_{|z|=1} \frac{dz}{2 \pi i z} (1-z^2) H[\CF^\flat](t; r; u, z)  u^{-2k m} x^{m}~,
\end{equation}
where $x$ is the fugacity of the topological symmetry associated with the $\U(1)$ gauge symmetry.

Each term in this expression deserves some explanation.  The D-term equations restrict the flux of the monopole operator to be of the form $(m,m; m)$, such that $m \in \BZ$, under the $\U(1)_{2k} \times \U(2)_{-k}$ gauge symmetry.   Let us refer to the monopole operator with this flux as $V_m$.   The R-charge of $V_m$ is zero. Under the gauge $\U(1)$ combination that we are considering (whose fugacity is $u$), $V_m$ carries charge $-2km$.  Also, under the corresponding topological symmetry, $V_m$ carries charge $m$; this is denoted by the term $x^m$ in the Hilbert series.  In order to form gauge-invariant fields, the monopole operator $V_m$ has to be dressed with the combination of the chiral field with $\U(1)$ gauge charge $2km$; this explains the factor $u^{-2km}$ as well as the rest of the terms in the Hilbert series.  Finally, we need to sum over all magnetic flux sectors; this explains the infinite sum over $m\in \BZ$. 

For simplicity, we report the result for $x=1$ as follows.
\begin{align} \label{HS3dunref}
&H[\text{Laufer}_k] (t;r;x=1)= \\
&  \frac{D^{-1}(t) \,t^{2k(r-5)}}{(1-t^{2k(r-5)})(t^k - t^{2k(r-5)})} \left[  t^3 +\sum_{m=1}^5 t^m - \sum_{j=0}^6 t^{j + k} - \sum_{l=0}^2 (-1)^{l+1} \, t^{3l + 2k(r-5)}\right] + \nonumber \\
&+ \frac{D^{-1}(t)\,t^{k(2r+3)}}{(1-t^{k(2r+3)})(t^k - t^{k(2r+3)})} \left[ \sum_{m=0}^6 t^m- \sum_{j=1}^5 t^{j + k} - \sum_{l=0}^2 (-1)^{l} \, t^{3l + k(2r+3)}\right]~, \nonumber
\end{align}
where the first line comes from the sum over $m\geq 0$, the second from the sum over $m<0$, and we defined
\begin{equation}
D(t) \equiv 1 - t^3 - t^4 + t^{10} + t^{11} - t^{14}~.
\end{equation}
Let us report the result for $k=1$:
\begin{equation}
\begin{split}
&H[\text{Laufer}_{k=1}] (t;r;x=1) = \\
& \frac{1 + t^4 + t^6 + t^7 + t^8 + t^9 + t^{10} + t^{11} + t^{12} + t^{14} + t^{18} + \
t^{13 - 2 r} + t^{5 + 2 r}}{ \left(1-t^{2 r+2}\right) \left(1-t^{2 r+3}\right)  \left(1-t^{10-2 r} \right)\left(1-t^{11-2 r}\right) }~.
\end{split}
\end{equation}
From \eref{HS3dunref}, we may compute the volume of the SE$_7$ base of the CY$_4$ as follows. (This is under the assumption that a SE metric on the base of the CY$_4$ exists. Observe that for Laufer's CY$_3$, this was proven to be the case in Section \ref{sub:reeb}.) First, let us set $t= \exp \left(- \frac{1}{4} s \right)$, where $1/4$ comes from the fact that $t$ keeps track of R-charge in the unit of $1/4$.  Then, we consider the limit $s \rightarrow 0$ and obtain the following expansion:
\begin{align}
H[\text{Laufer}_k] (t= e^{-\frac{1}{4}s} ;r;x=1) =&\ \frac{832}{k (r-5) (r+1) (2 r-11) (2 r+3) s^4}\ + \nonumber \\
&+\frac{832}{k (r-5) (r+1) (2 r-11) (2 r+3) s^3}\ +\\
& +\frac{-56 r^2+224 r+1359}{3 k (r-5) (r+1) (2 r-11) (2 r+3) s^2} + O(s^{-1})\ .\nonumber
\end{align}
Observe that the leading order is $s^{-4}$; this indicates that the moduli space is four complex dimensional, as expected.  The coefficient of $s^{-4}$, multiplied by the factor $\frac{\pi^4}{48}$, gives rise to a function of $r$:
\begin{equation} \label{volfuncLaufer}
V(r)=  \frac{52 \pi ^4}{3 k (r-5) (r+1) (2 r-11) (2 r+3)}~.
\end{equation}
The value of this function at its local minimum is the volume of the SE$_7$ base of the CY$_4$ \cite{martelli-sparks-yau-volmin}. $V(r)$ attains its local minimum at $r=2$, where it equals
\begin{equation} \label{volmin}
\Vol(\text{SE}_7) \equiv V(r=2) = \frac{52 \pi ^4}{1323 k} = \frac{\pi^4}{k} \, \frac{2^2 \, 13}{3^3 \, 7^2} ~.
\end{equation}
The value $r=2$ also sets the R-charges of $A$ and $B$ to be $1/2$.  To summarize, the R-charges of the quiver fields are
\begin{equation}
R[A] =1/2~, \quad R[B]= 1/2~, \quad R[\phi] = 1/2~, \quad R[\psi]=3/4~.
\end{equation}
Let us now examine the chiral ring of the theory with $k=1$. The Hilbert series is
\begin{equation}
\begin{split}
H[\text{Laufer}_{k=1} ] (t; r =2;x)
=& \PE \Big[ t^4 + \left(x+1+\frac{1}{x}\right) t^6 + \left(x+1+\frac{1}{x}\right) t^7 \ + \\
&+ \left(x+1+\frac{1}{x}\right) t^9 - t^{12} - \left(x+2+\frac{1}{x}\right) t^{13} -t^{14} + \ldots \Big]~.
\end{split}
\end{equation}
Note that there are infinitely many terms inside the plethystic exponential, and so the space is no longer a complete intersection (which was instead the case for Laufer's threefold).  The first positive terms inside the $\PE$ indicate that the generators of the moduli space are (indicating their order by $t^n$):
\begin{equation}
\begin{array}{lllll}
&t^4:  &\quad W=\Tr(\hat{\phi}^2)\ ;  & \\
&t^6: &\quad \Upsilon_+ = V_{+1}\epsilon^{\beta \gamma} A_\alpha A_\beta \hat{\phi}^\alpha_{~\gamma}~, \quad &Y=\Tr(\hat{\psi}^2)~, \quad & \Upsilon_- = V_{-1}\epsilon_{\beta \gamma} B^\alpha B^\beta \hat{\phi}_{~\alpha}^{\gamma}\ ; \\
&t^7:  &\quad V_{+1}\epsilon^{\beta \gamma} A_\alpha A_\beta \hat{\psi}^\alpha_{~\gamma}~, \quad &Z=\Tr(\hat{\psi} \hat{P})~, \quad &V_{-1}\epsilon_{\beta \gamma} B^\alpha B^\beta \hat{\psi}_{~\alpha}\ ; \\
&t^9:  &\quad V_{+1} \epsilon^{\delta \gamma} A_\delta A_\alpha \hat{\phi}^\alpha_{~\beta} \hat{\psi}^{\beta}_{~\gamma}~, \quad & X=\Tr(\hat{\psi} \hat{\phi} \hat{P})~, \quad & V_{-1} \epsilon_{\delta \gamma} B^\delta B^\alpha \hat{\phi}^\beta_{~\alpha} \hat{\psi}^\gamma_{~\beta}\ .
\end{array} \\
\end{equation}
There is a relation at order $t^{12}$ that follows from the quantum relation
\begin{equation}
V_{+1} V_{-1} = 1~;
\end{equation}
namely:
\begin{equation}
\begin{split}
\Upsilon_+ \Upsilon_ - 
&=  \left(\epsilon^{\beta \gamma} A_\alpha A_\beta \hat{\phi}^\alpha_{~\gamma} \right) \left( \epsilon_{\beta \gamma} B^\alpha B^\beta \hat{\phi}_{~\alpha}^{\gamma}\right)\\
&=  \frac{1}{2} \Tr(\hat{\phi}^2) (\Tr P)^2 - [\Tr (\hat{\phi} \hat{P})]^2 \\
&= \frac{1}{2} [\Tr(\hat{\phi}^2)]^3 - [\Tr (\hat{\psi}^2)]^2 \\
&= \frac{1}{2} W^3 - Y^2~,
\end{split}
\end{equation}
where in obtaining the second equality we have used the identity $\epsilon^{\alpha \beta} \epsilon_{\alpha' \beta'} = \delta^{\alpha}_{\alpha'} \delta^{\beta}_{\beta'} - \delta^{\beta}_{\alpha'} \delta^{\alpha}_{\beta'}$, and for the third equality we have used \eref{branchLaufer} and \eref{FtermsLaufer}.

\subsection{The model in Section \ref{C3Z2Z2}}
\label{appC2}
Let us consider the non-toric phase of the 4d model in Figure~\ref{fig:SD}, obtained via Seiberg duality. Consider the case $N=1$. We take the R-charges of the quiver fields to be as follows:
\begin{equation}
\begin{array}{lll}
R[Q_{12}]=R[Q_{21}]= r_2~, &\quad R[Q_{13}]=R[Q_{31}]= r_3~, &\quad R[Q_{14}]=R[Q_{41}]= r_4\\
R[\Phi_{22}]= 2-2r_2~, &\quad R[\Phi_{33}] =2-2r_3~, &\quad  R[\Phi_{44}] = 2-2r_4~.
\end{array}
\end{equation}
In the following we shall consider the branch of the moduli space on which
\begin{equation} \label{branchtoric}
\Tr(Q_{12}Q_{21})=0~, \ \Tr(Q_{13} Q_{31})=0~, \ \Tr(Q_{14} Q_{41})=0~,\ \Phi_{22}=\Phi_{33}=\Phi_{44}=0\ .
\end{equation}
The Master space Hilbert series is given by
\begin{equation}
\begin{split}
H[\fflat](t; u_1, \ldots, u_4, z)
= & \PE \Big[ -t^{2r_2}  -t^{2r_3} -t^{2r_4} +t^{r_2} \left(\frac{u_1 }{u_2}+\frac{u_2}{u_1 }\right)\left(z+\frac{1}{z}\right) + \\
& +t^{r_3} \left(\frac{u_1 }{u_3}+\frac{u_3}{u_1 }\right)\left(z+\frac{1}{z}\right) +t^{r_4} \left(\frac{u_1 }{u_4}+\frac{u_4}{u_1 }\right)\left(z+\frac{1}{z}\right) \Big]~,
\end{split}
\end{equation}
where $u_2$, $u_3$ and $u_4$ are the $\U(1)$ gauge fugacities associated with nodes $2$, $3$ and $4$; $u_1$ is the fugacity for the $\U(1) < \U(2)$ gauge group corresponding to node $1$; $z$ is the fugacity for the $\SU(2) < \U(2)$ gauge group corresponding to node $1$.   The fugacity $t$ keeps track of the R-charge.  The third and fourth terms correspond to the bifundamentals between nodes $1$ and $2$, $1$ and $3$, $1$ and $4$.  We also impose the condition $\Phi_{22}= \Phi_{33}= \Phi_{44}=0$, so that there is no contribution of these fields to the above Hilbert series. The terms with minus signs correspond to the relations \eref{branchtoric}.

We can obtain the Hilbert series of $\BC^3/(\BZ_2 \times \BZ_2)$, with the orbifold actions $(1,0,1)$ and $(0,1,1)$, by setting $r_2=r_3=r_4 \equiv \Delta$, so that all of the bifundamentals have equal R-charge $\Delta$, and integrating over the Haar measures of $\U(2) \times \U(1)^3$:
 \begin{equation}
 \begin{split}
&H[\BC^3/(\BZ_2 \times \BZ_2)](t)  \\
&=  \oint_{|z|=1} \frac{d z}{2\pi i z} (1-z^2) \left( \prod_{i=1}^4 \oint_{|u_i|=1} \frac{d u_i}{2 \pi i u_i} \right)H[\fflat](t; u_1, \ldots, u_4, z)|_{r_2=r_3=r_4=\Delta} \\
&= \PE \left[3 t^{4 \Delta} + t^{6 \Delta} - t^{12 \Delta} \right]~.
\end{split}
 \end{equation}

\subsubsection*{CS levels $(k,-2k,0,0)$}
Let us set
\begin{equation}
\begin{split}
&R[Q_{12}]=R[Q_{21}]=r_2= 1-\frac{1}{2} \Delta~, \\
&R[Q_{13}]=R[Q_{31}]=r_3= \frac{1}{4} \Delta~, \\ 
&R[Q_{14}]=R[Q_{41}]=r_4= \frac{1}{4}\Delta~.
\end{split}
\end{equation}
The Hilbert series for the required CY$_4$ can be computed by taking into account the contribution of the monopole operators:
\begin{equation}
\begin{split}
H(t;\Delta)= &\sum_{m \in \BZ} \oint_{|z|=1} \frac{d z}{2\pi i z} (1-z^2) \left( \prod_{i=1}^4 \oint_{|u_i|=1} \frac{d u_i}{2 \pi i u_i} \right) u_1^{-2km} u_2^{2km}\, \cdot \\
&\cdot H[\fflat](t; u_1, \ldots, u_4, z)|_{r_2 = 1-\frac{1}{2}\Delta, r_3 = \frac{1}{4}\Delta, r_4= \frac{1}{4}\Delta}   \\
=&\  \frac{1+ 3 t^{2-\Delta/2}+3 t^2+t^{4-\Delta/2}}{\left(1-t^{2-\Delta/2}\right)^3 \left(1-t^{\Delta }\right)}~,
\end{split}
\end{equation}
for $k=1$. With this CS assignment the CY$_4$ is not a complete intersection.

\subsubsection*{CS levels $(k,0,-k,-k)$}
Let us set
\begin{equation}
\begin{split}
&R[Q_{12}]=R[Q_{21}]=r_2=1-2 \Delta~, \\ 
&R[Q_{13}]=R[Q_{31}]=r_3= \Delta~, \\ 
&R[Q_{14}]=R[Q_{41}]=r_4= \Delta~.
\end{split}
\end{equation}
The Hilbert series for the corresponding CY$_4$ is given by
\begin{equation}
\begin{split}
H(t;\Delta)= &\sum_{m \in \BZ} \oint_{|z|=1} \frac{d z}{2\pi i z} (1-z^2) \left( \prod_{i=1}^4 \oint_{|u_i|=1} \frac{d u_i}{2 \pi i u_i} \right) \, \cdot\\
& \cdot H[\fflat](t; u_1, \ldots, u_4, z)|_{r_2 = 1-2\Delta, r_3 = \Delta, r_4=\Delta}  u_1^{-2km} u_3^{km}  u_4^{km} \\
=&\ \frac{(1-t^2) (1+t^{2-2\Delta})}{(1-t^{2 \Delta })^2 (1-t^{2-2\Delta })^3}\\ =&\ \PE \left[ 2t^{2\Delta} + 4t^{2-2\Delta} - t^2 -  t^{4-4\Delta} \right]~,
\end{split}
\end{equation}
for $k=1$. This is a four complex dimensional complete intersection.

\subsection{The model in Section \ref{genlaufer}}
\label{appC3}

We take the R-charges of the quiver fields to be
\begin{equation}
R[Q_{12}]= R[Q_{21}] =1-\Delta~, \ R[\Phi_{11}] =2\Delta~, \ R[\Phi_{22}] =\Delta~,\ R[\Psi_{22}] =1-\frac{1}{2}\Delta~.
\end{equation}
We focus on the branch on which
\begin{equation} \label{branchgenLaufer}
\Phi_{11}=\Tr(\Phi_{22}^2)~, \ \Tr(\Phi_{22}) = \Tr(\Psi_{22}) =0~,\ \Tr(\Phi_{22} \Psi_{22}) =0~, \ \Tr(Q_{12}Q_{21}) =0~.
\end{equation}
The Master space Hilbert series is
\begin{equation}
\begin{split}
H[\fflat](t; z, u) =& \PE \Big[\left( u+\frac{1}{u} \right) \left(z+\frac{1}{z}\right) t^{1-\Delta}+\left(z^2+\frac{1}{z^2}+1\right) t^\Delta \ + \\
& + \left(z^2+\frac{1}{z^2}+1\right) t^{1-\frac{\Delta}{2}}  -t^{2 \left(1-\frac{\Delta}{2}\right)}-t^{2 (1-\Delta)} -t^{\Delta+(1-\frac{\Delta}{2})} \Big]~.
\end{split}
\end{equation}
Integrating over the Haar measure of $\U(1) \times \SU(2)$, we obtain the Hilbert series of the corresponding CY$_3$:
\begin{equation} \label{CY3HS}
\begin{split}
H(t;\Delta) =&\oint_{|z|=1} \frac{d z}{2\pi i z} (1-z^2) \oint_{|u|=1} \frac{d u}{2 \pi i u} H[\fflat](t; z, u)\\
=& \PE \Big[ t^{2-\Delta} + t^{2\Delta}+t^{3-\frac{5}{2} \Delta}+ t^{3-\frac{3}{2} \Delta} -t^{6-3\Delta} \Big]~.
\end{split}
\end{equation}
The Hilbert series for the required CY$_4$ can be computed by taking into account the contribution of the monopole operators:
\begin{equation} \label{CY4HS}
H(t;\Delta) = \sum_{m \in \BZ} \oint_{|z|=1} \frac{d z}{2\pi i z} (1-z^2) \oint_{|u|=1} \frac{d u}{2 \pi i u} u^{-2km} H[\fflat](t; z, u)\ ,
\end{equation}
which equals
\begin{equation}
{\small \frac{1-t^{1-\frac{\Delta }{2}}+t^{3-\frac{3 \Delta }{2}}+t^{3-\frac{5 \Delta }{2}}-2 t^{5-\frac{7 \Delta }{2}}-t^{7-\frac{9 \Delta }{2}}+2 t^{2-\Delta }-t^{4-2 \Delta }-t^{4-3 \Delta }+t^{6-4 \Delta }}{\left(1-t^{3-{5 \Delta }/{2}}\right)^2 \left(1-t^{2-\Delta }\right) \left(1-t^{1-{\Delta }/{2}}\right) \left(1-t^{2 \Delta }\right)}}~,
\end{equation}
for $k=1$. The CY$_4$ is not a complete intersection.

\subsection{The model in Section \ref{case1}}
\label{appC4}

We will now explain why we are not able to provide the $N=1$ Hilbert series of the model in Section \ref{case1} with free energy \eref{FS3case1}. As in other cases, in order for this technique to work, we would first have to obtain the Master space Hilbert series, where the gauge group is taken to be $\U(1) \times \U(2)$, that reproduces correctly the Hilbert series of the corresponding CY$_3$.  The latter series is given in the third line of Table \ref{tab:4dmodels}.  As can be seen there, there is a generator of the moduli space with R-charge $2-\Delta$.  This generator must take the form of $Q_{12} \Phi_{22} \Phi_{22} Q_{21}$, with an appropriate contraction of the gauge indices.  Without loss of generality, we may focus on the traceless part of $\Phi_{22}$, namely $\hat{\Phi}_{22} = \Phi_{22} - \frac{1}{2} \Tr(\Phi_{22}) \mathbf{1}$. There are two possible gauge-invariant combinations one can consider:
\begin{equation} \label{twocombicase1}
(Q_{12})_\alpha \tr (\hat{\Phi}^2_{22}) (Q_{21})^\alpha~, \quad  (Q_{12})_\alpha (\hat{\Phi}_{22})^\alpha_\beta (\hat{\Phi}_{22})^\beta_\gamma  (Q_{21})^\gamma~,
\end{equation}
where $\alpha, \beta, \gamma =1,2$ are the $\SU(2) < \U(2)$ gauge indices.  Obviously, the first gauge-invariant combination cannot be a generator of the moduli space because it is constructed from a product of two gauge invariants with lower R-charges.  Let us consider the second one.  For a two-by-two traceless matrix $\hat{\Phi}_{22}$, we have
\begin{equation}
(Q_{12})_\alpha (\hat{\Phi}_{22})^\alpha_\beta (\hat{\Phi}_{22})^\beta_\gamma  (Q_{21})^\gamma= \frac{1}{2} (Q_{12})_\alpha (Q_{21})^\alpha \tr(\hat{\Phi}_{22}^2)\ ,
\end{equation}
but this turns out to be proportional to the first combination in \eref{twocombicase1}, so it cannot be a generator either.  The main idea is that when the gauge group is taken to be $\U(1) \times \U(2)$ we do not have access to certain generators of the required CY$_3$.  In other words, we cannot construct the required CY$_3$ moduli space of the given quiver description with gauge group $\U(1) \times \U(2)$.

Nevertheless, the Hilbert series for the corresponding CY$_3$ (i.e. the large-$N$ Hilbert series, from the field theory perspective) may be obtained using the path algebra associated with the quiver described in Section \ref{case1}, using a method proposed by \cite{ginzburg, BOCKLANDT200814} (see also \cite[Theorem 6.2]{Eager:2010yu}).  This produces the $H(t;\Delta)$ in the third line of Table \ref{tab:4dmodels}, which we can rewrite for convenience as
\begin{equation} \label{HS43}
H(t;\Delta) = \frac{1-t^{4-\Delta}}{(1-t^{2-\Delta/2})(1-t^{2-\Delta})(1-t^{2-3\Delta/2})(1-t^{2\Delta})}\ .
\end{equation}
We see that there are four generators satisfying a single relation of R-charge $4-\Delta$. Therefore the CY$_3$ is a complete intersection defined by one equation. Remembering \eqref{eq:Rtwonode}, we may write down the four generators associated with closed paths in the quiver which carry R-charges $2-\frac{1}{2}\Delta , 2-\Delta, 2- \frac{3}{2} \Delta, 2\Delta$ respectively, as
\begin{equation}
x \sim Q_{12} \Phi_{22}^3 Q_{21}\ , \quad y \sim Q_{12} \Phi_{22}^2 Q_{21}\ , \quad z \sim Q_{12} \Phi_{22} Q_{21}\ , \quad w \sim \Phi_{11}\ .
\end{equation}
From the Hilbert series \eref{HS43}, we see that these generators satisfy a single polynomial constraint (with charge $4-\Delta$) of the form\footnote{Indeed these are the only two monomials of the form $x^\alpha y^\beta w^\gamma z^\delta$ with $\{\alpha,\beta,\gamma,\delta\} \in \mathbb{N}_0$ such that $\alpha R[x] +\beta R[y] +\gamma R[w]+\delta R[t] = 4-\Delta$ with $0<\Delta \leq 1$ (i.e. requiring positive R-charges of the corresponding gauge-invariants).}
\begin{equation} \label{constr43}
x^2 + wz^2 = 0\ \subset\ \cc^4\ ,
\end{equation}
where the relative coefficient between the two terms can be absorbed upon redefining $x$, $w$ or $z$.  This is a non-isolated singularity, given the gradient of the above equation vanishes along the codimension-two locus $x=z=0$ inside $\cc^4$. It is worth pointing out that the generator $y$ does not appear in the constraint \eref{constr43}. When the ambient space is $\cc^3$ rather than $\cc^4$  (i.e. neglecting $y$ altogether), the surface is a well-known (non-normal) singularity known as \emph{Whitney umbrella} (see e.g. \cite{Collinucci:2008pf} for a physics application).

Finally we remark that, starting from \eref{HS43}, it is not possible to obtain the Hilbert series of the corresponding CY$_4$ in question using the method adopted in earlier sections.  The reason is that for such a method to work we would need to start from the Master space Hilbert series (i.e. for $N=1$) which, after integrating over the gauge fugacities, would be given by \eref{HS43}; cf. \eref{CY3HS}.  To obtain the Hilbert series of the CY$_4$ in question, one also needs to introduce a factor which is a contribution from the monopole operators, integrate over gauge fugacities, as well as sum over the magnetic fluxes; cf. \eref{CY4HS}. As can be seen from the previous subsection, one cannot obtain \eref{CY4HS} given \eref{CY3HS} if one does not consider the precise contribution from the dressed monopoles.

\subsection{The model in Section \ref{case2}}
\label{appC5}
Consider the 4d model with three gauge groups of Section \ref{case2}. Take $N=1$. 
We take the R-charges of the quiver fields to be
\begin{equation}
\begin{split}
&R[{Q_{21}}] = R[{Q_{12}}]=R[{Q_{32}}] = R[{Q_{23}}]
= 
1-\Delta, \\
&R[{\Phi_{11}}] = R[{\Phi_{33}}]  = 2 \Delta~, \quad R[{\Phi_{22}}] = \Delta~.
\end{split}
\end{equation} 
We focus on the branch of the moduli space on which
\begin{equation} \label{branch3nodes}
\begin{split}
&\Phi_{11}=\Phi_{33}=\Tr(\Phi_{22})=0~,  \quad \Tr(Q_{12}Q_{21}) = \Tr( Q_{23}Q_{32})=0~, \\
& \Tr(Q_{12}\Phi_{22}Q_{21}) = \Tr(Q_{32}\Phi_{22}Q_{23})~.
\end{split}
\end{equation}
The Master space Hilbert series is
\begin{equation}
\begin{split}
H[\fflat](t;z,u_1,u_2,u_3)
= &\PE \Big[-t^{2(1-\Delta)+\Delta }-2 t^{2(1- \Delta) }-t^{\Delta } \ + \\
& +  \left(\frac{u_1}{u_2}+\frac{u_2}{u_1}\right) \left(z+\frac{1}{z}\right) t^{1-\Delta }+\left(\frac{u_2}{u_3}+\frac{u_3}{u_2}\right) \left(z+\frac{1}{z}\right) t^{1-\Delta }\ + \\ &+ \left(z^2+\frac{1}{z^2}+2\right)t^{\Delta } \Big]~.
\end{split}
\end{equation}
The Hilbert series for the corresponding CY$_3$ is then
\begin{equation}
\begin{split}
H(t;\Delta) = & \oint_{|z|=1} \frac{d z}{2\pi i z} (1-z^2) \left( \prod_{i=1}^3 \oint_{|u_i|=1} \frac{d u_i}{2 \pi i u_i} \right) H[\fflat](t;z,u_1,u_2,u_3) \\
=&\PE\left[ t^{4 (1-\Delta )}+t^{2-\Delta }+t^{2 \Delta }+t^{4-3 \Delta } -t^{2(4-3 \Delta) } \right]~.
\end{split} 
\end{equation}
Since this is a four complex dimensional complete intersection, let us write down explicitly the generators and the relation of the moduli space.  For convenience, we denote
\begin{equation}
\begin{split}
(\hat{P}_1)^\beta_{\alpha} &\equiv (Q_{12})_\alpha (Q_{21})^\beta -\frac{1}{2}  (Q_{12})_\gamma (Q_{21})^\gamma \delta^\beta_\alpha  ~,\\
(\hat{P}_3)^\beta_{\alpha} &\equiv (Q_{32})_\alpha (Q_{23})^\beta -\frac{1}{2}  (Q_{32})_\gamma (Q_{23})^\gamma \delta^\beta_\alpha~,\\
\hat{\Phi} &\equiv \Phi_{22} - \frac{1}{2} \Tr(\Phi_{22}) \mathbf{1}_2~.
\end{split}
\end{equation}
where $\alpha, \beta,\gamma=1,2$ are $\SU(2) < \U(2)$ gauge indices.  Thanks to \eref{branch3nodes}, they also satisfy 
\begin{equation}
\Tr(\hat{\Phi} \hat{P}_1) = \Tr(\hat{\Phi} \hat{P}_3)~.
\end{equation}
The generators of the moduli space are therefore:
\begin{equation}
\begin{split}
t^{4-4\Delta}:  &\quad W=\Tr(\hat{P}_1 \hat{P}_3)\ ; \\
t^{2-\Delta }:  &\quad Y=\Tr(\hat{\Phi} \hat{P}_1)= \Tr(\hat{\Phi} \hat{P}_3)\ ; \\
t^{2\Delta }:  &\quad Z=\Tr(\hat{\Phi}^2)\ ; \\
t^{4-3 \Delta }: &\quad X=\Tr(\hat{P}_1 \hat{\Phi} \hat{P}_3)\ .
\end{split}
\end{equation}
We may redefine the variables $(W,Y, Z,X)$ to $(w,y,z,x)$ by appropriate numerical factors as in \eref{redefgenLaufer}.  The relation can then be written as
\begin{equation}\label{eq:case2CY3}
x^2+wy^2+w^2 z =0~.
\end{equation}

\subsubsection*{Adding CS interactions}
Now we turn on the CS levels $\vec k=(k_1, k_2, k_3)$ of the gauge groups $\U(1) \times \U(2) \times \U(1)$ respectively.  The Hilbert series for the corresponding CY$_4$ is given by
\begin{align} \label{HSkmkk}
H(t;\Delta) =&\sum_{m \in \BZ} \oint_{|z|=1} \frac{d z}{2\pi i z} (1-z^2) \left( \prod_{i=1}^3 \oint_{|u_i|=1} \frac{d u_i}{2 \pi i u_i} \right)  u_1^{k_1m} u_2^{2k_2m} u_3^{k_3m} \, \cdot \nonumber \\
&\cdot H[\fflat](t;z,u_1,u_2,u_3) \\
=& 
\begin{cases}
\frac{t^{2-\Delta }+3 t^{4-3 \Delta }+t^{4-4 \Delta }-t^{6-4 \Delta }-3 t^{6-5 \Delta }-t^{8-7 \Delta }-t^{10-8 \Delta }+1}{\left(1-t^{4 (1-\Delta )}\right)^2 \left(1-t^{2-\Delta }\right)^2 \left(1-t^{2 \Delta }\right)}~, & \vec k =(2,-1,0)~;\\
\PE\Big[ 3t^{2-\Delta }+ 2t^{2 (1-\Delta )}+ t^{2\Delta}-t^{4-3 \Delta } -t^{2(2-\Delta)}  \Big]~,  & \vec k =(1,-1,1) \ ;\\
\PE\Big[ 3t^{2-\Delta }+ 2t^{2 (1-\Delta )}+ t^{2\Delta}-t^{4-3 \Delta } -t^{2(2-\Delta)}  \Big]~,  & \vec k =(1,0,-1) \ .
\end{cases} \nonumber
\end{align}
Observe that the last two cases yield the same result. With the first CS assignment, the resulting CY$_4$ is not a complete intersection (whereas the last two cases are).

\paragraph{CS levels $(1,0,-1)$.}
Let us examine the generators and the moduli space of the CY$_4$ for the case $(1,0,-1)$.  Let us denote the monopole operator with flux $(m;m ,m;m)$ under $\U(1)_{1} \times \U(2)_{0} \times \U(1)_{-1}$ by $V_m$.  It carries zero R-charge, topological charge $m$, and charges $(-m,0,m)$ under the $\U(1) \times \U(1) \times \U(1)$ gauge symmetry, where the second $\U(1)$ is a subgroup of the $\U(2)$ gauge group.  The monopole operator satisfies the quantum relation
\begin{equation}
V_{+1} V_{-1} =1~.
\end{equation}
The generators of the moduli space are
\begin{equation}\label{eq:genCS1}
\begin{array}{llll}
t^{2-2\Delta}: &\quad \Xi_+ = V_{+1} Q_{12} Q_{23}~, & \quad & \quad \Xi_- = V_{-1} Q_{32} Q_{21} \ ;\\
t^{2-\Delta}: &\quad \Upsilon_+=V_{+1} Q_{12} Q_{22} Q_{23}~, & \quad Y~, &\quad  \Upsilon_-=V_{-1} Q_{32} Q_{22} Q_{21} \ ;\\
t^{2\Delta}: &\quad Z\ , \quad &\quad   &\quad
\end{array}
\end{equation}
where the $\SU(2) < \U(2)$ are contracted appropriately.  The relation at order $t^{4-3\Delta}$ can be written as
\begin{equation} \label{rel1}
 \Upsilon_+ \Xi_- = \Upsilon_- \Xi_+~.
\end{equation}
The relation at order $t^{4-2\Delta}$ can be written as
\begin{equation} \label{rel2}
  \Upsilon_+ \Upsilon_- + Z\Xi_+ \Xi_- + Y^2= 0 ~.
\end{equation}
Note that $W$ and $X$ satisfies the following relation
\begin{equation} \label{relWX}
W=  \Xi_+ \Xi_-~, \quad X= \Upsilon_+ \Xi_- = \Upsilon_- \Xi_+
\end{equation}
so they are composite of the dressed monopole operators.

\paragraph{CS levels $(1,-1,1)$.}
Now let us consider the case of $(1,-1,1)$.  Let us denote the monopole operator with flux $(m;m ,m;m)$ under $\U(1)_{1} \times \U(2)_{-1} \times \U(1)_{1}$ by $V_m$.  It carries zero R-charge, topological charge $m$, and charges $(-m,2m,-m)$ under the $\U(1) \times \U(1) \times \U(1)$ gauge symmetry, where the second $\U(1)$ is a subgroup of the $\U(2)$ gauge group.  The monopole operator satisfies the quantum relation
\begin{equation}
V_{+1} V_{-1} =1~.
\end{equation}
The generators of the moduli space are
\begin{equation}\label{eq:genCS2}
\begin{array}{llll}
t^{2-2\Delta}: &\quad \Xi_+ = V_{+1} Q_{12} Q_{32}~, & \quad & \quad \Xi_- = V_{-1} Q_{23} Q_{21}\ ; \\
t^{2-\Delta }: &\quad \Upsilon_+=V_{+1} Q_{12} Q_{22} Q_{32}~, & \quad Y~, &\quad  \Upsilon_-=V_{-1} Q_{23} Q_{22} Q_{21}\ ; \\
t^{2\Delta} : & \quad Z\ ,&\quad  &\quad \\
\end{array}
\end{equation}
where the $\SU(2) < \U(2)$ are contracted appropriately.  The relations \eref{rel1}, \eref{rel2} and \eref{relWX} still hold.

\subsection{The model in Section \ref{sub:laufer-n}}
\label{appc5}

We consider the gauge group $\U(1) \times \U(2)$, and take the R-charges of the quiver fields to be
\begin{equation}
\begin{split}
&R[A]= \frac{n}{n+1}~, \quad R[B]= \frac{n}{n+1}~, \\
& R[\Phi_{11}] = \frac{2}{n+1}~, \quad R[\Phi_{22}] = \frac{1}{n+1}~, \quad R[\Psi_{22}] = \frac{2n+1}{2n+2}~.
\end{split}
\end{equation}
In the above, we have taken $R[A]=R[B]$.  In fact, one may assume that $R[A]=\frac{1}{4}r$ and $R[B]=\frac{2n}{n+1}-\frac{1}{4}r$ and perform the volume minimization in the same way as described around \eref{volfuncLaufer} and \eref{volmin}.  As a result, one obtains $r= \frac{4n}{n+1}$ as expected.

We focus on the same branch of the moduli space similar to that described in \eref{branchgenLaufer}:
\begin{equation}
\begin{array}{lll}
\Phi_{11} =0~,  &\quad \Tr(\Phi_{22})=\Tr(\Psi_{22})=0~, &\quad \Tr P+ [\Tr(\hat{\Phi}_{22}^{2})]^n = 0~, \\
&\quad \Tr(\hat{\Phi}_{22} \hat{P}) + \Tr(\hat{\Psi}_{22}^2) =0~, &\quad \Tr(\hat{\Phi}_{22} \hat{\Psi}_{22})=0~,
\end{array}
\end{equation}
where $P$ is defined as in \eref{defP} and the hat denotes the traceless part of a given matrix. The Hilbert series of the Master space is
\begin{equation}
\begin{split}
H[\fflat](t; z, u) = &\PE \Big[ \left( u+\frac{1}{u} \right) \left(z+\frac{1}{z}\right) t^{\frac{n}{n+1}}+\left(z^2+\frac{1}{z^2}+1\right) t^{\frac{1}{n+1}}\ + \\
& +\left(z^2+\frac{1}{z^2}+1\right) t^{\frac{2 n+1}{2 n+2}}-t^{\frac{2 n+1}{n+1}} -t^{2 \left(\frac{n}{n+1}\right)}-t^{\frac{2 n+1}{2 n+2}+\frac{1}{n+1}} \Big]\ .
\end{split}
\end{equation}
The Hilbert series of the corresponding CY$_3$ is
\begin{equation}\label{eq:HSlaufer-n}
\begin{split}
H[\text{Laufer-$n$}](t)= &\oint_{|z|=1} \frac{d z}{2\pi i z} (1-z^2) \oint_{|u|=1} \frac{d u}{2 \pi i u} H[\fflat](t; z, u)\\
=& \PE \Big[ t^{\frac{2}{n+1}} +t^{\frac{2n+1}{n+1}}+ t^{\frac{6n+1}{2n+2}}+t^{\frac{6n+3}{2n+2}}-t^{\frac{6n+3}{n+1}}\Big]\ .
\end{split}
\end{equation}
Each positive term in the PE corresponds to the generators $w$, $y$, $z$, $x$, in the same way as described in \eref{genLaufer} and \eref{redefgenLaufer}, respectively.  The relation can be written as
\begin{equation}
x^2+y^3+wz^2+w^{2n+1} y =0~.
\end{equation}
Let us now turn on the CS levels $(2k,-k)$ to the gauge groups $\U(1)\times \U(2)$.  Taking into account the contribution of the monopole operators, the Hilbert series of the corresponding CY$_4$ is
\begin{equation}
H[\text{Laufer-$n$}_{k}](t) = \sum_{m \in \BZ} \oint_{|z|=1} \frac{d z}{2\pi i z} (1-z^2) \oint_{|u|=1} \frac{d u}{2 \pi i u} u^{-2km} H[\fflat](t; z, u)\ ,
\end{equation}
which gives
\begin{equation}
\frac{1+t^{\frac{2 n+1}{n+1}}-t^{\frac{6 n+2}{n+1}}-t^{\frac{8 n+3}{n+1}}+t^{\frac{6 n+1}{2 n+2}}+3 t^{\frac{6 n+3}{2 n+2}}-3 t^{\frac{10 n+3}{2 n+2}}-t^{\frac{10 n+5}{2 n+2}}}{\left(1-t^{\frac{2}{n+1}}\right) \left(1-t^{\frac{2 n+1}{n+1}}\right)^2 \left(1-t^{\frac{6 n+1}{2 n+2}}\right)^2}
\end{equation}
for $k=1$.

\bibliographystyle{ytphys}
\bibliography{ref}

\end{document}